\PassOptionsToPackage{x11names}{xcolor}
\documentclass[nolinenum]{jfp}
\usepackage[utf8]{inputenc}
\usepackage{graphicx}
\usepackage{idris2}
\usepackage{amssymb}
\usepackage{multicol}
\usepackage{subcaption}
\usepackage{natbib}
\usepackage{array}
\usepackage{amsmath}

\newcommand\ignore[1]{}

\newcommand\symbolendofaside{\raisebox{.02\baselineskip}{\includegraphics[height=.45\baselineskip]{endofproofwd.pdf}}}

\newenvironment{aside}{\paragraph*{Aside.} }{\hfill\symbolendofaside\par}
\newcommand\figref[1]{Fig.~\ref{#1}}

\begin{document}

\journaltitle{JFP}
\cpr{Cambridge University Press}
\jnlDoiYr{2023}

\doival{10.1017/xxxxx}

\totalpg{\pageref{lastpage01}}

\title{Idris TyRE: a dependently typed regex parser}
\begin{authgrp}
  \author{Ohad Kammar}
  \author{Katarzyna Marek}
  \affiliation{University of Edinburgh, Scotland\\
    \email{\{kmarek,ohad.kammar\}@ed.ac.uk}}
\end{authgrp}

\label{firstpage}

\begin{abstract}
Regular expressions --- regexes --- are widely used not only for
validating, but also for parsing textual data. Generally, regex
parsers output a loose structure, e.g. an unstructured list of
matches, leaving it up to the user to validate the output's properties
and transform it into the desired structure. Since the regex itself
carries information about the structure, this design leads to unnecessary
repetition.

Radanne introduced \emph{typed regexes} --- TyRE --- a type-indexed
combinator layer that can be added on top of an existing regex
engine. We extend Radanne's design, and implement a parser which
maintains type-safety throughout all layers: the user-facing regexes;
their internal, desugared, representation; its compiled finite-state
automaton; and the automaton's associated instruction-set for constructing the
parse-trees. We implemented TyRE in the
dependently-typed language Idris~2.
\end{abstract}

\maketitle
\section{Introduction}
Regular expressions are an old and well researched subject. They have
become a standard way to validate or parse textual data for a class of
use cases. Yet, in statically typed languages regex parsers do not
have the same type-induced safety as handwritten parsers or combinator-based
parsing may offer. Even though the pattern --- the regex --- is known
at compile time, the type system takes no advantage of the information
contained in the regex to constrain the result. For example, take the
regex parser in the standard library of the programming language Scala
2. Parsing a string using a regex results in a tuple of matches, which
the user can reshape by pattern matching. However, it is up to the
user to make sure that the tuple multiplicity is correct, even though
all the necessary information to check this multiplicity is provided
at compile time.

Here we describe a regex parser which produces a result of a
pattern-dependent type. This design eliminates errors such as tuple
multiplicity mistakes. As with parser combinators, the shape of the
result reflects the structure of the parser. This typing provides a
level of safety both for correctness of the parser itself and
correctness of the pattern. Additionally, the user can write
result-data transformations, as with combinators, to obtain the
desired data structure, without any repetition of the information
contained in the pattern. But unlike parser combinator designs, the
parser is guaranteed to be linear in time and space complexity
with respect to the length of the parsed string.

This work is based on Radanne's \citeyearpar{Radanne2019TypedEngines}
design of a typed regex parser. Radanne takes advantage of an existing
regex engine and adds a typed layer on top of it. This layer adds
a level of safety and allows for convenient data transformation. He
implements this design in the OCaml programming language~\citep{ocaml} and uses
Vouillon's regex engine\footnote{J\'er\^ome Vouillon's \texttt{ocaml-re} is available from \url{https://github.com/ocaml/ocaml-re} .}.

Idris~2~\citep{brady:idris2} is a new strict dependently-typed programming
language. Idris~2 does not have a standard regex
parser library, providing an opportunity to revisit Radanne's
design. The new design maintains type safety throughout all the
layers, which makes the type constraints become an additional safety
check for the parser itself.

We make the following contributions:
\begin{itemize}
\item a typed instruction language for a Moore machine;
\item a compilation scheme into this machine language accompanying
  Thompson's regex-to-non-deterministic finite-state automaton
  construction;
\item a user-facing regex literal layer that uses a type-level parser; and
\item a use-case stream editing library --- \emph{sedris}
  --- that uses the new TyRE library for its search and replace
  operations.
\end{itemize}
The source-code for both the regex\footnote{TyRE is available
  from \url{https://github.com/kasiamarek/tyre} .} and stream editing\footnote{Sedris is available from
  \url{https://github.com/kasiamarek/sedris} .} libraries is publicly available.

We proceed as follows.
Sec.~\ref{lib_overview} briefly presents the core functionality of the
library.
Sec.~\ref{architecture_overview} overviews our design and how its
implementation layers fit.
Sec.~\ref{typed_regexes} describes the regex layer and its
implementation in detail.
Sec.~\ref{moore_machine} describes the automata-based parser and its
Moore-machine typed instruction language.
Sec.~\ref{evaluation} evaluates the library: while space and time
complexity is linear w.r.t~the length of the parsed stream, the size
of the regex literal and its structure make the size of the parser
scale non-linearly. We measure TyRE's performance compared with
Idris~2's parser combinator library on discriminating, but
pathological, examples. We also describe our experience in using TyRE
when developing the stream editing library sedris.
Sec.~\ref{related_work} surveys immediately
related work. Sec.~\ref{conclusion} concludes.

\begin{aside}
  Idris~2 is a relatively new language. To make this manuscript
  self-contained but brief, we explain Idris's syntax
  and semantics as the need arises, formatted like so.
\end{aside}

\ignore{
\begin{code}
\IdrisKeyword{import}\KatlaSpace{}\IdrisModule{Data.Regex}\KatlaNewline{}
\IdrisKeyword{import}\KatlaSpace{}\IdrisModule{Data.Stream}\KatlaNewline{}
\IdrisKeyword{import}\KatlaSpace{}\IdrisModule{Data.SortedSet}\KatlaNewline{}
\IdrisKeyword{import}\KatlaSpace{}\IdrisModule{Data.List.Elem}\KatlaNewline{}
\IdrisKeyword{import}\KatlaSpace{}\IdrisModule{Sedris}\KatlaNewline{}
\end{code}
}

\ignore{
\begin{code}
\IdrisKeyword{import}\KatlaSpace{}\IdrisModule{Data.Regex}\KatlaNewline{}
\IdrisKeyword{import}\KatlaSpace{}\IdrisModule{Data.Stream}\KatlaNewline{}
\end{code}
}

\section{Library overview}
\label{lib_overview}
TyRE provides two kinds of constructors for typed regexes: string
literals, which implicitly define the shape of the output parse tree;
and combinator-style regex constructors, that give a fine-grained
control over the shape of the parse tree. To see TyRE's string
literals in action, we'll look at a regex for time-stamps in the
format \textit{hh}:\textit{mm}.

There are two related ways to use regexes:
\begin{itemize}
\item \emph{matching}: determining whether the string matches the regex:
\ignore{
\begin{code}
\IdrisKeyword{namespace}\KatlaSpace{}\IdrisNamespace{Hidden}\KatlaNewline{}
\end{code}
}
\begin{code}
\KatlaSpace{}\KatlaSpace{}\IdrisFunction{match}\KatlaSpace{}\IdrisKeyword{:}\KatlaSpace{}\IdrisType{TyRE}\KatlaSpace{}\IdrisBound{a}\KatlaSpace{}\IdrisKeyword{\KatlaDash{}>}\KatlaSpace{}\IdrisType{String}\KatlaSpace{}\IdrisKeyword{\KatlaDash{}>}\KatlaSpace{}\IdrisType{Bool}\KatlaNewline{}
\end{code}
\item \emph{parsing}: returning a parse-tree when the string matches the regex:
\ignore{
\begin{code}
\IdrisKeyword{namespace}\KatlaSpace{}\IdrisNamespace{Hidden}\KatlaNewline{}
\end{code}
}
\begin{code}
\KatlaSpace{}\KatlaSpace{}\IdrisFunction{parse}\KatlaSpace{}\IdrisKeyword{:}\KatlaSpace{}\IdrisType{TyRE}\KatlaSpace{}\IdrisBound{a}\KatlaSpace{}\IdrisKeyword{\KatlaDash{}>}\KatlaSpace{}\IdrisType{String}\KatlaSpace{}\IdrisKeyword{\KatlaDash{}>}\KatlaSpace{}\IdrisType{Maybe}\KatlaSpace{}\IdrisBound{a}\KatlaNewline{}
\end{code}
\end{itemize}
The type of the result parse-tree, \IdrisBound{a} in both cases, is
only relevant for parsing, not matching.

\begin{aside}
Idris expects type declarations for all top-level
definitions, such as \IdrisFunction{match} and
\IdrisFunction{parse}. If you're reading this manuscript in colour, we
use semantic highlighting distinguishing between:
\IdrisFunction{defined} functions or constants;
\IdrisBound{bound/binding} occurrences of variables;
\IdrisType{type constructors}; and
\IdrisData{data constructors} (not shown).
\end{aside}

The following code defines a regex \emph{literal}: a string encoding
of the regular expression, in a variation on the standard notation
present in Unix-like systems:
\begin{code}
\KatlaSpace{}\KatlaSpace{}\KatlaSpace{}\KatlaSpace{}\KatlaSpace{}\KatlaSpace{}\KatlaSpace{}\KatlaSpace{}\KatlaSpace{}\KatlaSpace{}\KatlaSpace{}\KatlaSpace{}\KatlaSpace{}\KatlaSpace{}\KatlaSpace{}\KatlaSpace{}\KatlaSpace{}\KatlaSpace{}\IdrisComment{\KatlaDash{}\KatlaDash{}\KatlaSpace{}(range\KatlaSpace{}from\KatlaSpace{}0\KatlaSpace{}to\KatlaSpace{}9)}\KatlaNewline{}
\IdrisFunction{time1}\KatlaSpace{}\IdrisKeyword{:}\KatlaSpace{}\IdrisType{TyRE}\KatlaSpace{}\IdrisType{Unit}\KatlaSpace{}\IdrisComment{\KatlaDash{}\KatlaDash{}\KatlaSpace{}v\KatlaSpace{}\KatlaSpace{}\KatlaSpace{}\KatlaSpace{}\KatlaSpace{}\KatlaSpace{}\KatlaSpace{}\KatlaSpace{}\KatlaSpace{}\KatlaSpace{}\KatlaSpace{}\KatlaSpace{}\KatlaSpace{}v\KatlaDash{}(matches\KatlaSpace{}exactly\KatlaSpace{}`:')}\KatlaNewline{}
\IdrisFunction{time1}\KatlaSpace{}\IdrisKeyword{=}\KatlaSpace{}\IdrisFunction{r}\KatlaSpace{}\IdrisData{"(([01][0\KatlaDash{}9])|([2][0\KatlaDash{}3])):[0\KatlaDash{}5][0\KatlaDash{}9]"}\KatlaNewline{}
\IdrisComment{\KatlaDash{}\KatlaDash{}\KatlaSpace{}\KatlaSpace{}\KatlaSpace{}\KatlaSpace{}\KatlaSpace{}\KatlaSpace{}^\KatlaSpace{}\KatlaSpace{}\KatlaSpace{}\KatlaSpace{}\KatlaSpace{}^\KatlaSpace{}\KatlaSpace{}\KatlaSpace{}\KatlaSpace{}\KatlaSpace{}\KatlaSpace{}\KatlaSpace{}\KatlaSpace{}^(alternation\KatlaSpace{}of\KatlaSpace{}expressions)}\KatlaNewline{}
\IdrisComment{\KatlaDash{}\KatlaDash{}\KatlaSpace{}\KatlaSpace{}\KatlaSpace{}\KatlaSpace{}\KatlaSpace{}\KatlaSpace{}|\KatlaSpace{}\KatlaSpace{}\KatlaSpace{}\KatlaSpace{}\KatlaSpace{}(one\KatlaSpace{}of\KatlaSpace{}the\KatlaSpace{}characters\KatlaSpace{}'0'\KatlaSpace{}and\KatlaSpace{}'1'\})}\KatlaNewline{}
\IdrisComment{\KatlaDash{}\KatlaDash{}\KatlaSpace{}\KatlaSpace{}\KatlaSpace{}\KatlaSpace{}\KatlaSpace{}\KatlaSpace{}(smart\KatlaSpace{}constructor\KatlaSpace{}for\KatlaSpace{}typed\KatlaSpace{}regex\KatlaSpace{}from\KatlaSpace{}a\KatlaSpace{}string\KatlaSpace{}literal)}\KatlaNewline{}
\end{code}
\begin{aside}
Comments begin with a double-dash (\IdrisComment{--}).
\end{aside}

The comments point at the syntax with ASCII-art arrows
(\IdrisComment{\string^,v}). Regex literals match characters verbatim,
with a few key characters having a special role:
\begin{itemize}
\item Brackets (\IdrisData{[]}) enumerate alternatives, either through:
\begin{itemize}
\item exhaustive enumeration (\IdrisData{[01]}); or
\item a character range (\IdrisData{[0-9]}).
\end{itemize}
\item Pipes (\IdrisData{|}) specify alternation between two sub-regexes; and
\item Parentheses (\IdrisData{()}) help the regex-literal parser parse the
  literal.
\end{itemize}
To validate a chosen string against the regex, use \IdrisFunction{match}:
\begin{code}
\IdrisFunction{validateTimestamp}\KatlaSpace{}\IdrisKeyword{:}\KatlaSpace{}\IdrisType{String}\KatlaSpace{}\IdrisKeyword{\KatlaDash{}>}\KatlaSpace{}\IdrisType{Bool}\KatlaNewline{}
\IdrisFunction{validateTimestamp}\KatlaSpace{}\IdrisBound{str}\KatlaSpace{}\IdrisKeyword{=}\KatlaSpace{}\IdrisFunction{match}\KatlaSpace{}\IdrisFunction{time1}\KatlaSpace{}\IdrisBound{str}\KatlaNewline{}
\end{code}

Fig.~\ref{fig:structured time tyre} includes a \IdrisType{TyRE} for
parsing a timestamp into a more structured format: a pair of arbitrary
precision integer value representing the hours and minutes.
\begin{figure}
\begin{code}
\IdrisComment{\KatlaDash{}\KatlaDash{}\KatlaSpace{}\KatlaSpace{}\KatlaSpace{}\KatlaSpace{}\KatlaSpace{}\KatlaSpace{}\KatlaSpace{}\KatlaSpace{}\KatlaSpace{}\KatlaSpace{}\KatlaSpace{}v\KatlaDash{}\KatlaDash{}\KatlaDash{}\KatlaDash{}\KatlaDash{}\KatlaDash{}\KatlaDash{}\KatlaDash{}\KatlaDash{}\KatlaDash{}\KatlaDash{}\KatlaDash{}\KatlaDash{}\KatlaDash{}\KatlaDash{}\KatlaDash{}v\KatlaDash{}(the\KatlaSpace{}shape\KatlaSpace{}of\KatlaSpace{}the\KatlaSpace{}result\KatlaSpace{}parse\KatlaSpace{}tree)}\KatlaNewline{}
\IdrisFunction{time2}\KatlaSpace{}\IdrisKeyword{:}\KatlaSpace{}\IdrisType{TyRE}\KatlaSpace{}\IdrisKeyword{(}\IdrisType{Integer,}\KatlaSpace{}\IdrisType{Integer}\IdrisKeyword{)}\KatlaNewline{}
\IdrisFunction{time2}\KatlaSpace{}\IdrisKeyword{=}\KatlaNewline{}
\KatlaSpace{}\KatlaSpace{}\IdrisComment{\KatlaDash{}\KatlaDash{}\KatlaSpace{}\KatlaSpace{}(map\KatlaSpace{}:\KatlaSpace{}(a\KatlaSpace{}\KatlaDash{}>\KatlaSpace{}\KatlaSpace{}b)\KatlaSpace{}\KatlaDash{}>\KatlaSpace{}\KatlaSpace{}TyRE\KatlaSpace{}a\KatlaSpace{}\KatlaDash{}>\KatlaSpace{}\KatlaSpace{}TyRE\KatlaSpace{}b)}\KatlaNewline{}
\KatlaSpace{}\KatlaSpace{}\IdrisComment{\KatlaDash{}\KatlaDash{}\KatlaSpace{}\KatlaSpace{}v\KatlaSpace{}\KatlaSpace{}\KatlaSpace{}\KatlaSpace{}\KatlaSpace{}\KatlaSpace{}\KatlaSpace{}\KatlaSpace{}\KatlaSpace{}\KatlaSpace{}\KatlaSpace{}\KatlaSpace{}\KatlaSpace{}\KatlaSpace{}\KatlaSpace{}\KatlaSpace{}\KatlaSpace{}\KatlaSpace{}\KatlaSpace{}\KatlaSpace{}v\KatlaDash{}(TyRE's\KatlaSpace{}notation\KatlaSpace{}for\KatlaSpace{}a\KatlaSpace{}capture\KatlaSpace{}group)}\KatlaNewline{}
\KatlaSpace{}\KatlaSpace{}\KatlaSpace{}\KatlaSpace{}\KatlaSpace{}\KatlaSpace{}\IdrisFunction{map}\KatlaSpace{}\IdrisFunction{f}\KatlaSpace{}\IdrisKeyword{(}\IdrisFunction{r}\KatlaSpace{}\IdrisData{"([01][0\KatlaDash{}9])!"}\KatlaSpace{}\KatlaSpace{}\KatlaSpace{}\KatlaSpace{}\KatlaSpace{}\KatlaSpace{}\IdrisFunction{`or`}\KatlaSpace{}\IdrisFunction{r}\KatlaSpace{}\IdrisData{"([2][0\KatlaDash{}3])!"}\IdrisKeyword{)}\KatlaNewline{}
\KatlaSpace{}\KatlaSpace{}\IdrisData{<*>}\KatlaSpace{}\IdrisFunction{map}\KatlaSpace{}\IdrisFunction{f}\KatlaSpace{}\IdrisKeyword{(}\IdrisFunction{r}\KatlaSpace{}\IdrisData{":([0\KatlaDash{}5][0\KatlaDash{}9])!"}\IdrisKeyword{)}\KatlaSpace{}\IdrisComment{\KatlaDash{}\KatlaDash{}\KatlaSpace{}^(or\KatlaSpace{}:\KatlaSpace{}TyRE\KatlaSpace{}a\KatlaSpace{}\KatlaDash{}>\KatlaSpace{}\KatlaSpace{}TyRE\KatlaSpace{}a\KatlaSpace{}\KatlaDash{}>\KatlaSpace{}\KatlaSpace{}TyRE\KatlaSpace{}a)}\KatlaNewline{}
\KatlaSpace{}\KatlaSpace{}\IdrisComment{\KatlaDash{}\KatlaDash{}^((<*>)\KatlaSpace{}:\KatlaSpace{}TyRE\KatlaSpace{}a\KatlaSpace{}\KatlaDash{}>\KatlaSpace{}\KatlaSpace{}TyRE\KatlaSpace{}b\KatlaSpace{}\KatlaDash{}>\KatlaSpace{}\KatlaSpace{}TyRE\KatlaSpace{}(a,\KatlaSpace{}b))}\KatlaNewline{}
\KatlaNewline{}
\KatlaSpace{}\KatlaSpace{}\IdrisKeyword{where}\KatlaSpace{}\IdrisFunction{digit}\KatlaSpace{}\IdrisKeyword{:}\KatlaSpace{}\IdrisType{Char}\KatlaSpace{}\IdrisKeyword{\KatlaDash{}>}\KatlaSpace{}\IdrisType{Integer}\KatlaNewline{}
\KatlaSpace{}\KatlaSpace{}\KatlaSpace{}\KatlaSpace{}\KatlaSpace{}\KatlaSpace{}\KatlaSpace{}\KatlaSpace{}\IdrisFunction{digit}\KatlaSpace{}\IdrisBound{c}\KatlaSpace{}\IdrisKeyword{=}\KatlaSpace{}\IdrisFunction{cast}\KatlaSpace{}\IdrisBound{c}\KatlaSpace{}\IdrisFunction{\KatlaDash{}}\KatlaSpace{}\IdrisFunction{cast}\KatlaSpace{}\IdrisData{'0'}\KatlaNewline{}
\KatlaNewline{}
\KatlaSpace{}\KatlaSpace{}\KatlaSpace{}\KatlaSpace{}\KatlaSpace{}\KatlaSpace{}\KatlaSpace{}\KatlaSpace{}\IdrisFunction{f}\KatlaSpace{}\IdrisKeyword{:}\KatlaSpace{}\IdrisKeyword{(}\IdrisType{Char,}\KatlaSpace{}\IdrisType{Char}\IdrisKeyword{)}\KatlaSpace{}\IdrisKeyword{\KatlaDash{}>}\KatlaSpace{}\IdrisType{Integer}\KatlaNewline{}
\KatlaSpace{}\KatlaSpace{}\KatlaSpace{}\KatlaSpace{}\KatlaSpace{}\KatlaSpace{}\KatlaSpace{}\KatlaSpace{}\IdrisFunction{f}\KatlaSpace{}\IdrisKeyword{(}\IdrisBound{c1}\IdrisData{,}\KatlaSpace{}\IdrisBound{c2}\IdrisKeyword{)}\KatlaSpace{}\IdrisKeyword{=}\KatlaSpace{}\IdrisData{10}\KatlaSpace{}\IdrisFunction{*}\KatlaSpace{}\IdrisFunction{digit}\KatlaSpace{}\IdrisBound{c1}\KatlaSpace{}\IdrisFunction{+}\KatlaSpace{}\IdrisFunction{digit}\KatlaSpace{}\IdrisBound{c2}\KatlaNewline{}
\end{code}
\caption{A structured timestamp parser.}
\label{fig:structured time tyre}
\end{figure}
\begin{aside}
Idris pairs
\IdrisKeyword{(}\IdrisBound{c1}\IdrisData{, }\IdrisBound{c2}\IdrisKeyword{)}
and pair types
\IdrisKeyword{(}\IdrisType{Char}\IdrisType{, }\IdrisType{Char}\IdrisKeyword{)}%
, like Haskell's, use a bracketed sequence
notation. The \IdrisType{TyRE} type-constructor has the structure of
a \IdrisType{Functor}, providing the function \IdrisFunction{map} (see
below). Infix operators, are either sequences of lexically-designated operator
characters (\IdrisData{<*>}) or any function name surrounded by
back-ticks (\IdrisFunction{\`{}or\`{}}). Idris supports \IdrisKeyword{where}
clauses for local definitions.
\end{aside}

The exclamation mark (\IdrisData{!}) in this regex literal is a
TyRE-specific special character that works as a typed capture
group. We will produce parse-trees for sub-strings matching the
sub-regexes marked with the exclamation mark, while other sub-strings
will parse as the \IdrisType{Unit}
value \IdrisData{()}. The regex \IdrisFunction{time1} is of
type \IdrisType{TyRE Unit} exactly because it has no capture groups.  The
TyRE library parses the string literal and computes its \emph{shape}:
its parse-trees' type.  Smart constructors mixed
with regex literals exert fine-grained control over the shape:
\begin{itemize}
\item
We transform the result using the \IdrisData{Functor} instance
of \IdrisType{TyRE}:
\ignore{
\begin{code}
\IdrisKeyword{namespace}\KatlaSpace{}\IdrisNamespace{Hide}\KatlaNewline{}
\end{code}
}
\begin{code}
\KatlaSpace{}\KatlaSpace{}\IdrisFunction{map}\KatlaSpace{}\IdrisKeyword{:}\KatlaSpace{}\IdrisKeyword{(}\IdrisBound{a}\KatlaSpace{}\IdrisKeyword{\KatlaDash{}>}\KatlaSpace{}\IdrisBound{b}\IdrisKeyword{)}\KatlaSpace{}\IdrisKeyword{\KatlaDash{}>}\KatlaSpace{}\IdrisType{TyRE}\KatlaSpace{}\IdrisBound{a}\KatlaSpace{}\IdrisKeyword{\KatlaDash{}>}\KatlaSpace{}\IdrisType{TyRE}\KatlaSpace{}\IdrisBound{b}\KatlaNewline{}
\end{code}
\item Uniform alternation works the same as alternation, but both sub-regexes have the same
  shape, and the result does not track which alternative regex
  matched:
\begin{code}
\KatlaSpace{}\KatlaSpace{}\IdrisFunction{or}\KatlaSpace{}\IdrisKeyword{:}\KatlaSpace{}\IdrisType{TyRE}\KatlaSpace{}\IdrisBound{a}\KatlaSpace{}\IdrisKeyword{\KatlaDash{}>}\KatlaSpace{}\IdrisType{TyRE}\KatlaSpace{}\IdrisBound{a}\KatlaSpace{}\IdrisKeyword{\KatlaDash{}>}\KatlaSpace{}\IdrisType{TyRE}\KatlaSpace{}\IdrisBound{a}\KatlaNewline{}
\end{code}
\end{itemize}

We pass the regex to \IdrisFunction{parse}, which returns either \IdrisData{Just} a parse tree consisting of a pair of integers,
if the string matches the regex, or \IdrisData{Nothing} otherwise:
\begin{code}
\IdrisFunction{parsedTimestamp}\KatlaSpace{}\IdrisKeyword{:}\KatlaSpace{}\IdrisType{String}\KatlaSpace{}\IdrisKeyword{\KatlaDash{}>}\KatlaSpace{}\IdrisType{Maybe}\KatlaSpace{}\IdrisKeyword{(}\IdrisType{Integer,}\KatlaSpace{}\IdrisType{Integer}\IdrisKeyword{)}\KatlaNewline{}
\IdrisFunction{parsedTimestamp}\KatlaSpace{}\IdrisBound{str}\KatlaSpace{}\IdrisKeyword{=}\KatlaSpace{}\IdrisFunction{parse}\KatlaSpace{}\IdrisFunction{time2}\KatlaSpace{}\IdrisBound{str}\KatlaNewline{}
\end{code}

\ignore{
\begin{code}
\IdrisKeyword{namespace}\KatlaSpace{}\IdrisNamespace{Hide}\KatlaNewline{}
\end{code}
}

TyRE provides other string-parsing functions:
\begin{itemize}
  \item \emph{Prefix parsing}: parse the longest or shortest matching
    prefix, depending on the \IdrisBound{greedy} flag, also returning
    the unparsed remainder of the string:
\begin{code}
\KatlaSpace{}\KatlaSpace{}\IdrisFunction{parsePrefix}\KatlaSpace{}\IdrisKeyword{:}\KatlaSpace{}\IdrisType{TyRE}\KatlaSpace{}\IdrisBound{a}\KatlaSpace{}\IdrisKeyword{\KatlaDash{}>}\KatlaNewline{}
\KatlaSpace{}\KatlaSpace{}\KatlaSpace{}\KatlaSpace{}\KatlaSpace{}\KatlaSpace{}\KatlaSpace{}\KatlaSpace{}\KatlaSpace{}\KatlaSpace{}\KatlaSpace{}\KatlaSpace{}\KatlaSpace{}\KatlaSpace{}\KatlaSpace{}\KatlaSpace{}\IdrisType{String}\KatlaSpace{}\IdrisKeyword{\KatlaDash{}>}\KatlaSpace{}\IdrisKeyword{(}\IdrisBound{greedy}\KatlaSpace{}\IdrisKeyword{:}\KatlaSpace{}\IdrisType{Bool}\IdrisKeyword{)}\KatlaSpace{}\IdrisKeyword{\KatlaDash{}>}\KatlaSpace{}\IdrisKeyword{(}\IdrisType{Maybe}\KatlaSpace{}\IdrisBound{a}\IdrisType{,}\KatlaSpace{}\IdrisType{String}\IdrisKeyword{)}\KatlaNewline{}
\end{code}

\item \emph{Disjoint matching:} some text processing tasks require finding
all matching occurrences of a regex in a string:
\begin{code}
\KatlaSpace{}\KatlaSpace{}\IdrisFunction{disjointMatches}\KatlaSpace{}\IdrisKeyword{:}\KatlaSpace{}\IdrisKeyword{(}\IdrisBound{tyre}\KatlaSpace{}\IdrisKeyword{:}\KatlaSpace{}\IdrisType{TyRE}\KatlaSpace{}\IdrisBound{a}\IdrisKeyword{)}\KatlaSpace{}\IdrisKeyword{\KatlaDash{}>}\KatlaNewline{}
\KatlaSpace{}\KatlaSpace{}\KatlaSpace{}\KatlaSpace{}\IdrisKeyword{\{auto}\KatlaSpace{}\IdrisKeyword{0}\KatlaSpace{}\IdrisBound{consuming}\KatlaSpace{}\IdrisKeyword{:}\KatlaSpace{}\IdrisType{IsConsuming}\KatlaSpace{}\IdrisBound{tyre}\IdrisKeyword{\}}\KatlaSpace{}\IdrisKeyword{\KatlaDash{}>}\KatlaSpace{}\IdrisType{String}\KatlaSpace{}\IdrisKeyword{\KatlaDash{}>}\KatlaNewline{}
\KatlaSpace{}\KatlaSpace{}\KatlaSpace{}\KatlaSpace{}\IdrisKeyword{(}\IdrisBound{greedy}\KatlaSpace{}\IdrisKeyword{:}\KatlaSpace{}\IdrisType{Bool}\IdrisKeyword{)}\KatlaSpace{}\IdrisKeyword{\KatlaDash{}>}\KatlaSpace{}\IdrisKeyword{(}\IdrisType{List}\KatlaSpace{}\IdrisKeyword{(}\IdrisType{List}\KatlaSpace{}\IdrisType{Char,}\KatlaSpace{}\IdrisBound{a}\IdrisKeyword{)}\IdrisType{,}\KatlaSpace{}\IdrisType{List}\KatlaSpace{}\IdrisType{Char}\IdrisKeyword{)}\KatlaNewline{}
\end{code}
  The given \IdrisBound{tyre} must be \IdrisBound{consuming} --- only
  match non-empty strings, and we require a (runtime-erased) witness
  of type \IdrisData{IsConsuming }\IdrisBound{tyre}.

  \begin{aside}\label{auto-exposition}
  The \IdrisKeyword{auto} argument in braces (\IdrisKeyword{\{\}})
  means Idris will search for this argument using a process
  called \emph{proof search}. It follows user-defined hints, attempts
  visible data constructors, and prioritises bound variables of
  appropriate types. Idris implements type-classes by desugaring
  instance resolution to proof search.
  \end{aside}

  \item \emph{Substitution:} replace all found disjoint matches in a
    string:
\begin{code}
\KatlaSpace{}\KatlaSpace{}\IdrisFunction{substitute}\KatlaSpace{}\IdrisKeyword{:}\KatlaSpace{}\IdrisKeyword{(}\IdrisBound{tyre}\IdrisKeyword{:}\KatlaSpace{}\IdrisType{TyRE}\KatlaSpace{}\IdrisBound{a}\IdrisKeyword{)}\KatlaSpace{}\IdrisKeyword{\KatlaDash{}>}\KatlaNewline{}
\KatlaSpace{}\KatlaSpace{}\KatlaSpace{}\KatlaSpace{}\IdrisKeyword{\{auto}\KatlaSpace{}\IdrisKeyword{0}\KatlaSpace{}\IdrisBound{consuming}\KatlaSpace{}\IdrisKeyword{:}\KatlaSpace{}\IdrisType{IsConsuming}\KatlaSpace{}\IdrisBound{tyre}\IdrisKeyword{\}}\KatlaSpace{}\IdrisKeyword{\KatlaDash{}>}\KatlaSpace{}\IdrisKeyword{(}\IdrisBound{a}\KatlaSpace{}\IdrisKeyword{\KatlaDash{}>}\KatlaSpace{}\IdrisType{String}\IdrisKeyword{)}\KatlaSpace{}\IdrisKeyword{\KatlaDash{}>}\KatlaNewline{}
\KatlaSpace{}\KatlaSpace{}\KatlaSpace{}\KatlaSpace{}\IdrisType{String}\KatlaSpace{}\IdrisKeyword{\KatlaDash{}>}\KatlaSpace{}\IdrisType{String}\KatlaNewline{}
\end{code}
For example, we change the format in which the time is given in some
text.
\ignore{
\begin{code}
\IdrisKeyword{namespace}\KatlaSpace{}\IdrisNamespace{Example}\KatlaNewline{}
\end{code}
}
\begin{code}
\KatlaSpace{}\KatlaSpace{}\IdrisFunction{changeTimeFormat}\KatlaSpace{}\IdrisKeyword{:}\KatlaSpace{}\IdrisType{String}\KatlaSpace{}\IdrisKeyword{\KatlaDash{}>}\KatlaSpace{}\IdrisType{String}\KatlaNewline{}
\KatlaSpace{}\KatlaSpace{}\IdrisFunction{changeTimeFormat}\KatlaSpace{}\IdrisKeyword{=}\KatlaSpace{}\IdrisFunction{substitute}\KatlaSpace{}\IdrisFunction{time2}\KatlaNewline{}
\KatlaSpace{}\KatlaSpace{}\KatlaSpace{}\KatlaSpace{}\KatlaSpace{}\KatlaSpace{}\IdrisKeyword{(\textbackslash{}case}\KatlaSpace{}\IdrisKeyword{(}\IdrisBound{h}\IdrisData{,}\KatlaSpace{}\IdrisBound{m}\IdrisKeyword{)}\KatlaSpace{}\IdrisKeyword{=>}\KatlaSpace{}\IdrisFunction{show}\KatlaSpace{}\IdrisBound{m}\KatlaSpace{}\IdrisFunction{++}\KatlaSpace{}\IdrisData{"\KatlaSpace{}past\KatlaSpace{}"}\KatlaSpace{}\IdrisFunction{++}\KatlaSpace{}\IdrisFunction{show}\KatlaSpace{}\IdrisBound{h}\IdrisKeyword{)}\KatlaNewline{}
\KatlaSpace{}\KatlaSpace{}\KatlaSpace{}\IdrisComment{\KatlaDash{}\KatlaDash{}\KatlaSpace{}\KatlaSpace{}^\KatlaDash{}(pattern\KatlaSpace{}matching\KatlaSpace{}anonymous\KatlaSpace{}function)}\KatlaNewline{}
\end{code}
\begin{aside}
  The keyword \IdrisKeyword{\textbackslash{}case} expresses an anonymous
  function that immediately matches on its argument.
\end{aside}
The call \IdrisFunction{changeTimeFormat}
\IdrisData{"Look, it is 11:15."} evaluates to \IdrisData{"Look, it is 15
  past 11."}.

\item \emph{Streams:}
The TyRE library can be used not only for parsing strings but also for tokenizing
a \IdrisType{Stream}\KatlaSpace{}\IdrisType{Char} using the function:
\begin{code}
\KatlaSpace{}\KatlaSpace{}\IdrisFunction{getToken}\KatlaSpace{}\IdrisKeyword{:}\KatlaSpace{}\IdrisKeyword{(}\IdrisBound{greedy}\KatlaSpace{}\IdrisKeyword{:}\KatlaSpace{}\IdrisType{Bool}\IdrisKeyword{)}\KatlaSpace{}\IdrisKeyword{\KatlaDash{}>}\KatlaSpace{}\IdrisType{TyRE}\KatlaSpace{}\IdrisBound{a}\KatlaSpace{}\IdrisKeyword{\KatlaDash{}>}\KatlaSpace{}\IdrisType{Stream}\KatlaSpace{}\IdrisType{Char}\KatlaSpace{}\IdrisKeyword{\KatlaDash{}>}\KatlaNewline{}
\KatlaSpace{}\KatlaSpace{}\KatlaSpace{}\KatlaSpace{}\IdrisKeyword{(}\IdrisType{Maybe}\KatlaSpace{}\IdrisBound{a}\IdrisType{,}\KatlaSpace{}\IdrisType{Stream}\KatlaSpace{}\IdrisType{Char}\IdrisKeyword{)}\KatlaNewline{}
\end{code}
\begin{aside}
  Idris \IdrisType{Stream}s are infinite lazy lists.
\end{aside}
Here is a simple lexer, tokenizing a semicolon-separated stream of digits:
\ignore{
\begin{code}
\IdrisKeyword{namespace}\KatlaSpace{}\IdrisNamespace{Example2}\KatlaNewline{}
\end{code}
}
\begin{code}
\KatlaSpace{}\KatlaSpace{}\IdrisFunction{getDigits}\KatlaSpace{}\IdrisKeyword{:}\KatlaSpace{}\IdrisType{Stream}\KatlaSpace{}\IdrisType{Char}\KatlaSpace{}\IdrisKeyword{\KatlaDash{}>}\KatlaSpace{}\IdrisType{Stream}\KatlaSpace{}\IdrisKeyword{(}\IdrisType{Maybe}\KatlaSpace{}\IdrisType{Integer}\IdrisKeyword{)}\KatlaNewline{}
\KatlaSpace{}\KatlaSpace{}\IdrisFunction{getDigits}\KatlaSpace{}\IdrisKeyword{=}\KatlaSpace{}\KatlaSpace{}\KatlaSpace{}\KatlaSpace{}\KatlaSpace{}\KatlaSpace{}\KatlaSpace{}\KatlaSpace{}\KatlaSpace{}\KatlaSpace{}\KatlaSpace{}\KatlaSpace{}\KatlaSpace{}\KatlaSpace{}\KatlaSpace{}\IdrisComment{\KatlaDash{}\KatlaDash{}\KatlaSpace{}\KatlaSpace{}v\KatlaDash{}(match\KatlaSpace{}:\KatlaSpace{}Char\KatlaSpace{}\KatlaDash{}>\KatlaSpace{}\KatlaSpace{}TyRE\KatlaSpace{}())}\KatlaNewline{}
\KatlaSpace{}\KatlaSpace{}\KatlaSpace{}\KatlaSpace{}\KatlaSpace{}\KatlaSpace{}\IdrisFunction{unfoldr}\KatlaSpace{}\IdrisKeyword{(}\IdrisFunction{getToken}\KatlaSpace{}\IdrisData{True}\KatlaSpace{}\IdrisKeyword{(}\IdrisFunction{match}\KatlaSpace{}\IdrisData{';'}\KatlaSpace{}\IdrisFunction{*>}\KatlaSpace{}\IdrisFunction{digit}\IdrisKeyword{))}\KatlaNewline{}
\KatlaSpace{}\KatlaSpace{}\IdrisComment{\KatlaDash{}\KatlaDash{}((*>)\KatlaSpace{}:\KatlaSpace{}TyRE\KatlaSpace{}a\KatlaSpace{}\KatlaDash{}>\KatlaSpace{}\KatlaSpace{}TyRE\KatlaSpace{}b\KatlaSpace{}\KatlaDash{}>\KatlaSpace{}\KatlaSpace{}TyRE\KatlaSpace{}b)\KatlaDash{}^\KatlaSpace{}\KatlaSpace{}^(digit\KatlaSpace{}:\KatlaSpace{}TyRE\KatlaSpace{}Integer)}\KatlaNewline{}
\end{code}
We used three smart constructors:
\begin{itemize}
\item match exactly the chosen character:
\begin{code}
\KatlaSpace{}\KatlaSpace{}\IdrisFunction{match}\KatlaSpace{}\IdrisKeyword{:}\KatlaSpace{}\IdrisType{Char}\KatlaSpace{}\IdrisKeyword{\KatlaDash{}>}\KatlaSpace{}\IdrisType{TyRE}\KatlaSpace{}\IdrisType{()}\KatlaNewline{}
\end{code}
\item a digit TyRE:
\begin{code}
\KatlaSpace{}\KatlaSpace{}\IdrisFunction{digit}\KatlaSpace{}\IdrisKeyword{:}\KatlaSpace{}\IdrisType{TyRE}\KatlaSpace{}\IdrisType{Integer}\KatlaNewline{}
\end{code}
\item a regex combinator that discards the first parse-tree and returns only the second:
\begin{code}
\KatlaSpace{}\KatlaSpace{}\IdrisFunction{(*>)}\KatlaSpace{}\IdrisKeyword{:}\KatlaSpace{}\IdrisType{TyRE}\KatlaSpace{}\IdrisBound{a}\KatlaSpace{}\IdrisKeyword{\KatlaDash{}>}\KatlaSpace{}\IdrisType{TyRE}\KatlaSpace{}\IdrisBound{b}\KatlaSpace{}\IdrisKeyword{\KatlaDash{}>}\KatlaSpace{}\IdrisType{TyRE}\KatlaSpace{}\IdrisBound{b}\KatlaNewline{}
\end{code}
\end{itemize}
\end{itemize}

\ignore{
\begin{code}
\IdrisKeyword{import}\KatlaSpace{}\IdrisModule{Data.Regex}\KatlaNewline{}
\IdrisKeyword{import}\KatlaSpace{}\IdrisModule{Data.Stream}\KatlaNewline{}
\end{code}
}

\newcommand\ExampleSnocList{%
  \IdrisData{[<}\KatlaSpace{}%
    \IdrisBound{x1}\IdrisData{,}\KatlaSpace{}%
    \IdrisBound{x2}\IdrisData{,}\KatlaSpace{}%
    $\ldots$\IdrisData{,}\KatlaSpace{}%
    \IdrisBound{xn}\IdrisData{]}\KatlaNewline{}%
}
\newcommand\ExampleSnocListDesugared{%
  \IdrisKeyword{(((}\IdrisData{Lin}\KatlaSpace{}\IdrisData{:<}\KatlaSpace{}%
  \IdrisBound{x1}\IdrisKeyword{)}\KatlaSpace{}\IdrisData{:<}\KatlaSpace{}%
  \IdrisBound{x2}\IdrisKeyword{)}\KatlaSpace{}%
  $\cdots$\IdrisKeyword{)}\KatlaSpace{}%
  \IdrisData{:<}\KatlaSpace{}\IdrisBound{xn}\KatlaNewline{}%
}
\newcommand\ExampleSingleton{%
  \IdrisData{[<}\KatlaSpace{}\IdrisType{Char}\IdrisData{]}%
  \KatlaSpace{}\IdrisKeyword{:}\KatlaSpace{}%
  \IdrisType{SnocList}\KatlaSpace{\IdrisType{Type}}%
}

\section{Architecture}\label{architecture_overview}
\begin{figure}
  \centering
  \includegraphics[width=\textwidth]{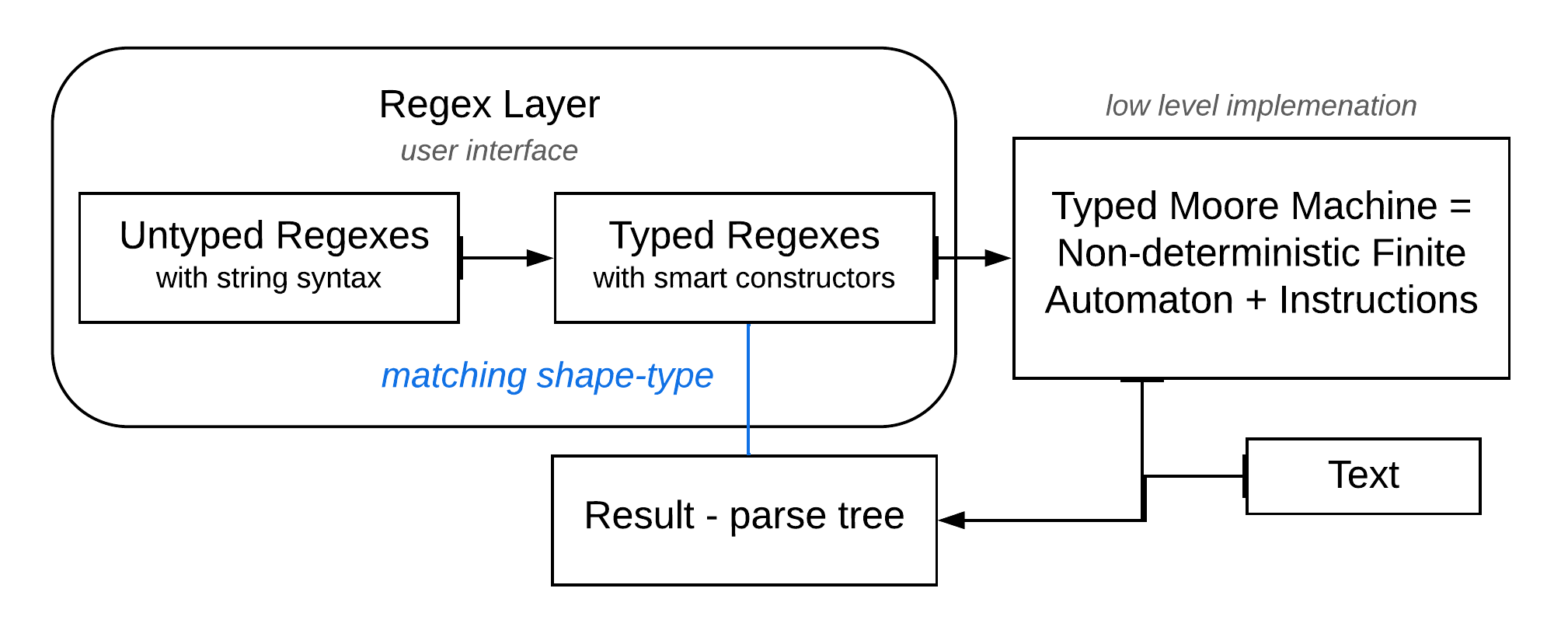}
  \caption{Overview of the parser's architecture.}
  \label{fig:arch}
\end{figure}

TyRE is composed of two main parts:
\begin{itemize}
\item A regex layer, which contains typed regular expressions and
  user-facing interface:
  \begin{itemize}
  \item smart constructors;
  \item regexes with computed/inferred shape, which we call
    \emph{untyped regexes}; and
  \item regex literals.
\end{itemize}
\item
  A typed Moore machine that builds the parse-tree while executing a
  nondeterministic finite-state automaton for the regex.
\end{itemize}
\figref{fig:arch} depicts TyRE's architecture. Users may
construct typed or untyped regexes directly, or as regex literals,
and the regex layer will translate them into the core typed regex
representation. A variant of Thompson's construction then generates a
typed Moore machine, i.e., a non-deterministic finite state automaton
whose transitions also emit action in a typed language of low-level
\emph{routines}. Given the input text, we execute this Moore machine,
either reporting failure to match the regex, or a result parse tree.

\begin{figure}
  \begin{subfigure}{.5\textwidth}
      \includegraphics[width=1\textwidth]{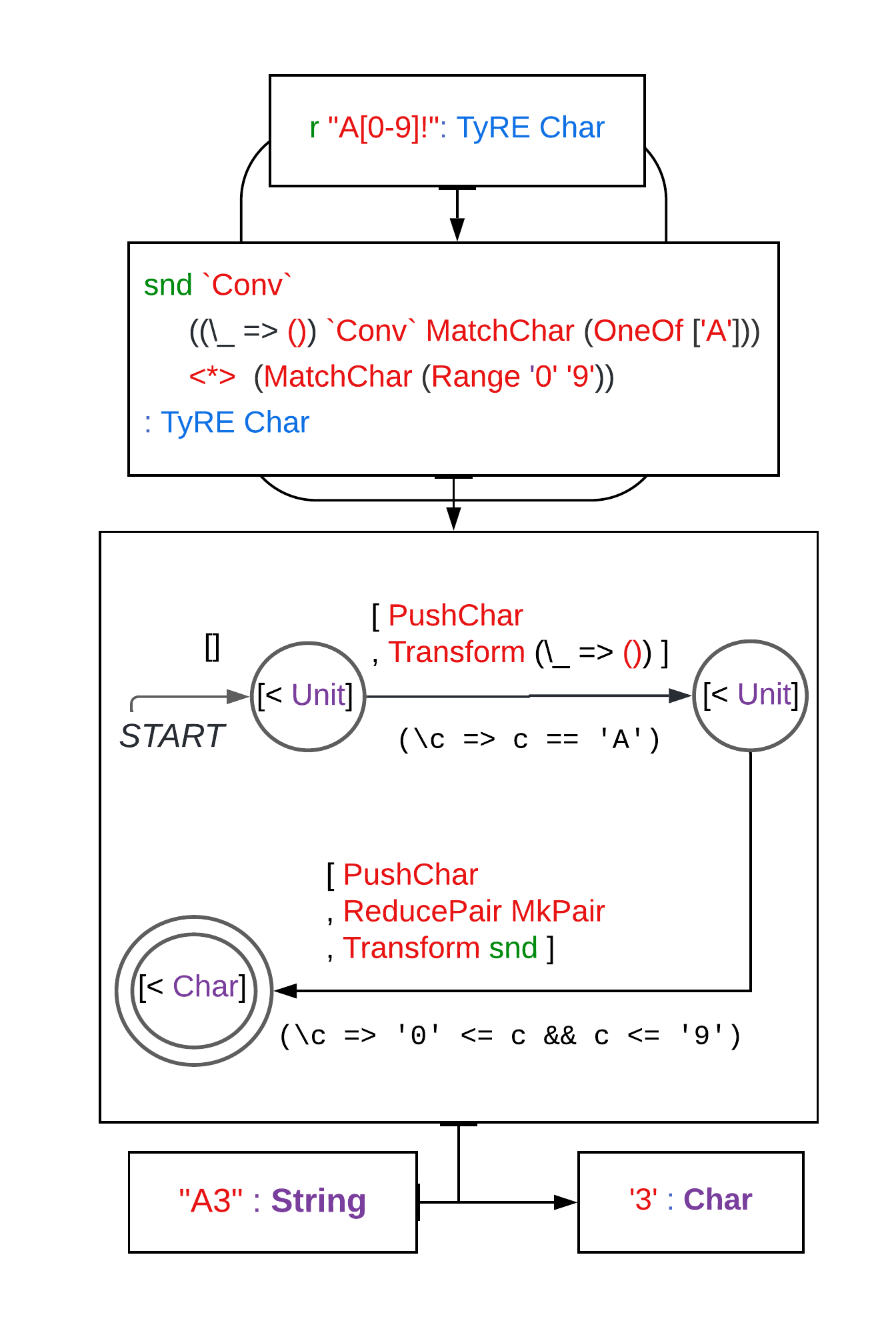}
    \end{subfigure}
    \begin{subfigure}{.5\textwidth}
      \begin{tabular}[b]{l l}
  ROUTINE & STACK \\
  \hline
  & [$<$ ] \\
  \hline
  \multicolumn{2}{l}{\IdrisComment{Current character: 'A'.}}\\
  \multicolumn{2}{l}{\IdrisComment{Condition check: $'A' == 'A'$}}\\
  \hline
  \IdrisData{PushChar} & [ $<$ \IdrisData{'A'}]\\
  \IdrisData{Transform} ($\backslash \_ =>$ \IdrisData{()}) & [$<$ \IdrisData{()}]\\
  \hline
  \multicolumn{2}{l}{\IdrisComment{Current character: '3'.}}\\
  \multicolumn{2}{l}{\IdrisComment{Condition check:}}\\
  \multicolumn{2}{r}{\IdrisComment{'0' <= '3' \&\& '3' <= '9'}}\\
  \hline
  \IdrisData{PushChar} & [$<$ \IdrisData{()}, \IdrisData{'3'}]\\
  \IdrisData{ReducePair MkPair} & [$<$ (\IdrisData{()}, \IdrisData{'3'})]\\
  \IdrisData{Transfrom} \IdrisFunction{snd} & [$<$ \IdrisData{'3'}]\\
  \hline
  \multicolumn{2}{l}{\IdrisComment{Accept.}}\\
\end{tabular}
\\~
      \vspace{3cm}
    \end{subfigure}\\
    \begin{subfigure}{.5\textwidth}
      \caption{Translation}
      \label{fig:arch_example_translation}
    \end{subfigure}
    \begin{subfigure}{.5\textwidth}
      \caption{Execution stack-trace}
      \label{fig:arch_example_stacktrace}
    \end{subfigure}
  \caption{parsing "A3" with the regex literal \texttt{r "A[0-9]!"}.}
  \label{fig:arch_example}
\end{figure}

\figref{fig:arch_example} demonstrates the connections between the
components and their responsibilities.  We parse the string
\IdrisData{"A3"} using \IdrisFunction{r} \IdrisData{"A[0-9]!"}. TyRE
first parses the regex string literal to a typed regex
\IdrisFunction{$r_0$}: \IdrisType{TyRE Char}. From there, we build a
typed Moore machine using a variant of Thompson's construction%
~\citep{mcnaughton-yamada:nfa-construction,thompson:nfa-construction}.
\figref{fig:arch_example_translation} shows the
resulting typed machine. It is a non-deterministic finite automaton
for the regex literal \texttt{"A[0-9]"}: the labels under the arrows
are a \IdrisType{Char}-predicate determining when to take this
transition.
The labels over the arrows are code-routines to execute in each
transition that will result in a correct parse-tree when we reach an
accepting state.

We feed the Moore machine with our chosen string
\IdrisData{"A3"}. As we traverse the NFA, we collect parts of the
parse tree on a stack\footnotemark{}.
\figref{fig:arch_example_stacktrace} traces the
stack during the execution of this Moore machine, interleaving the
current NFA state --- the currently read character and the condition
guarding the next transition --- with the Moore machine code over this
transition and its effect on the auxiliary stack.
\footnotetext{Although the Moore-machine uses a stack in its auxiliary
  state, this stack does not affect the control-flow of execution,
  which is governed solely by the current state and current character.
  Thus the model of computation is still a regular non-deterministic
  finite state automaton, and not a push-down automaton.%
}
To ensure the content of the stack can produce a well-formed
parse-tree, we associate to each state a \emph{stack shape}.  It
describes what must be the shape of the stack at this point.
Concretely, the stack shape is a sequence of types, describing
which parts of the result parse-tree have already been
constructed on the stack.  We implement the stack shape by a
\IdrisType{SnocList} of types.  Each code-routine instruction is typed
by the assumed stack shapes before and after its execution, thus its
typing judgements are a form of a Hoare triple. The stack shape we
associate to the unique accepting state is the singleton
\ExampleSingleton{}, matching the input shape \IdrisType{TyRE Char}.
Thus a successful path through the Moore machines guarantees we can
extract a parse-tree of the correct shape.

\aside
Idris supports the following syntactic sugar for left-nested lists, where:
\begin{center}
  \hfill
  \ExampleSnocList{} desugars into \ExampleSnocListDesugared{}
  \endaside
\end{center}

\section{Regex layer}\label{typed_regexes}
\ignore{
\begin{code}
\IdrisKeyword{import}\KatlaSpace{}\IdrisModule{Data.Regex}\KatlaNewline{}
\IdrisKeyword{import}\KatlaSpace{}\IdrisModule{Data.Stream}\KatlaNewline{}
\end{code}
}
The regex layer provides the user-facing abstractions that we will now
describe in detail. When users use the smart-constructor
  \IdrisFunction{r},
\begin{itemize}
\item the \emph{regex literal} is parsed at compile into an
\item \emph{untyped regex} (\IdrisType{RE}), whose inferred
  shape is also computed at compile time, and then translated into a
\item \emph{typed regex} (\IdrisType{TyRE}).
\end{itemize}

\subsection{Regex literals}
TyRE provides string literals to create typed regexes. Taking
advantage of Idris's type-level computation we parse string literals
into typed regexes at compile time. This feature both ensures that the
string is a valid regex expression and calculates the type of the
result parse tree. We parse regex literals in two steps. First, we
parse the string literal into an untyped regex data structure
(\IdrisType{RE}) using a parser combinator library\footnotemark{}.
\footnotetext{%
  The formal language of well-formed regex literals is context-free,
  and not regular, and so we need the additional expressiveness parser
  combinators offer.%
}%
In the second step, we translate this untyped regex to typed ones
using one-to-one mappings between parse-trees and shape simplification
rules.

Table~\ref{table:string-syntax} shows the mapping between the string
regex and \IdrisType{RE} and defines an approximation to the expected
untyped regex's shape. However, the actual regex's shape calculation
is more complex:
\begin{itemize}
\item Every sub-regex that is not wrapped in the \IdrisData{Keep}
  constructor, marked by an exclamation mark (\IdrisData{!}) in regex
  literals, gets flattened into a \IdrisType{Unit}, for example:
\begin{center}
  \IdrisFunction{r}
\IdrisData{"((([a-z])+)!)|(hj)"} has the shape \IdrisType{Either}
\IdrisKeyword{(}\IdrisType{List} \IdrisType{Char}\IdrisKeyword{)}
\IdrisType{Unit}
\end{center}
\item We eliminate redundancy \IdrisData{Unit} types in kept sub-regexes.
  Concretely, we simplify:
  \begin{itemize}
  \item a pair of \IdrisType{Unit}s to a \IdrisType{Unit};
  \item a \IdrisType{List Unit} to a \IdrisType{Nat}, representing the
    list's length;
  \item an \IdrisType{Either Unit Unit} to a \IdrisType{Bool}ean.
  \end{itemize}
  For example:
  \begin{center}
    \begin{tabular}{l@{ has shape }l}
      \IdrisFunction{r}\IdrisData{"((([a-z])+)!)|(hj)"}
      &
      \IdrisType{List} \IdrisType{Char}
    \\
      \IdrisFunction{r}\IdrisData{"((ab*[vkw]([a-z])+)|(hj))!"}
      &
      \IdrisType{Maybe} \IdrisKeyword{(}\IdrisType{Nat,}
      \IdrisKeyword{(}\IdrisType{Char,} \IdrisType{List}
      \IdrisType{Char}\IdrisKeyword{))}
    \end{tabular}
  \end{center}
\end{itemize}

To parse regex literals at compile time, we use a
copy of the parser combinator library in Idris's standard library.
In our copy of the parser combinator library, all the functions are qualified as
\IdrisKeyword{public export}. Thus the type-checker
evaluates well-formed regex literals into a value
of type \IdrisType{RE}.
\begin{table}
  \begin{center}
    \begin{tabular}{| p{0.14\textwidth} | p{0.22\textwidth} | p{0.24\textwidth} | p{0.25\textwidth} |}
  \hline
  String regex
    & \IdrisType{RE}
    & \IdrisFunction{Shape}
    & Keep Shape (\IdrisFunction{KS})
  \\ \hline \hline
      \IdrisFunction{r} \IdrisData{"a"}
    & \IdrisData{Exactly} \IdrisBound{a}
    & \IdrisType{Unit}
    & \IdrisType{Unit}
  \\\hline
      \IdrisFunction{r} \IdrisData{"[ab]"}
    & \IdrisData{OneOf ['a', 'b']}
    & \IdrisType{Unit}
    & \IdrisType{Char}
  \\\hline
      \IdrisFunction{r} \IdrisData{"[a-c]"}
    & \IdrisData{To 'a' 'c'}
    & \IdrisType{Unit}
    & \IdrisType{Char}
  \\\hline
      \IdrisFunction{r} \IdrisData{"."}
    & \IdrisData{Any}
    & \IdrisType{Unit}
    & \IdrisType{Char}
  \\\hline
      \IdrisFunction{r} \texttt{"AB"}
    & \IdrisData{Concat} \IdrisBound{A} \IdrisBound{B}
    & \IdrisKeyword{(}\IdrisFunction{Shape} \IdrisBound{A}\IdrisType{,} \IdrisFunction{Shape} \IdrisBound{B}\IdrisKeyword{)}
    & \IdrisKeyword{(}\IdrisFunction{KS} \IdrisBound{A}\IdrisType{,} \IdrisFunction{KS} \IdrisBound{B}\IdrisKeyword{)}
  \\\hline
      \IdrisFunction{r} \texttt{"A$\vert$B"}
    & \IdrisData{Alt} \IdrisBound{A} \IdrisBound{B}
      & \IdrisType{Either}
      \begin{tabular}[t]{@{}l@{}}
        \IdrisKeyword{(}\IdrisFunction{Shape} \IdrisBound{A}\IdrisKeyword{)}
        \\
        \IdrisKeyword{(}\IdrisFunction{Shape} \IdrisBound{B}\IdrisKeyword{)}
      \end{tabular}
    & \IdrisType{Either} \IdrisKeyword{(}\IdrisFunction{KS} \IdrisBound{A}\IdrisKeyword{)} \IdrisKeyword{(}\IdrisFunction{KS} \IdrisBound{B}\IdrisKeyword{)}
  \\\hline
      \IdrisFunction{r} \texttt{"A?"}
    & \IdrisData{Maybe} \IdrisBound{A}
    & \IdrisType{Maybe} \IdrisKeyword{(}\IdrisFunction{Shape} \IdrisBound{A}\IdrisKeyword{)}
    & \IdrisType{Maybe} \IdrisKeyword{(}\IdrisFunction{KS} \IdrisBound{A}\IdrisKeyword{)}
  \\\hline
      \IdrisFunction{r} \texttt{"A*"}
    & \IdrisData{Rep0} \IdrisBound{A}
    & \IdrisType{List} \IdrisKeyword{(}\IdrisFunction{Shape} \IdrisBound{A}\IdrisKeyword{)}
    & \IdrisType{List} \IdrisKeyword{(}\IdrisFunction{KS} \IdrisBound{A}\IdrisKeyword{)}
  \\\hline
      \IdrisFunction{r} \texttt{"A+"}
    & \IdrisData{Rep1} \IdrisBound{A}
    & \IdrisType{List} \IdrisKeyword{(}\IdrisFunction{Shape} \IdrisBound{A}\IdrisKeyword{)}
    & \IdrisType{List} \IdrisKeyword{(}\IdrisFunction{KS} \IdrisBound{A}\IdrisKeyword{)}
  \\\hline
    \IdrisFunction{r} \texttt{"A!"}
    & \IdrisData{Keep} \IdrisBound{A}
    & \IdrisFunction{KS} \IdrisBound{A}
    & \IdrisFunction{KS} \IdrisBound{A}
  \\\hline
\end{tabular}

  \end{center}
  \caption{Regex literal syntax, \IdrisType{RE} and associated shape}
  \label{table:string-syntax}
\end{table}

\aside{}
Idris's definitions have \emph{visibility} qualifiers:
\begin{itemize}
\item \IdrisKeyword{export}: client modules can use the definition
  with its associated type;
\item \IdrisKeyword{public export}: client modules types can also
  rely on the body of this definition;
\item \IdrisKeyword{private} (the default): client modules may not
  refer to this definition directly.
  \endaside{}
\end{itemize}

\subsection{Typed Regexes}
Typed regexes express the parse-tree shape explicitly.  For a typed
regex of type \IdrisType{TyRE} \IdrisBound{a} the result of parsing
will be a parse tree of type \IdrisBound{a} or nothing
(\IdrisType{Maybe} \IdrisBound{a}). \figref{fig:tyre definition}
defines the \IdrisType{TyRE} data structure.  It has constructors of
two kinds.

The first kind are the usual Kleene-algebra operations: empty string
regex, conditional character regex --- where \figref{fig:character
  regex} defines the operations on single-character conditional,
concatenation, alternation and repetition. There is no constructor for
an empty language. We can express the empty language by passing an
impossible predicate to the \IdrisData{CharPred} constructor,
e.g.~\IdrisData{CharPred} \IdrisKeyword{(}\IdrisData{Pred}
\IdrisKeyword{(}\IdrisKeyword{\textbackslash{}} \IdrisBound{\_}
\IdrisKeyword{=>} \IdrisData{False}\IdrisKeyword{))}.

The second kind of constructors manipulate the parse-tree
shape without changing the regex. The \IdrisData{Conv}ersion
constructor transforms the parse-tree according to the given
function. It is this constructor that allows us to implement the
\IdrisType{Functor TyRE} instance we mentioned in
Sec.~\ref{lib_overview}. The \IdrisData{Group} constructor forgets the
parse-tree shape and will result in simply returning the sub-string of
matching characters. By forfeiting structured-parsing, we replace
regex parsing with regex matching. Matching enables automata
minimisation techniques, resulting in faster parsers.

\ignore{
\begin{code}
\IdrisKeyword{namespace}\KatlaSpace{}\IdrisNamespace{TyREDef}\KatlaNewline{}
\KatlaSpace{}\KatlaSpace{}\IdrisKeyword{\%hide}\KatlaSpace{}TyRE\KatlaNewline{}
\KatlaSpace{}\KatlaSpace{}\IdrisKeyword{\%hide}\KatlaSpace{}CharCond\KatlaNewline{}
\end{code}
}
\begin{figure}
\begin{code}
\KatlaSpace{}\KatlaSpace{}\IdrisKeyword{data}\KatlaSpace{}\IdrisType{CharCond}\KatlaSpace{}\IdrisKeyword{:}\KatlaSpace{}\IdrisType{Type}\KatlaSpace{}\IdrisKeyword{where}\KatlaNewline{}
\KatlaSpace{}\KatlaSpace{}\KatlaSpace{}\KatlaSpace{}\IdrisComment{|||\KatlaSpace{}Matches\KatlaSpace{}if\KatlaSpace{}the\KatlaSpace{}character\KatlaSpace{}is\KatlaSpace{}contained\KatlaSpace{}in\KatlaSpace{}the\KatlaSpace{}set}\KatlaNewline{}
\KatlaSpace{}\KatlaSpace{}\KatlaSpace{}\KatlaSpace{}\IdrisData{OneOf}\KatlaSpace{}\IdrisKeyword{:}\KatlaSpace{}\IdrisType{SortedSet}\KatlaSpace{}\IdrisType{Char}\KatlaSpace{}\IdrisKeyword{\KatlaDash{}>}\KatlaSpace{}\IdrisType{CharCond}\KatlaNewline{}
\KatlaSpace{}\KatlaSpace{}\KatlaSpace{}\KatlaSpace{}\IdrisComment{|||\KatlaSpace{}Matches\KatlaSpace{}if\KatlaSpace{}the\KatlaSpace{}character\KatlaSpace{}is\KatlaSpace{}in\KatlaSpace{}the\KatlaSpace{}range\KatlaSpace{}(inclusive)}\KatlaNewline{}
\KatlaSpace{}\KatlaSpace{}\KatlaSpace{}\KatlaSpace{}\IdrisData{Range}\KatlaSpace{}\IdrisKeyword{:}\KatlaSpace{}\IdrisKeyword{(}\IdrisType{Char,}\KatlaSpace{}\IdrisType{Char}\IdrisKeyword{)}\KatlaSpace{}\IdrisKeyword{\KatlaDash{}>}\KatlaSpace{}\IdrisType{CharCond}\KatlaNewline{}
\KatlaSpace{}\KatlaSpace{}\KatlaSpace{}\KatlaSpace{}\IdrisComment{|||\KatlaSpace{}Matches\KatlaSpace{}if\KatlaSpace{}the\KatlaSpace{}character\KatlaSpace{}matches\KatlaSpace{}the\KatlaSpace{}predicate}\KatlaNewline{}
\KatlaSpace{}\KatlaSpace{}\KatlaSpace{}\KatlaSpace{}\IdrisData{Pred}\KatlaSpace{}\IdrisKeyword{:}\KatlaSpace{}\IdrisKeyword{(}\IdrisType{Char}\KatlaSpace{}\IdrisKeyword{\KatlaDash{}>}\KatlaSpace{}\IdrisType{Bool}\IdrisKeyword{)}\KatlaSpace{}\IdrisKeyword{\KatlaDash{}>}\KatlaSpace{}\IdrisType{CharCond}\KatlaNewline{}
\end{code}
\caption{Single character conditionals.}
\label{fig:character regex}
\end{figure}
\begin{figure}
\begin{code}
\KatlaSpace{}\KatlaSpace{}\IdrisKeyword{data}\KatlaSpace{}\IdrisType{TyRE}\KatlaSpace{}\IdrisKeyword{:}\KatlaSpace{}\IdrisType{Type}\KatlaSpace{}\IdrisKeyword{\KatlaDash{}>}\KatlaSpace{}\IdrisType{Type}\KatlaSpace{}\IdrisKeyword{where}\KatlaNewline{}
\KatlaSpace{}\KatlaSpace{}\KatlaSpace{}\KatlaSpace{}\IdrisComment{|||\KatlaSpace{}Regex\KatlaSpace{}for\KatlaSpace{}an\KatlaSpace{}empty\KatlaSpace{}word\KatlaSpace{}(epsilon)}\KatlaNewline{}
\KatlaSpace{}\KatlaSpace{}\KatlaSpace{}\KatlaSpace{}\IdrisData{Empty}\KatlaSpace{}\KatlaSpace{}\KatlaSpace{}\KatlaSpace{}\KatlaSpace{}\IdrisKeyword{:}\KatlaSpace{}\IdrisType{TyRE}\KatlaSpace{}\IdrisType{()}\KatlaNewline{}
\KatlaSpace{}\KatlaSpace{}\KatlaSpace{}\KatlaSpace{}\IdrisComment{|||\KatlaSpace{}Single\KatlaSpace{}character\KatlaSpace{}regex}\KatlaNewline{}
\KatlaSpace{}\KatlaSpace{}\KatlaSpace{}\KatlaSpace{}\IdrisData{MatchChar}\KatlaSpace{}\IdrisKeyword{:}\KatlaSpace{}\IdrisType{CharCond}\KatlaSpace{}\IdrisKeyword{\KatlaDash{}>}\KatlaSpace{}\IdrisType{TyRE}\KatlaSpace{}\IdrisType{Char}\KatlaNewline{}
\KatlaSpace{}\KatlaSpace{}\KatlaSpace{}\KatlaSpace{}\IdrisComment{|||\KatlaSpace{}Concatenation.\KatlaSpace{}A\KatlaSpace{}word\KatlaSpace{}matches\KatlaSpace{}when\KatlaSpace{}it\KatlaSpace{}can\KatlaSpace{}be\KatlaSpace{}divided\KatlaSpace{}into}\KatlaNewline{}
\IdrisComment{\KatlaSpace{}\KatlaSpace{}\KatlaSpace{}\KatlaSpace{}|||\KatlaSpace{}two\KatlaSpace{}subsequent\KatlaSpace{}parts\KatlaSpace{}matching\KatlaSpace{}the\KatlaSpace{}sub\KatlaDash{}regexes\KatlaSpace{}in\KatlaSpace{}order.}\KatlaNewline{}
\KatlaSpace{}\KatlaSpace{}\KatlaSpace{}\KatlaSpace{}\IdrisData{(<*>)}\KatlaSpace{}\KatlaSpace{}\KatlaSpace{}\KatlaSpace{}\KatlaSpace{}\IdrisKeyword{:}\KatlaSpace{}\IdrisType{TyRE}\KatlaSpace{}\IdrisBound{a}\KatlaSpace{}\IdrisKeyword{\KatlaDash{}>}\KatlaSpace{}\IdrisType{TyRE}\KatlaSpace{}\IdrisBound{b}\KatlaSpace{}\IdrisKeyword{\KatlaDash{}>}\KatlaSpace{}\IdrisType{TyRE}\KatlaSpace{}\IdrisKeyword{(}\IdrisBound{a}\IdrisType{,}\KatlaSpace{}\IdrisBound{b}\IdrisKeyword{)}\KatlaNewline{}
\KatlaSpace{}\KatlaSpace{}\KatlaSpace{}\KatlaSpace{}\IdrisComment{|||\KatlaSpace{}Alternation.\KatlaSpace{}A\KatlaSpace{}word\KatlaSpace{}matches\KatlaSpace{}when\KatlaSpace{}it\KatlaSpace{}matches\KatlaSpace{}either\KatlaSpace{}sub\KatlaDash{}regex.}\KatlaNewline{}
\KatlaSpace{}\KatlaSpace{}\KatlaSpace{}\KatlaSpace{}\IdrisData{(<|>)}\KatlaSpace{}\KatlaSpace{}\KatlaSpace{}\KatlaSpace{}\KatlaSpace{}\IdrisKeyword{:}\KatlaSpace{}\IdrisType{TyRE}\KatlaSpace{}\IdrisBound{a}\KatlaSpace{}\IdrisKeyword{\KatlaDash{}>}\KatlaSpace{}\IdrisType{TyRE}\KatlaSpace{}\IdrisBound{b}\KatlaSpace{}\IdrisKeyword{\KatlaDash{}>}\KatlaSpace{}\IdrisType{TyRE}\KatlaSpace{}\IdrisKeyword{(}\IdrisType{Either}\KatlaSpace{}\IdrisBound{a}\KatlaSpace{}\IdrisBound{b}\IdrisKeyword{)}\KatlaNewline{}
\KatlaSpace{}\KatlaSpace{}\KatlaSpace{}\KatlaSpace{}\IdrisComment{|||\KatlaSpace{}Kleene\KatlaSpace{}star.\KatlaSpace{}A\KatlaSpace{}word\KatlaSpace{}matches\KatlaSpace{}when\KatlaSpace{}it\KatlaSpace{}can\KatlaSpace{}be\KatlaSpace{}divided}\KatlaNewline{}
\IdrisComment{\KatlaSpace{}\KatlaSpace{}\KatlaSpace{}\KatlaSpace{}|||\KatlaSpace{}into\KatlaSpace{}n\KatlaSpace{}subsequent\KatlaSpace{}parts\KatlaSpace{}each\KatlaSpace{}matching\KatlaSpace{}the\KatlaSpace{}argument.}\KatlaNewline{}
\KatlaSpace{}\KatlaSpace{}\KatlaSpace{}\KatlaSpace{}\IdrisData{Rep}\KatlaSpace{}\KatlaSpace{}\KatlaSpace{}\KatlaSpace{}\KatlaSpace{}\KatlaSpace{}\KatlaSpace{}\IdrisKeyword{:}\KatlaSpace{}\IdrisType{TyRE}\KatlaSpace{}\IdrisBound{a}\KatlaSpace{}\IdrisKeyword{\KatlaDash{}>}\KatlaSpace{}\IdrisType{TyRE}\KatlaSpace{}\IdrisKeyword{(}\IdrisType{SnocList}\KatlaSpace{}\IdrisBound{a}\IdrisKeyword{)}\KatlaNewline{}
\KatlaSpace{}\KatlaSpace{}\KatlaSpace{}\KatlaSpace{}\IdrisComment{|||\KatlaSpace{}Shape\KatlaSpace{}transformation.}\KatlaNewline{}
\IdrisComment{\KatlaSpace{}\KatlaSpace{}\KatlaSpace{}\KatlaSpace{}|||\KatlaSpace{}Maintain\KatlaSpace{}the\KatlaSpace{}underlying\KatlaSpace{}regex,\KatlaSpace{}and\KatlaSpace{}converse\KatlaSpace{}the\KatlaSpace{}result\KatlaSpace{}parse\KatlaDash{}tree}\KatlaNewline{}
\IdrisComment{\KatlaSpace{}\KatlaSpace{}\KatlaSpace{}\KatlaSpace{}|||\KatlaSpace{}according\KatlaSpace{}to\KatlaSpace{}the\KatlaSpace{}given\KatlaSpace{}function.}\KatlaNewline{}
\KatlaSpace{}\KatlaSpace{}\KatlaSpace{}\KatlaSpace{}\IdrisData{Conv}\KatlaSpace{}\KatlaSpace{}\KatlaSpace{}\KatlaSpace{}\KatlaSpace{}\KatlaSpace{}\IdrisKeyword{:}\KatlaSpace{}\IdrisType{TyRE}\KatlaSpace{}\IdrisBound{a}\KatlaSpace{}\IdrisKeyword{\KatlaDash{}>}\KatlaSpace{}\IdrisKeyword{(}\IdrisBound{a}\KatlaSpace{}\IdrisKeyword{\KatlaDash{}>}\KatlaSpace{}\IdrisBound{b}\IdrisKeyword{)}\KatlaSpace{}\IdrisKeyword{\KatlaDash{}>}\KatlaSpace{}\IdrisType{TyRE}\KatlaSpace{}\IdrisBound{b}\KatlaNewline{}
\KatlaSpace{}\KatlaSpace{}\KatlaSpace{}\KatlaSpace{}\IdrisComment{|||\KatlaSpace{}Forfeit\KatlaSpace{}structured\KatlaSpace{}parsing\KatlaSpace{}and\KatlaSpace{}return\KatlaSpace{}matched\KatlaSpace{}sub\KatlaDash{}string.}\KatlaNewline{}
\IdrisComment{\KatlaSpace{}\KatlaSpace{}\KatlaSpace{}\KatlaSpace{}|||\KatlaSpace{}The\KatlaSpace{}resulting\KatlaSpace{}parser\KatlaSpace{}only\KatlaSpace{}collects\KatlaSpace{}the\KatlaSpace{}consumed\KatlaSpace{}characters.\KatlaSpace{}May}\KatlaNewline{}
\IdrisComment{\KatlaSpace{}\KatlaSpace{}\KatlaSpace{}\KatlaSpace{}|||\KatlaSpace{}generate\KatlaSpace{}a\KatlaSpace{}smaller\KatlaSpace{}(=\KatlaSpace{}faster)\KatlaSpace{}parser.}\KatlaNewline{}
\KatlaSpace{}\KatlaSpace{}\KatlaSpace{}\KatlaSpace{}\IdrisData{Group}\KatlaSpace{}\KatlaSpace{}\KatlaSpace{}\KatlaSpace{}\KatlaSpace{}\IdrisKeyword{:}\KatlaSpace{}\IdrisType{TyRE}\KatlaSpace{}\IdrisBound{a}\KatlaSpace{}\IdrisKeyword{\KatlaDash{}>}\KatlaSpace{}\IdrisType{TyRE}\KatlaSpace{}\IdrisType{String}\KatlaNewline{}
\end{code}
\caption{Typed Regexes}
\label{fig:tyre definition}
\end{figure}
\ignore{
\begin{code}
\IdrisKeyword{\%unhide}\KatlaSpace{}TyRE.Core.TyRE\KatlaNewline{}
\end{code}
}

\section{Moore machine}\label{moore_machine}
\newcommand\Powerset{\mathcal P}
\ignore{
\begin{code}
\IdrisKeyword{import}\KatlaSpace{}\IdrisModule{Data.Regex}\KatlaNewline{}
\IdrisKeyword{import}\KatlaSpace{}\IdrisModule{TyRE.Parser.SM}\KatlaNewline{}
\IdrisKeyword{import}\KatlaSpace{}\IdrisModule{Data.Stream}\KatlaNewline{}
\IdrisKeyword{import}\KatlaSpace{}\IdrisModule{Data.SortedSet}\KatlaNewline{}
\IdrisKeyword{import}\KatlaSpace{}\IdrisModule{Data.List.Elem}\KatlaNewline{}
\IdrisKeyword{import}\KatlaSpace{}\IdrisModule{Data.DPair}\KatlaNewline{}
\end{code}
}
We parse by executing a typed Moore machine.
Typed machine code maintains semantic invariants that may otherwise
require substantial proofs to
guarantee~\citep{morisett-walker-crary-neal-from-system-f-to-typed-assembly-language}.
In a dependently-typed language, it is straightforward to embed such
typed machine-languages~\citep{mckinna2006type}.
We divide the machine
into two conceptually-distinct parts:
\begin{itemize}
  \item \emph{NFA:} a non-deterministic finite automaton executing the matching; and
  \item \emph{routines:} lists of instructions labelling the
    transitions of the NFA, and executing these instructions along an
    accepting path is guaranteed to construct a valid parse-tree.
\end{itemize}
The NFA component governs be the control flow of the Moore machine,
and its routines component governs the machine's cumulative side-effects.

\subsection{Abstractions}\label{moore layer abstractions}
The Moore-machine layer manipulates several key programming
abstractions. We describe these abstractions in this subsection, and the data
types implementing them in \S\ref{moore datatypes}.

\paragraph*{NFAs.}
A \emph{non-deterministic finite-state automaton} over a set of input
symbols $\Sigma$ --- called the \emph{alphabet} --- is a tuple $(Q, S,
q_{acc}, next, P)$ consisting of:
\begin{itemize}
    \item $Q$: a finite set of \emph{states};
    \item $S \subseteq Q$: a set of \emph{starting/initial} states;
    \item $\mathrm{acc} \in Q$: a distinguished \emph{accepting} state;
    \item $next : (Q \setminus \{\mathrm{acc}\}) \times \Sigma \rightarrow
      \Powerset Q$: a relation, the \emph{transition} relation.\\
      (We denote the power set by $\Powerset$.)
\end{itemize}
Each $q_2 \in next(q_1, a)$ represents a transition $q_1
\xrightarrow{a} q_2$ from state $q_1$ to $q_2$ labelled by the symbol
$a$.  In this formulation, the accepting state has no arrows out of
it, and we disallow silent transitions, often known as epsilon
transitions. This choice does not affect the class of parsers we can
express. We restrict attention to automata over
\IdrisType{Char}acters.

\paragraph*{Runtime model.}
As we traverse the NFA, we build parts of the parse tree, collecting
them on a heterogeneous stack.  More precisely:
\begin{itemize}
\item
  To ensure that the stack's values form a valid parse-tree upon
  acceptance, we index types by the \emph{stack shape} --- a left-nested list of
  types, one for each stack position: \ignore{
\begin{code}
\IdrisKeyword{namespace}\KatlaSpace{}\IdrisNamespace{Foo}\KatlaNewline{}
\KatlaSpace{}\IdrisKeyword{public}\KatlaSpace{}\IdrisKeyword{export}\KatlaNewline{}
\end{code}
}
\begin{code}
\KatlaSpace{}\IdrisFunction{Shape}\KatlaSpace{}\IdrisKeyword{:}\KatlaSpace{}\IdrisType{Type}\KatlaNewline{}
\KatlaSpace{}\IdrisFunction{Shape}\KatlaSpace{}\IdrisKeyword{=}\KatlaSpace{}\IdrisType{SnocList}\KatlaSpace{}\IdrisType{Type}\KatlaNewline{}
\end{code}
\item
  We execute a pool of threads transitioning through the NFA state:
\ignore{
\begin{code}
\IdrisKeyword{data}\KatlaNewline{}
\end{code}
}
\begin{code}
\KatlaSpace{}\IdrisType{ThreadData}\KatlaSpace{}\IdrisKeyword{:}\KatlaSpace{}\IdrisFunction{Shape}\KatlaSpace{}\IdrisKeyword{\KatlaDash{}>}\KatlaSpace{}\IdrisType{Type}\KatlaNewline{}
\end{code}
\item
  Each thread-data may record the currently matched sub-string, and
  maintains a \emph{stack} of the partially constructed parse-tree:
\ignore{
\begin{code}
\IdrisKeyword{data}\KatlaNewline{}
\end{code}
}
\begin{code}
\KatlaSpace{}\IdrisType{Stack}\KatlaSpace{}\KatlaSpace{}\KatlaSpace{}\KatlaSpace{}\KatlaSpace{}\KatlaSpace{}\IdrisKeyword{:}\KatlaSpace{}\IdrisFunction{Shape}\KatlaSpace{}\IdrisKeyword{\KatlaDash{}>}\KatlaSpace{}\IdrisType{Type}\KatlaNewline{}
\end{code}
\end{itemize}

\paragraph*{Routines.}
A routine is a list of instructions, each affecting the thread data. To
ensure that instructions produce parse-trees of the correct type, we
index each instruction, and therefore each routine, by an assumption
about the stack shape before its execution and the stack shape it
guarantees after its execution:  \ignore{
\begin{code}
\IdrisKeyword{data}\KatlaNewline{}
\end{code}
}
\begin{code}
\KatlaSpace{}\IdrisType{Instruction}\KatlaSpace{}\IdrisKeyword{:}\KatlaSpace{}\IdrisKeyword{(}\IdrisBound{pre}\IdrisKeyword{,}\KatlaSpace{}\IdrisBound{post}\KatlaSpace{}\IdrisKeyword{:}\KatlaSpace{}\IdrisFunction{Shape}\IdrisKeyword{)}\KatlaSpace{}\IdrisKeyword{\KatlaDash{}>}\KatlaSpace{}\IdrisType{Type}\KatlaNewline{}
\KatlaSpace{}\IdrisComment{\KatlaDash{}\KatlaDash{}\KatlaSpace{}\KatlaSpace{}\KatlaSpace{}\KatlaSpace{}\KatlaSpace{}\KatlaSpace{}\KatlaSpace{}\KatlaSpace{}\KatlaSpace{}\KatlaSpace{}\KatlaSpace{}\KatlaSpace{}\KatlaSpace{}^\KatlaDash{}\KatlaDash{}\KatlaDash{}\KatlaDash{}\KatlaDash{}\KatlaDash{}\KatlaDash{}\KatlaDash{}\KatlaDash{}\KatlaDash{}\KatlaDash{}\KatlaDash{}\KatlaDash{}\KatlaDash{}\KatlaDash{}^}\KatlaNewline{}
\end{code}
\ignore{
\begin{code}
\IdrisKeyword{\%hide}\KatlaSpace{}Main.Instruction\KatlaNewline{}
\IdrisKeyword{data}\KatlaNewline{}
\end{code}
}
\vspace{-2\baselineskip}
\vspace{-1pt}

\begin{code}
\KatlaSpace{}\IdrisComment{\KatlaDash{}\KatlaDash{}\KatlaSpace{}(stack\KatlaSpace{}shape\KatlaSpace{}before\KatlaSpace{}and\KatlaSpace{}after\KatlaSpace{}executing\KatlaSpace{}the\KatlaSpace{}instruction/routine)}\KatlaNewline{}
\KatlaSpace{}\IdrisComment{\KatlaDash{}\KatlaDash{}\KatlaSpace{}\KatlaSpace{}\KatlaSpace{}\KatlaSpace{}\KatlaSpace{}\KatlaSpace{}\KatlaSpace{}\KatlaSpace{}\KatlaSpace{}\KatlaSpace{}\KatlaSpace{}\KatlaSpace{}\KatlaSpace{}v\KatlaDash{}\KatlaDash{}\KatlaDash{}\KatlaDash{}\KatlaDash{}\KatlaDash{}\KatlaDash{}\KatlaDash{}\KatlaDash{}\KatlaDash{}\KatlaDash{}\KatlaDash{}\KatlaDash{}\KatlaDash{}\KatlaDash{}v}\KatlaNewline{}
\KatlaSpace{}\IdrisType{Routine}\KatlaSpace{}\KatlaSpace{}\KatlaSpace{}\KatlaSpace{}\KatlaSpace{}\IdrisKeyword{:}\KatlaSpace{}\IdrisKeyword{(}\IdrisBound{pre}\IdrisKeyword{,}\KatlaSpace{}\IdrisBound{post}\KatlaSpace{}\IdrisKeyword{:}\KatlaSpace{}\IdrisFunction{Shape}\IdrisKeyword{)}\KatlaSpace{}\IdrisKeyword{\KatlaDash{}>}\KatlaSpace{}\IdrisType{Type}\KatlaNewline{}
\end{code}
One of the instructions --- pushing the last-read symbol onto the
stack --- is not valid before the machine has consumed a symbol. We therefore
also designate which instructions can execute when we initialise the machine:
\ignore{
\begin{code}
\IdrisKeyword{\%hide}\KatlaSpace{}Main.Routine\KatlaNewline{}
\IdrisKeyword{data}\KatlaNewline{}
\end{code}
}
\begin{code}
\KatlaSpace{}\IdrisType{IsInitRoutine}\KatlaSpace{}\IdrisKeyword{:}\KatlaSpace{}\IdrisType{Routine}\KatlaSpace{}\IdrisBound{xs}\KatlaSpace{}\IdrisBound{ys}\KatlaSpace{}\IdrisKeyword{\KatlaDash{}>}\KatlaSpace{}\IdrisType{Type}\KatlaNewline{}
\end{code}

\paragraph*{Typed Moore Machines.}
Since our NFAs include a single, distinguished, accepting state, we
will implement the states of the Moore machine using the
\IdrisType{Maybe} type constructor, with the distinguished state
\IdrisData{Nothing} representing the accepting state. We will index
Moore machines by their result parse-tree type:
\ignore{
\begin{code}
\IdrisKeyword{data}\KatlaNewline{}
\end{code}
}
\begin{code}
\KatlaSpace{}\IdrisType{MooreMachine}\KatlaSpace{}\IdrisKeyword{:}\KatlaSpace{}\IdrisKeyword{(}\IdrisBound{t}\KatlaSpace{}\IdrisKeyword{:}\KatlaSpace{}\IdrisType{Type}\IdrisKeyword{)}\KatlaSpace{}\IdrisKeyword{\KatlaDash{}>}\KatlaSpace{}\IdrisType{Type}\KatlaNewline{}
\end{code}
Each \IdrisType{MooreMachine} assigns expected stack shapes to each
non-accepting state:
\ignore{
\begin{code}
\IdrisKeyword{namespace}\KatlaSpace{}\IdrisNamespace{Hide}\KatlaNewline{}
\end{code}
}
\begin{code}
\KatlaSpace{}\IdrisFunction{lookup}\KatlaSpace{}\IdrisKeyword{:}\KatlaSpace{}\IdrisBound{s}\KatlaSpace{}\IdrisKeyword{\KatlaDash{}>}\KatlaSpace{}\IdrisFunction{Shape}\KatlaNewline{}
\end{code}
and so together with the result parse-tree type \IdrisBound{t}, we
associate a stack shape to \emph{every} state:\label{shapeOf location}
\ignore{
\begin{code}
\IdrisKeyword{namespace}\KatlaSpace{}\IdrisNamespace{MM2}\KatlaNewline{}
\end{code}
}
\begin{code}
\KatlaSpace{}\IdrisFunction{shapeOf}\KatlaSpace{}\IdrisKeyword{:}\KatlaSpace{}\IdrisKeyword{(}\IdrisBound{lookup}\KatlaSpace{}\IdrisKeyword{:}\KatlaSpace{}\IdrisBound{s}\KatlaSpace{}\IdrisKeyword{\KatlaDash{}>}\KatlaSpace{}\IdrisFunction{Shape}\IdrisKeyword{)}\KatlaSpace{}\IdrisKeyword{\KatlaDash{}>}\KatlaSpace{}\IdrisKeyword{(}\IdrisBound{t}\KatlaSpace{}\IdrisKeyword{:}\KatlaSpace{}\IdrisType{Type}\IdrisKeyword{)}\KatlaSpace{}\IdrisKeyword{\KatlaDash{}>}\KatlaNewline{}
\KatlaSpace{}\KatlaSpace{}\KatlaSpace{}\KatlaSpace{}\KatlaSpace{}\KatlaSpace{}\KatlaSpace{}\KatlaSpace{}\KatlaSpace{}\KatlaSpace{}\KatlaSpace{}\IdrisType{Maybe}\KatlaSpace{}\IdrisBound{s}\KatlaSpace{}\IdrisKeyword{\KatlaDash{}>}\KatlaSpace{}\IdrisFunction{Shape}\KatlaNewline{}
\KatlaSpace{}\IdrisFunction{shapeOf}\KatlaSpace{}\IdrisBound{lookup}\KatlaSpace{}\IdrisBound{t}\KatlaSpace{}\IdrisKeyword{(}\IdrisData{Just}\KatlaSpace{}\IdrisBound{s}\IdrisKeyword{)}\KatlaSpace{}\IdrisKeyword{=}\KatlaSpace{}\IdrisBound{lookup}\KatlaSpace{}\IdrisBound{s}\KatlaSpace{}\KatlaSpace{}\KatlaSpace{}\IdrisComment{\KatlaDash{}\KatlaDash{}\KatlaSpace{}non\KatlaDash{}accepting\KatlaSpace{}state}\KatlaNewline{}
\KatlaSpace{}\IdrisFunction{shapeOf}\KatlaSpace{}\IdrisBound{lookup}\KatlaSpace{}\IdrisBound{t}\KatlaSpace{}\IdrisData{Nothing}\KatlaSpace{}\IdrisKeyword{=}\KatlaSpace{}\IdrisData{[<}\KatlaSpace{}\IdrisBound{t}\KatlaSpace{}\IdrisData{]}\KatlaSpace{}\KatlaSpace{}\KatlaSpace{}\KatlaSpace{}\KatlaSpace{}\KatlaSpace{}\IdrisComment{\KatlaDash{}\KatlaDash{}\KatlaSpace{}\KatlaSpace{}\KatlaSpace{}\KatlaSpace{}\KatlaSpace{}accepting\KatlaSpace{}state}\KatlaNewline{}
\end{code}

\subsection{Aside: Idris preliminaries}
We need some more Idris features to discuss the Moore machine layer
implementation.

\paragraph*{Quantities.}
A substantial innovation in Idris~2~\citep{brady:idris2} is its
\emph{quantity annotations}, following McBride's (unpublished) and
Atkey's~\citeyearpar{Atkey2018SyntaxTheory} \emph{quantitative type
  theory}. For example, we may define the following \emph{semi-erased}
sum type:
\begin{code}
\IdrisKeyword{data}\KatlaSpace{}\IdrisType{EitherErased}\KatlaSpace{}\IdrisKeyword{:}\KatlaSpace{}\IdrisType{Type}\KatlaSpace{}\IdrisKeyword{\KatlaDash{}>}\KatlaSpace{}\IdrisType{Type}\KatlaSpace{}\IdrisKeyword{\KatlaDash{}>}\KatlaSpace{}\IdrisType{Type}\KatlaSpace{}\IdrisKeyword{where}\KatlaNewline{}
\KatlaSpace{}\KatlaSpace{}\IdrisData{Left}\KatlaSpace{}\KatlaSpace{}\IdrisKeyword{:}\KatlaSpace{}\IdrisKeyword{(}\KatlaSpace{}\KatlaSpace{}\IdrisBound{x}\KatlaSpace{}\IdrisKeyword{:}\KatlaSpace{}\IdrisBound{a}\IdrisKeyword{)}\KatlaSpace{}\IdrisKeyword{\KatlaDash{}>}\KatlaSpace{}\IdrisType{EitherErased}\KatlaSpace{}\IdrisBound{a}\KatlaSpace{}\IdrisBound{b}\KatlaNewline{}
\KatlaSpace{}\KatlaSpace{}\IdrisData{Right}\KatlaSpace{}\IdrisKeyword{:}\KatlaSpace{}\IdrisKeyword{(0}\KatlaSpace{}\IdrisBound{y}\KatlaSpace{}\IdrisKeyword{:}\KatlaSpace{}\IdrisBound{b}\IdrisKeyword{)}\KatlaSpace{}\IdrisKeyword{\KatlaDash{}>}\KatlaSpace{}\IdrisType{EitherErased}\KatlaSpace{}\IdrisBound{a}\KatlaSpace{}\IdrisBound{b}\KatlaNewline{}
\end{code}
Its \IdrisData{Left} constructor is similar to the more familiar,
standard, constructor of the \IdrisType{Either} type
constructor. However, the argument to its \IdrisData{Right}
constructor is annotated with the quantity \IdrisKeyword{0}, meaning
this argument will be \emph{erased} at runtime. Thus, \IdrisData{Right
}\IdrisBound{y} is represented at runtime as a tag, meaning values of
\IdrisType{EitherErased }\IdrisBound{a b} have the same runtime
representation as values of the type
\IdrisType{Maybe }\IdrisBound{a}. Idris 2 supports the quantities:
\begin{itemize}
\item runtime-erased (\IdrisKeyword{0});
\item linear (\IdrisKeyword{1}): the variable must be used exactly
  once at run-time; and
\item any (unannotated, the default): the variable may be used any
  number of times.
\end{itemize}

\paragraph*{Records.}
Idris 2's \emph{records} desugar into one-constructor \IdrisKeyword{data}
declarations together with projection functions for each
\emph{field}. For example, here is the definition of a \emph{dependent
  pair}:
\ignore{
\begin{code}
\IdrisKeyword{namespace}\KatlaSpace{}\IdrisNamespace{Presugar}\KatlaNewline{}
\end{code}
}
\begin{code}
\KatlaSpace{}\IdrisKeyword{record}\KatlaSpace{}\IdrisType{DPair}\KatlaSpace{}\IdrisBound{a}\KatlaSpace{}\IdrisKeyword{(}\IdrisBound{b}\KatlaSpace{}\IdrisKeyword{:}\KatlaSpace{}\IdrisBound{a}\KatlaSpace{}\IdrisKeyword{\KatlaDash{}>}\KatlaSpace{}\IdrisType{Type}\IdrisKeyword{)}\KatlaSpace{}\IdrisKeyword{where}\KatlaNewline{}
\KatlaSpace{}\KatlaSpace{}\KatlaSpace{}\IdrisKeyword{constructor}\KatlaSpace{}\IdrisData{MkDPair}\KatlaNewline{}
\KatlaSpace{}\KatlaSpace{}\KatlaSpace{}\IdrisFunction{fst}\KatlaSpace{}\IdrisKeyword{:}\KatlaSpace{}\IdrisBound{a}\KatlaNewline{}
\KatlaSpace{}\KatlaSpace{}\KatlaSpace{}\IdrisFunction{snd}\KatlaSpace{}\IdrisKeyword{:}\KatlaSpace{}\IdrisBound{b}\KatlaSpace{}\IdrisBound{fst}\KatlaNewline{}
\end{code}
which desugars\footnotemark{} into a data declaration:
\protect\footnotetext{Agda cognoscenti might expect records to support
  $\eta$-equality, meaning the judgemental equality:\\
\KatlaSpace{}\KatlaSpace{}\IdrisBound{rec}\KatlaSpace{}\IdrisKeyword{=}\KatlaSpace{}\IdrisKeyword{(}\IdrisData{MkDPair}\KatlaSpace{}\IdrisBound{rec}\IdrisFunction{.fst}\KatlaSpace{}\IdrisBound{rec}\IdrisFunction{.snd}\IdrisKeyword{)}\KatlaNewline{}\\
Since ordinary \IdrisKeyword{data} declarations don't support such
$\eta$-equality judgementally, Agda records can't be simply desugared
into single-constructor \IdrisKeyword{data} declaractions.
While Idris may in the future support $\eta$-equality for records, it
does not do so at the moment, and desugars records accordingly.
}
\ignore{
\begin{code}
\IdrisKeyword{namespace}\KatlaSpace{}\IdrisNamespace{Desugar}\KatlaNewline{}
\KatlaSpace{}\IdrisKeyword{\%hide}\KatlaSpace{}Builtin.DPair.DPair\KatlaNewline{}
\KatlaSpace{}\IdrisKeyword{\%hide}\KatlaSpace{}Main.Presugar.DPair\KatlaNewline{}
\end{code}
}
\begin{code}
\KatlaSpace{}\IdrisKeyword{data}\KatlaSpace{}\IdrisType{DPair}\KatlaSpace{}\IdrisKeyword{:}\KatlaSpace{}\IdrisKeyword{(}\IdrisBound{a}\KatlaSpace{}\IdrisKeyword{:}\KatlaSpace{}\IdrisType{Type}\IdrisKeyword{)}\KatlaSpace{}\IdrisKeyword{\KatlaDash{}>}\KatlaSpace{}\IdrisKeyword{(}\IdrisBound{b}\KatlaSpace{}\IdrisKeyword{:}\KatlaSpace{}\IdrisBound{a}\KatlaSpace{}\IdrisKeyword{\KatlaDash{}>}\KatlaSpace{}\IdrisType{Type}\IdrisKeyword{)}\KatlaSpace{}\IdrisKeyword{\KatlaDash{}>}\KatlaSpace{}\IdrisType{Type}\KatlaSpace{}\IdrisKeyword{where}\KatlaNewline{}
\KatlaSpace{}\KatlaSpace{}\KatlaSpace{}\IdrisData{MkDPair}\KatlaSpace{}\IdrisKeyword{:}\KatlaSpace{}\IdrisKeyword{(}\IdrisBound{fst}\KatlaSpace{}\IdrisKeyword{:}\KatlaSpace{}\IdrisBound{a}\IdrisKeyword{)}\KatlaSpace{}\IdrisKeyword{\KatlaDash{}>}\KatlaSpace{}\IdrisKeyword{(}\IdrisBound{snd}\KatlaSpace{}\IdrisKeyword{:}\KatlaSpace{}\IdrisBound{b}\KatlaSpace{}\IdrisBound{fst}\IdrisKeyword{)}\KatlaSpace{}\IdrisKeyword{\KatlaDash{}>}\KatlaSpace{}\IdrisType{DPair}\KatlaSpace{}\IdrisBound{a}\KatlaSpace{}\IdrisBound{b}\KatlaNewline{}
\end{code}
and two definitions, called \emph{projections}, that use Idris's
post-fix notation:
\begin{code}
\KatlaSpace{}\IdrisFunction{(.fst)}\KatlaSpace{}\IdrisKeyword{:}\KatlaSpace{}\IdrisType{DPair}\KatlaSpace{}\IdrisBound{a}\KatlaSpace{}\IdrisBound{b}\KatlaSpace{}\IdrisKeyword{\KatlaDash{}>}\KatlaSpace{}\IdrisBound{a}\KatlaNewline{}
\KatlaSpace{}\IdrisKeyword{(}\IdrisData{MkDPair}\KatlaSpace{}\IdrisBound{x}\KatlaSpace{}\IdrisBound{y}\IdrisKeyword{)}\IdrisFunction{.fst}\KatlaSpace{}\IdrisKeyword{=}\KatlaSpace{}\IdrisBound{x}\KatlaNewline{}
\KatlaSpace{}\IdrisFunction{(.snd)}\KatlaSpace{}\IdrisKeyword{:}\KatlaSpace{}\IdrisKeyword{(}\IdrisBound{rec}\KatlaSpace{}\IdrisKeyword{:}\KatlaSpace{}\IdrisType{DPair}\KatlaSpace{}\IdrisBound{a}\KatlaSpace{}\IdrisBound{b}\IdrisKeyword{)}\KatlaSpace{}\IdrisKeyword{\KatlaDash{}>}\KatlaSpace{}\IdrisBound{b}\KatlaSpace{}\IdrisBound{rec}\IdrisFunction{.fst}\KatlaNewline{}
\KatlaSpace{}\IdrisKeyword{(}\IdrisData{MkDPair}\KatlaSpace{}\IdrisBound{x}\KatlaSpace{}\IdrisBound{y}\IdrisKeyword{)}\IdrisFunction{.snd}\KatlaSpace{}\IdrisKeyword{=}\KatlaSpace{}\IdrisBound{y}\KatlaNewline{}
\end{code}
\ignore{
\begin{code}
\IdrisKeyword{\%unhide}\KatlaSpace{}Builtin.DPair.DPair\KatlaNewline{}
\end{code}
}
Idris includes special syntax for dependent pairs, where the type:
\ignore{
\begin{code}
\IdrisKeyword{\%hide}\KatlaSpace{}Main.Stack\KatlaNewline{}
\IdrisFunction{BlahBlah}\KatlaSpace{}\IdrisKeyword{:}\KatlaNewline{}
\end{code}
}
\begin{code}
\KatlaSpace{}\IdrisType{(}\IdrisBound{shape}\KatlaSpace{}\IdrisKeyword{:}\KatlaSpace{}\IdrisFunction{Shape}\KatlaSpace{}\IdrisType{**}\KatlaSpace{}\IdrisType{Stack}\KatlaSpace{}\IdrisBound{shape}\IdrisType{)}\KatlaNewline{}
\end{code}
desugars into the dependent pair type:
\ignore{
\begin{code}
\IdrisFunction{BlahBlahDesugar}\KatlaSpace{}\IdrisKeyword{:}\KatlaNewline{}
\end{code}
}
\begin{code}
\KatlaSpace{}\IdrisType{DPair}\KatlaSpace{}\IdrisFunction{Shape}\KatlaSpace{}\IdrisKeyword{(\textbackslash{}}\IdrisBound{shape}\KatlaSpace{}\IdrisKeyword{=>}\KatlaSpace{}\IdrisType{Stack}\KatlaSpace{}\IdrisBound{shape}\IdrisKeyword{)}\KatlaNewline{}
\end{code}
and the tuple of a \IdrisFunction{Shape} and an appropriately-shaped \IdrisType{Stack}:
\ignore{
\begin{code}
\IdrisFunction{BlahBlah}\KatlaSpace{}\IdrisKeyword{=}\KatlaNewline{}
\end{code}
}
\begin{code}
\KatlaSpace{}(\IdrisData{[<}\KatlaSpace{}\IdrisType{Nat}\IdrisData{,}\KatlaSpace{}\IdrisType{Bool}\IdrisData{]}\KatlaSpace{}\IdrisData{**}\KatlaSpace{}\IdrisData{[<}\KatlaSpace{}\IdrisData{42,}\KatlaSpace{}\IdrisData{True]})\KatlaNewline{}
\end{code}
desugars into the dependent pair:
\ignore{
\begin{code}
\IdrisFunction{BlahBlahDesugar}\KatlaSpace{}\IdrisKeyword{=}\KatlaNewline{}
\end{code}
}
\begin{code}
\KatlaSpace{}\IdrisData{MkDPair}\KatlaSpace{}\IdrisData{[<}\KatlaSpace{}\IdrisType{Nat}\IdrisData{,}\KatlaSpace{}\IdrisType{Bool}\IdrisData{]}\KatlaSpace{}\IdrisData{[<}\KatlaSpace{}\IdrisData{42,}\KatlaSpace{}\IdrisData{True]}\KatlaNewline{}
\end{code}

Record fields may include quantity annotations, for example, Idris's standard
library includes records where one field is erased at runtime:
\ignore{
\begin{code}
\IdrisKeyword{namespace}\KatlaSpace{}\IdrisNamespace{Desugar}\KatlaNewline{}
\end{code}
}
\begin{code}
\KatlaSpace{}\IdrisKeyword{record}\KatlaSpace{}\IdrisType{Subset}\KatlaSpace{}\IdrisBound{a}\KatlaSpace{}\IdrisKeyword{(}\IdrisBound{b}\KatlaSpace{}\IdrisKeyword{:}\KatlaSpace{}\IdrisBound{a}\KatlaSpace{}\IdrisKeyword{\KatlaDash{}>}\KatlaSpace{}\IdrisType{Type}\IdrisKeyword{)}\KatlaSpace{}\IdrisKeyword{where}\KatlaNewline{}
\KatlaSpace{}\KatlaSpace{}\KatlaSpace{}\IdrisKeyword{constructor}\KatlaSpace{}\IdrisData{Element}\KatlaNewline{}
\KatlaSpace{}\KatlaSpace{}\KatlaSpace{}\IdrisFunction{fst}\KatlaSpace{}\IdrisKeyword{:}\KatlaSpace{}\IdrisBound{a}\KatlaNewline{}
\KatlaSpace{}\KatlaSpace{}\KatlaSpace{}\IdrisKeyword{0}\KatlaSpace{}\IdrisFunction{snd}\KatlaSpace{}\IdrisKeyword{:}\KatlaSpace{}\IdrisBound{b}\KatlaSpace{}\IdrisBound{fst}\KatlaNewline{}
\end{code}
We will use this dependent pair to exclude some values from a given
type, by requiring a dependent pair whose (\IdrisFunction{.snd}) field
contains a proof the value is allowed. For example, the type
\IdrisType{IsInitInstruction} of initialisation-safe
instructions. Since this field is erased at run-time, we will only be
passing around the instruction, not its safety proof.

\paragraph*{Implicit arguments.}
Idris supports two kinds of implicit arguments, which it tries to
construct during elaboration:
\begin{itemize}
\item \emph{Ordinary} implicit arguments will be elaborated using its
  dynamic pattern unification
  algorithm~\citep{miller:unification,reed:unification,gundry:thesis}. We
  declare them using braces, for example, we will later define the
  runtime state of a thread to be the following record, namely
  its NFA state and associated \IdrisType{ThreadData}: \ignore{
\begin{code}
\IdrisKeyword{namespace}\KatlaSpace{}\IdrisNamespace{ImplicitArg}\KatlaNewline{}
\KatlaSpace{}\IdrisKeyword{\%hide}\KatlaSpace{}Main.MooreMachine\KatlaNewline{}
\KatlaSpace{}\IdrisFunction{shapeOf}\KatlaSpace{}\IdrisKeyword{:}\KatlaSpace{}\IdrisKeyword{(}\IdrisBound{lookup}\KatlaSpace{}\IdrisKeyword{:}\KatlaSpace{}\IdrisBound{s}\KatlaSpace{}\IdrisKeyword{\KatlaDash{}>}\KatlaSpace{}\IdrisFunction{Shape}\IdrisKeyword{)}\KatlaSpace{}\IdrisKeyword{\KatlaDash{}>}\KatlaSpace{}\IdrisKeyword{(}\IdrisBound{t}\KatlaSpace{}\IdrisKeyword{:}\KatlaSpace{}\IdrisType{Type}\IdrisKeyword{)}\KatlaSpace{}\IdrisKeyword{\KatlaDash{}>}\KatlaNewline{}
\KatlaSpace{}\KatlaSpace{}\KatlaSpace{}\KatlaSpace{}\KatlaSpace{}\KatlaSpace{}\KatlaSpace{}\KatlaSpace{}\KatlaSpace{}\KatlaSpace{}\KatlaSpace{}\IdrisType{Maybe}\KatlaSpace{}\IdrisBound{s}\KatlaSpace{}\IdrisKeyword{\KatlaDash{}>}\KatlaSpace{}\IdrisFunction{Shape}\KatlaNewline{}
\KatlaSpace{}\IdrisFunction{MooreMachine}\KatlaSpace{}\IdrisKeyword{:}\KatlaSpace{}\IdrisType{Type}\KatlaSpace{}\IdrisKeyword{\KatlaDash{}>}\KatlaSpace{}\IdrisType{Type}\KatlaNewline{}
\KatlaSpace{}\IdrisFunction{MooreMachine}\KatlaSpace{}\IdrisBound{t}\KatlaSpace{}\IdrisKeyword{=}\KatlaSpace{}\IdrisType{SM}\KatlaSpace{}\IdrisBound{t}\KatlaNewline{}
\end{code}
}
\begin{code}
\KatlaSpace{}\IdrisKeyword{record}\KatlaSpace{}\IdrisType{Thread}\KatlaSpace{}\IdrisKeyword{\{}\IdrisBound{t}\KatlaSpace{}\IdrisKeyword{:}\KatlaSpace{}\IdrisType{Type}\IdrisKeyword{\}}\KatlaSpace{}\IdrisKeyword{(}\IdrisBound{machine}\KatlaSpace{}\IdrisKeyword{:}\KatlaSpace{}\IdrisFunction{MooreMachine}\KatlaSpace{}\IdrisBound{t}\IdrisKeyword{)}\KatlaSpace{}\IdrisKeyword{where}\KatlaNewline{}
\KatlaSpace{}\KatlaSpace{}\KatlaSpace{}\IdrisKeyword{constructor}\KatlaSpace{}\IdrisData{MkThread}\KatlaNewline{}
\KatlaSpace{}\KatlaSpace{}\KatlaSpace{}\IdrisFunction{state}\KatlaSpace{}\IdrisKeyword{:}\KatlaSpace{}\IdrisType{Maybe}\KatlaSpace{}\IdrisBound{machine}\IdrisFunction{.s}\KatlaNewline{}
\KatlaSpace{}\KatlaSpace{}\KatlaSpace{}\IdrisFunction{tddata}\KatlaSpace{}\IdrisKeyword{:}\KatlaSpace{}\IdrisType{ThreadData}\KatlaSpace{}\IdrisKeyword{(}\IdrisFunction{shapeOf}\KatlaSpace{}\IdrisBound{machine}\IdrisFunction{.lookup}\KatlaSpace{}\IdrisBound{t}\KatlaSpace{}\IdrisBound{state}\IdrisKeyword{)}\KatlaNewline{}
\end{code}
Users need only mention \IdrisType{Thread }\IdrisBound{machine}, since the
elaborator can deduce the expected result parse-tree type
\IdrisBound{t} from the type of the given Moore machine
\IdrisBound{machine}. Idris adopts by default two conventions for implicit
arguments:
\begin{itemize}
\item the keyword \IdrisKeyword{forall} \IdrisBound{a}\IdrisKeyword{.}
  binds an implicit argument called \IdrisBound{a} at quantity
  \IdrisKeyword{0}, trying to infer its type using unification;
\item unbound implicit arguments in type declarations are implicitly
  bound as implicit arguments at quantity \IdrisKeyword{0} in the
  beginning of the declaration. This feature, which may be turned off,
  can produce more succinct type-declaration, similar to the ones
  found in languages with prenex polymorphism.
\end{itemize}
\item \emph{Proof-search} implicit arguments will be elaborated, in
  addition to unification, by following user-declared
  \emph{unification hints}~\citep{asperti-et-al:hints-in-unification},
  attempting to use data/type constructors and record projections, and
  to plug-in bound variables. We declare proof-search implicit
  arguments using the \IdrisKeyword{auto} keyword we mentioned on
  page~\pageref{auto-exposition}. For example, we will later define the Moore
  machines as a record with a proof-search implicit argument:
\ignore{
\begin{code}
\IdrisKeyword{namespace}\KatlaSpace{}\IdrisNamespace{AutoImplicit}\KatlaNewline{}
\end{code}
}
\begin{code}
\KatlaSpace{}\IdrisKeyword{record}\KatlaSpace{}\IdrisType{MooreMachine}\KatlaSpace{}\IdrisKeyword{(}\IdrisBound{shape}\KatlaSpace{}\IdrisKeyword{:}\KatlaSpace{}\IdrisType{Type}\IdrisKeyword{)}\KatlaSpace{}\IdrisKeyword{where}\KatlaNewline{}
\KatlaSpace{}\KatlaSpace{}\KatlaSpace{}\IdrisKeyword{constructor}\KatlaSpace{}\IdrisData{MkMooreMachine}\KatlaNewline{}
\KatlaSpace{}\KatlaSpace{}\KatlaSpace{}\IdrisKeyword{0}\KatlaSpace{}\IdrisFunction{s}\KatlaSpace{}\IdrisKeyword{:}\KatlaSpace{}\IdrisType{Type}\KatlaNewline{}
\KatlaSpace{}\KatlaSpace{}\KatlaSpace{}\IdrisKeyword{0}\KatlaSpace{}\IdrisFunction{lookup}\KatlaSpace{}\IdrisKeyword{:}\KatlaSpace{}\IdrisBound{s}\KatlaSpace{}\IdrisKeyword{\KatlaDash{}>}\KatlaSpace{}\IdrisFunction{Shape}\KatlaNewline{}
\KatlaSpace{}\KatlaSpace{}\KatlaSpace{}\IdrisKeyword{\{auto}\KatlaSpace{}\IdrisFunction{isEq}\KatlaSpace{}\IdrisKeyword{:}\KatlaSpace{}\IdrisType{Eq}\KatlaSpace{}\IdrisBound{s}\IdrisKeyword{\}}\KatlaNewline{}
\KatlaSpace{}\KatlaSpace{}\KatlaSpace{}\IdrisFunction{init}\KatlaSpace{}\IdrisKeyword{:}\KatlaSpace{}\IdrisFunction{InitStatesType}\KatlaSpace{}\IdrisBound{shape}\KatlaSpace{}\IdrisBound{s}\KatlaSpace{}\IdrisBound{lookup}\KatlaNewline{}
\KatlaSpace{}\KatlaSpace{}\KatlaSpace{}\IdrisFunction{next}\KatlaSpace{}\IdrisKeyword{:}\KatlaSpace{}\IdrisFunction{TransitionRelation}\KatlaSpace{}\IdrisBound{shape}\KatlaSpace{}\IdrisBound{s}\KatlaSpace{}\IdrisBound{lookup}\KatlaNewline{}
\end{code}
\ignore{
\begin{code}
\KatlaSpace{}\IdrisKeyword{\%hide}\KatlaSpace{}TransitionRelation\KatlaNewline{}
\end{code}
}
The type \IdrisType{Eq }\IdrisBound{s} is a record with a field for an
equality-predicate on the machine's state type \IdrisBound{s}:
\begin{code}
\KatlaSpace{}\IdrisFunction{(==)}\KatlaSpace{}\IdrisKeyword{:}\KatlaSpace{}\IdrisBound{s}\KatlaSpace{}\IdrisKeyword{\KatlaDash{}>}\KatlaSpace{}\IdrisBound{s}\KatlaSpace{}\IdrisKeyword{\KatlaDash{}>}\KatlaSpace{}\IdrisType{Bool}\KatlaNewline{}
\end{code}
By designating a proof-search argument to the record's constructor,
the elaborator will search for values of \IdrisType{Eq t}, expressing
a form of type-class resolution using proof-search
implicits~\citep{sozeau-oury:first-class-type-classes}.
\end{itemize}

\subsection{Implementation data types}\label{moore datatypes}
We can now define the data-structures we use to represent Moore
machines in TyRE.

\paragraph*{Stacks.} We use heterogeneous stacks, indexed by their shape:
\ignore{
\begin{code}
\IdrisKeyword{namespace}\KatlaSpace{}\IdrisNamespace{StackImplementation}\KatlaNewline{}
\KatlaSpace{}\IdrisKeyword{\%hide}\KatlaSpace{}TyRE.Parser.SM.Stack.Stack\KatlaNewline{}
\end{code}
}
\begin{code}
\KatlaSpace{}\IdrisKeyword{data}\KatlaSpace{}\IdrisType{Stack}\KatlaSpace{}\IdrisKeyword{:}\KatlaSpace{}\IdrisFunction{Shape}\KatlaSpace{}\IdrisKeyword{\KatlaDash{}>}\KatlaSpace{}\IdrisType{Type}\KatlaSpace{}\IdrisKeyword{where}\KatlaNewline{}
\KatlaSpace{}\KatlaSpace{}\KatlaSpace{}\IdrisData{Lin}\KatlaSpace{}\KatlaSpace{}\IdrisKeyword{:}\KatlaSpace{}\IdrisType{Stack}\KatlaSpace{}\IdrisData{[<]}\KatlaNewline{}
\KatlaSpace{}\KatlaSpace{}\KatlaSpace{}\IdrisData{(:<)}\KatlaSpace{}\IdrisKeyword{:}\KatlaSpace{}\IdrisType{Stack}\KatlaSpace{}\IdrisBound{tps}\KatlaSpace{}\IdrisKeyword{\KatlaDash{}>}\KatlaSpace{}\IdrisKeyword{(}\IdrisBound{e}\KatlaSpace{}\IdrisKeyword{:}\KatlaSpace{}\IdrisBound{t}\IdrisKeyword{)}\KatlaSpace{}\IdrisKeyword{\KatlaDash{}>}\KatlaSpace{}\IdrisType{Stack}\KatlaSpace{}\IdrisKeyword{(}\IdrisBound{tps}\KatlaSpace{}\IdrisData{:<}\KatlaSpace{}\IdrisBound{t}\IdrisKeyword{)}\KatlaNewline{}
\end{code}
Thus a stack is either empty, or a snoc-cell consisting of a stack of
a given shape (\IdrisBound{tps}) with a value on top.

\begin{aside}
  Idris's desugaring convention for snoc-list notation is
  purely syntactic\footnotemark{}, and we can use snoc-list notation to implicitly
  left-nest sequences of any constructors or functions called
  \IdrisData{Lin} and snoc (\IdrisData{:<}).
  \footnotetext{
    If you're reading this manuscript in colour, you might enjoy this puzzle:
    construct the following term so that each character has the designated
    semantic role:
    \IdrisType{[} \IdrisBound{n} \IdrisData{,} \IdrisFunction{m} \IdrisType{,} \IdrisBound{n} \IdrisData{]},
    i.e.: type constructor, bound variable, data constructor, definition, type constructor,
    bound variable, and data constructor, respectively.
  }
\end{aside}
\paragraph*{Thread-local state.}
The state of each thread consists of a stack indexed by the thread
data's shape, together with a bit, \IdrisFunction{rec} that determines
whether the thread needs to \IdrisFunction{rec}ord the currently
parsed sub-string. In that case, the update the field dedicated to containing the characters
\IdrisFunction{recorded} so far:
\ignore{
\begin{code}
\IdrisKeyword{namespace}\KatlaSpace{}\IdrisNamespace{ThreadDataImplementation}\KatlaNewline{}
\KatlaSpace{}\IdrisKeyword{\%unhide}\KatlaSpace{}TyRE.Parser.SM.Stack.Stack\KatlaNewline{}
\end{code}
}
\begin{code}
\KatlaSpace{}\IdrisKeyword{record}\KatlaSpace{}\IdrisType{ThreadData}\KatlaSpace{}\IdrisKeyword{(}\IdrisBound{shape}\KatlaSpace{}\IdrisKeyword{:}\KatlaSpace{}\IdrisFunction{Shape}\IdrisKeyword{)}\KatlaSpace{}\IdrisKeyword{where}\KatlaNewline{}
\KatlaSpace{}\KatlaSpace{}\IdrisKeyword{constructor}\KatlaSpace{}\IdrisData{MkThreadData}\KatlaNewline{}
\KatlaSpace{}\KatlaSpace{}\IdrisFunction{stack}\KatlaSpace{}\IdrisKeyword{:}\KatlaSpace{}\IdrisType{Stack}\KatlaSpace{}\IdrisBound{shape}\KatlaNewline{}
\KatlaSpace{}\KatlaSpace{}\IdrisFunction{recorded}\KatlaSpace{}\IdrisKeyword{:}\KatlaSpace{}\IdrisType{SnocList}\KatlaSpace{}\IdrisType{Char}\KatlaNewline{}
\KatlaSpace{}\KatlaSpace{}\IdrisFunction{rec}\KatlaSpace{}\IdrisKeyword{:}\KatlaSpace{}\IdrisType{Bool}\KatlaNewline{}
\end{code}

\paragraph*{Instruction set.}
\begin{figure}
\ignore{
\begin{code}
\IdrisKeyword{namespace}\KatlaSpace{}\IdrisNamespace{InstructionImplementation}\KatlaNewline{}
\KatlaSpace{}\IdrisKeyword{\%hide}\KatlaSpace{}TyRE.Parser.SM.Routine\KatlaNewline{}
\KatlaSpace{}\IdrisKeyword{\%hide}\KatlaSpace{}TyRE.Parser.SM.Instruction\KatlaNewline{}
\end{code}
}
\begin{code}
\KatlaSpace{}\IdrisKeyword{data}\KatlaSpace{}\IdrisType{Instruction}\KatlaSpace{}\IdrisKeyword{:}\KatlaSpace{}\IdrisKeyword{(}\IdrisBound{pre}\IdrisKeyword{,}\KatlaSpace{}\IdrisBound{post}\KatlaSpace{}\IdrisKeyword{:}\KatlaSpace{}\IdrisFunction{Shape}\IdrisKeyword{)}\KatlaSpace{}\IdrisKeyword{\KatlaDash{}>}\KatlaSpace{}\IdrisType{Type}\KatlaSpace{}\IdrisKeyword{where}\KatlaNewline{}
\KatlaSpace{}\KatlaSpace{}\KatlaSpace{}\KatlaSpace{}\IdrisComment{|||\KatlaSpace{}Pushes\KatlaSpace{}chosen\KatlaSpace{}symbol\KatlaSpace{}on\KatlaSpace{}the\KatlaSpace{}stack}\KatlaNewline{}
\KatlaSpace{}\KatlaSpace{}\KatlaSpace{}\KatlaSpace{}\IdrisData{Push}\KatlaSpace{}\IdrisKeyword{:}\KatlaSpace{}\IdrisKeyword{\{0}\KatlaSpace{}\IdrisBound{x}\KatlaSpace{}\IdrisKeyword{:}\KatlaSpace{}\IdrisType{Type}\IdrisKeyword{\}}\KatlaSpace{}\IdrisKeyword{\KatlaDash{}>}\KatlaSpace{}\IdrisBound{x}\KatlaSpace{}\IdrisKeyword{\KatlaDash{}>}\KatlaSpace{}\IdrisType{Instruction}\KatlaSpace{}\IdrisBound{tps}\KatlaSpace{}\IdrisKeyword{(}\IdrisBound{tps}\KatlaSpace{}\IdrisData{:<}\KatlaSpace{}\IdrisBound{x}\IdrisKeyword{)}\KatlaNewline{}
\KatlaSpace{}\KatlaSpace{}\KatlaSpace{}\KatlaSpace{}\IdrisComment{|||\KatlaSpace{}Pushes\KatlaSpace{}currently\KatlaSpace{}consumed\KatlaSpace{}character\KatlaSpace{}on\KatlaSpace{}the\KatlaSpace{}stack}\KatlaNewline{}
\KatlaSpace{}\KatlaSpace{}\KatlaSpace{}\KatlaSpace{}\IdrisData{PushChar}\KatlaSpace{}\IdrisKeyword{:}\KatlaSpace{}\IdrisType{Instruction}\KatlaSpace{}\IdrisBound{tps}\KatlaSpace{}\IdrisKeyword{(}\IdrisBound{tps}\KatlaSpace{}\IdrisData{:<}\KatlaSpace{}\IdrisType{Char}\IdrisKeyword{)}\KatlaNewline{}
\KatlaSpace{}\KatlaSpace{}\KatlaSpace{}\KatlaSpace{}\IdrisComment{|||\KatlaSpace{}Reduces\KatlaSpace{}two\KatlaSpace{}last\KatlaSpace{}symbols\KatlaSpace{}from\KatlaSpace{}the\KatlaSpace{}stack}\KatlaNewline{}
\IdrisComment{\KatlaSpace{}\KatlaSpace{}\KatlaSpace{}\KatlaSpace{}|||\KatlaSpace{}according\KatlaSpace{}to\KatlaSpace{}the\KatlaSpace{}specified\KatlaSpace{}function}\KatlaNewline{}
\KatlaSpace{}\KatlaSpace{}\KatlaSpace{}\KatlaSpace{}\IdrisData{ReducePair}\KatlaSpace{}\IdrisKeyword{:}\KatlaSpace{}\IdrisKeyword{\{0}\KatlaSpace{}\IdrisBound{x}\IdrisKeyword{,}\KatlaSpace{}\IdrisBound{y}\IdrisKeyword{,}\KatlaSpace{}\IdrisBound{z}\KatlaSpace{}\IdrisKeyword{:}\KatlaSpace{}\IdrisType{Type}\IdrisKeyword{\}}\KatlaSpace{}\IdrisKeyword{\KatlaDash{}>}\KatlaSpace{}\IdrisKeyword{(}\IdrisBound{x}\KatlaSpace{}\IdrisKeyword{\KatlaDash{}>}\KatlaSpace{}\IdrisBound{y}\KatlaSpace{}\IdrisKeyword{\KatlaDash{}>}\KatlaSpace{}\IdrisBound{z}\IdrisKeyword{)}\KatlaNewline{}
\KatlaSpace{}\KatlaSpace{}\KatlaSpace{}\KatlaSpace{}\KatlaSpace{}\KatlaSpace{}\KatlaSpace{}\KatlaSpace{}\KatlaSpace{}\KatlaSpace{}\KatlaSpace{}\KatlaSpace{}\KatlaSpace{}\KatlaSpace{}\IdrisKeyword{\KatlaDash{}>}\KatlaSpace{}\IdrisType{Instruction}\KatlaSpace{}\IdrisKeyword{(}\IdrisBound{tps}\KatlaSpace{}\IdrisData{:<}\KatlaSpace{}\IdrisBound{x}\KatlaSpace{}\IdrisData{:<}\KatlaSpace{}\IdrisBound{y}\IdrisKeyword{)}\KatlaSpace{}\IdrisKeyword{(}\IdrisBound{tps}\KatlaSpace{}\IdrisData{:<}\KatlaSpace{}\IdrisBound{z}\IdrisKeyword{)}\KatlaNewline{}
\KatlaSpace{}\KatlaSpace{}\KatlaSpace{}\KatlaSpace{}\IdrisComment{|||\KatlaSpace{}Maps\KatlaSpace{}the\KatlaSpace{}top\KatlaSpace{}element\KatlaSpace{}of\KatlaSpace{}stack}\KatlaNewline{}
\KatlaSpace{}\KatlaSpace{}\KatlaSpace{}\KatlaSpace{}\IdrisData{Transform}\KatlaSpace{}\IdrisKeyword{:}\KatlaSpace{}\IdrisKeyword{\{0}\KatlaSpace{}\IdrisBound{x}\IdrisKeyword{,}\KatlaSpace{}\IdrisBound{y}\KatlaSpace{}\IdrisKeyword{:}\KatlaSpace{}\IdrisType{Type}\IdrisKeyword{\}}\KatlaSpace{}\IdrisKeyword{\KatlaDash{}>}\KatlaSpace{}\IdrisKeyword{(}\IdrisBound{x}\KatlaSpace{}\IdrisKeyword{\KatlaDash{}>}\KatlaSpace{}\IdrisBound{y}\IdrisKeyword{)}\KatlaNewline{}
\KatlaSpace{}\KatlaSpace{}\KatlaSpace{}\KatlaSpace{}\KatlaSpace{}\KatlaSpace{}\KatlaSpace{}\KatlaSpace{}\KatlaSpace{}\KatlaSpace{}\KatlaSpace{}\KatlaSpace{}\KatlaSpace{}\KatlaSpace{}\IdrisKeyword{\KatlaDash{}>}\KatlaSpace{}\IdrisType{Instruction}\KatlaSpace{}\IdrisKeyword{(}\IdrisBound{tps}\KatlaSpace{}\IdrisData{:<}\KatlaSpace{}\IdrisBound{x}\IdrisKeyword{)}\KatlaSpace{}\IdrisKeyword{(}\IdrisBound{tps}\KatlaSpace{}\IdrisData{:<}\KatlaSpace{}\IdrisBound{y}\IdrisKeyword{)}\KatlaNewline{}
\KatlaSpace{}\KatlaSpace{}\KatlaSpace{}\KatlaSpace{}\IdrisComment{|||\KatlaSpace{}Pushes\KatlaSpace{}collected\KatlaSpace{}string\KatlaSpace{}on\KatlaSpace{}the\KatlaSpace{}stack}\KatlaNewline{}
\KatlaSpace{}\KatlaSpace{}\KatlaSpace{}\KatlaSpace{}\IdrisData{EmitString}\KatlaSpace{}\IdrisKeyword{:}\KatlaSpace{}\IdrisType{Instruction}\KatlaSpace{}\IdrisBound{tps}\KatlaSpace{}\IdrisKeyword{(}\IdrisBound{tps}\KatlaSpace{}\IdrisData{:<}\KatlaSpace{}\IdrisType{String}\IdrisKeyword{)}\KatlaNewline{}
\KatlaSpace{}\KatlaSpace{}\KatlaSpace{}\KatlaSpace{}\IdrisComment{|||\KatlaSpace{}Raises\KatlaSpace{}record\KatlaSpace{}flag\KatlaSpace{}and\KatlaSpace{}starts\KatlaSpace{}collecting\KatlaSpace{}characters}\KatlaNewline{}
\KatlaSpace{}\KatlaSpace{}\KatlaSpace{}\KatlaSpace{}\IdrisData{Record}\KatlaSpace{}\IdrisKeyword{:}\KatlaSpace{}\IdrisType{Instruction}\KatlaSpace{}\IdrisBound{tps}\KatlaSpace{}\IdrisBound{tps}\KatlaNewline{}
\end{code}
  \caption{Moore machine instruction set}
  \label{fig:Moore machine instruction set}
\end{figure}

\figref{fig:Moore machine instruction set} presents the Moore machine
instructions, i.e., the instructions that may appear in labels for the
NFA's transition routines. These instructions will operate on the appropriate
\IdrisType{ThreadData} record. As we traverse the Moore machine we
usually collect the information about the parse tree on the
stack. There is one exception to this rule. When executing the Moore
machine we can enter a special mode enabled by raising the
\IdrisFunction{rec}ord flag using the \IdrisData{Record}
instruction. When the \IdrisFunction{rec}ord flag is raised, we will
also record the consumed characters in the dedicated buffer
\IdrisFunction{recorded}. The \IdrisFunction{recorded} characters can
be pushed on the stack using the \IdrisData{EmitString} instruction,
which also lowers the \IdrisFunction{rec}ord flag, reverts the machine
to the non-recording mode, and flushes the buffer. We will use the recording mode when
creating the parse tree for a \IdrisData{Group} sub-regex. All the
other instructions change the stack by either pushing onto it or
transforming it.
We define the predicate \IdrisType{IsInitInstruction} to include all
instructions but \IdrisData{Push}.

The stack shapes \IdrisBound{pre} and \IdrisBound{post} that index
each instruction rely on the thread's stack to have an appropriate
shape before execution, and then guarantee a given shape after
execution. We can then execute an instruction on a stack with the
appropriate shape by implementing a function with this type:
\ignore{
\begin{code}
\IdrisKeyword{namespace}\KatlaSpace{}\IdrisNamespace{ExecInstructionImpl}\KatlaNewline{}
\KatlaSpace{}\IdrisKeyword{\%unhide}\KatlaSpace{}TyRE.Parser.SM.Instruction\KatlaNewline{}
\KatlaSpace{}\IdrisKeyword{\%unhide}\KatlaSpace{}Main.ThreadData\KatlaNewline{}
\KatlaSpace{}\IdrisKeyword{\%hide}\KatlaSpace{}Main.ThreadDataImplementation.ThreadData\KatlaNewline{}
\KatlaSpace{}\IdrisKeyword{\%hide}\KatlaSpace{}Main.EitherErased\KatlaNewline{}
\end{code}
}
\begin{code}
\KatlaSpace{}\IdrisFunction{execInstruction}\KatlaSpace{}\IdrisKeyword{:}\KatlaSpace{}\IdrisKeyword{(}\IdrisBound{r}\KatlaSpace{}\IdrisKeyword{:}\KatlaSpace{}\IdrisType{Instruction}\KatlaSpace{}\IdrisBound{pre}\KatlaSpace{}\IdrisBound{post}\IdrisKeyword{)}\KatlaSpace{}\IdrisKeyword{\KatlaDash{}>}\KatlaNewline{}
\KatlaSpace{}\KatlaSpace{}\KatlaSpace{}\KatlaSpace{}\KatlaSpace{}\KatlaSpace{}\KatlaSpace{}\KatlaSpace{}\KatlaSpace{}\KatlaSpace{}\KatlaSpace{}\KatlaSpace{}\KatlaSpace{}\KatlaSpace{}\KatlaSpace{}\KatlaSpace{}\KatlaSpace{}\KatlaSpace{}\KatlaSpace{}\IdrisType{EitherErased}\KatlaSpace{}\IdrisType{Char}\KatlaSpace{}\IdrisKeyword{(}\IdrisType{IsInitInstruction}\KatlaSpace{}\IdrisBound{r}\IdrisKeyword{)}\KatlaSpace{}\IdrisKeyword{\KatlaDash{}>}\KatlaNewline{}
\KatlaSpace{}\KatlaSpace{}\KatlaSpace{}\KatlaSpace{}\KatlaSpace{}\KatlaSpace{}\KatlaSpace{}\KatlaSpace{}\KatlaSpace{}\KatlaSpace{}\KatlaSpace{}\KatlaSpace{}\KatlaSpace{}\KatlaSpace{}\KatlaSpace{}\KatlaSpace{}\KatlaSpace{}\KatlaSpace{}\KatlaSpace{}\IdrisType{ThreadData}\KatlaSpace{}\IdrisKeyword{(}\IdrisBound{p}\KatlaSpace{}\IdrisFunction{++}\KatlaSpace{}\IdrisBound{pre}\IdrisKeyword{)}\KatlaSpace{}\IdrisKeyword{\KatlaDash{}>}\KatlaNewline{}
\KatlaSpace{}\KatlaSpace{}\KatlaSpace{}\KatlaSpace{}\KatlaSpace{}\KatlaSpace{}\KatlaSpace{}\KatlaSpace{}\KatlaSpace{}\KatlaSpace{}\KatlaSpace{}\KatlaSpace{}\KatlaSpace{}\KatlaSpace{}\KatlaSpace{}\KatlaSpace{}\KatlaSpace{}\KatlaSpace{}\KatlaSpace{}\IdrisType{ThreadData}\KatlaSpace{}\IdrisKeyword{(}\IdrisBound{p}\KatlaSpace{}\IdrisFunction{++}\KatlaSpace{}\IdrisBound{post}\IdrisKeyword{)}\KatlaNewline{}
\end{code}
It requires either the last-read character, or a runtime-erased proof
that the instruction does not require this character. It will then
take a \IdrisType{ThreadData} whose stack's top layer adheres to the
instruction stack shape pre-condition, and will produce
\IdrisType{ThreadData} guaranteeing the post-condition.

We label the Moore machine's transition with \emph{routines}:
cons-lists of instructions, collapsing the intermediate stack shapes
telescopically in their indices:
\begin{code}
\KatlaSpace{}\IdrisKeyword{data}\KatlaSpace{}\IdrisType{Routine}\KatlaSpace{}\IdrisKeyword{:}\KatlaSpace{}\IdrisKeyword{(}\IdrisBound{pre}\IdrisKeyword{,}\KatlaSpace{}\IdrisBound{post}\KatlaSpace{}\IdrisKeyword{:}\KatlaSpace{}\IdrisFunction{Shape}\IdrisKeyword{)}\KatlaSpace{}\IdrisKeyword{\KatlaDash{}>}\KatlaSpace{}\IdrisType{Type}\KatlaSpace{}\IdrisKeyword{where}\KatlaNewline{}
\KatlaSpace{}\KatlaSpace{}\KatlaSpace{}\KatlaSpace{}\IdrisComment{|||\KatlaSpace{}Empty\KatlaSpace{}routine}\KatlaNewline{}
\KatlaSpace{}\KatlaSpace{}\KatlaSpace{}\KatlaSpace{}\IdrisData{Nil}\KatlaSpace{}\IdrisKeyword{:}\KatlaSpace{}\IdrisType{Routine}\KatlaSpace{}\IdrisBound{shape}\KatlaSpace{}\IdrisBound{shape}\KatlaNewline{}
\KatlaSpace{}\KatlaSpace{}\KatlaSpace{}\KatlaSpace{}\IdrisComment{|||\KatlaSpace{}Prepend\KatlaSpace{}an\KatlaSpace{}instruction\KatlaSpace{}to\KatlaSpace{}a\KatlaSpace{}routine}\KatlaNewline{}
\KatlaSpace{}\KatlaSpace{}\KatlaSpace{}\KatlaSpace{}\IdrisData{(::)}\KatlaSpace{}\IdrisKeyword{:}\KatlaSpace{}\IdrisType{Instruction}\KatlaSpace{}\IdrisBound{pre}\KatlaSpace{}\IdrisBound{mid}\KatlaSpace{}\IdrisKeyword{\KatlaDash{}>}\KatlaSpace{}\IdrisType{Routine}\KatlaSpace{}\IdrisBound{mid}\KatlaSpace{}\IdrisBound{post}\KatlaSpace{}\IdrisKeyword{\KatlaDash{}>}\KatlaNewline{}
\KatlaSpace{}\KatlaSpace{}\KatlaSpace{}\KatlaSpace{}\KatlaSpace{}\KatlaSpace{}\KatlaSpace{}\KatlaSpace{}\KatlaSpace{}\KatlaSpace{}\KatlaSpace{}\IdrisType{Routine}\KatlaSpace{}\IdrisBound{pre}\KatlaSpace{}\IdrisBound{post}\KatlaNewline{}
\end{code}
It is straightforward to extend execution from instructions to
routines inductively.

\paragraph*{Typed Moore machines.}
To represent the NFA abstraction from \S\ref{moore layer abstractions}
while easing our type-safety proofs, we pair each transition in the
NFA with its labelling routine. We will represent transition relations,
using a list to represent the set of outgoing edges, as follows:
\ignore{
\begin{code}
\IdrisKeyword{namespace}\KatlaSpace{}\IdrisNamespace{MooreMachineImpl}\KatlaNewline{}
\KatlaSpace{}\IdrisKeyword{\%hide}\KatlaSpace{}Main.Desugar.Subset\KatlaNewline{}
\KatlaSpace{}\IdrisKeyword{\%hide}\KatlaSpace{}Main.ImplicitArg.shapeOf\KatlaNewline{}
\KatlaSpace{}\IdrisKeyword{\%hide}\KatlaSpace{}TyRE.Parser.SM.InitStatesType\KatlaNewline{}
\KatlaSpace{}\IdrisKeyword{\%unhide}\KatlaSpace{}Main.Routine\KatlaNewline{}
\KatlaSpace{}\IdrisFunction{IsInitRoutine}\KatlaSpace{}\IdrisKeyword{:}\KatlaSpace{}\IdrisType{Routine}\KatlaSpace{}\IdrisBound{xs}\KatlaSpace{}\IdrisBound{ys}\KatlaSpace{}\IdrisKeyword{\KatlaDash{}>}\KatlaSpace{}\IdrisType{Type}\KatlaNewline{}
\KatlaSpace{}\IdrisFunction{shapeOf}\KatlaSpace{}\IdrisKeyword{:}\KatlaSpace{}\IdrisKeyword{(}\IdrisBound{lookup}\KatlaSpace{}\IdrisKeyword{:}\KatlaSpace{}\IdrisBound{s}\KatlaSpace{}\IdrisKeyword{\KatlaDash{}>}\KatlaSpace{}\IdrisFunction{Shape}\IdrisKeyword{)}\KatlaSpace{}\IdrisKeyword{\KatlaDash{}>}\KatlaSpace{}\IdrisKeyword{(}\IdrisBound{t}\KatlaSpace{}\IdrisKeyword{:}\KatlaSpace{}\IdrisType{Type}\IdrisKeyword{)}\KatlaSpace{}\IdrisKeyword{\KatlaDash{}>}\KatlaNewline{}
\KatlaSpace{}\KatlaSpace{}\KatlaSpace{}\KatlaSpace{}\KatlaSpace{}\KatlaSpace{}\KatlaSpace{}\KatlaSpace{}\KatlaSpace{}\KatlaSpace{}\IdrisType{Maybe}\KatlaSpace{}\IdrisBound{s}\KatlaSpace{}\IdrisKeyword{\KatlaDash{}>}\KatlaSpace{}\IdrisFunction{Shape}\KatlaNewline{}
\end{code}
}
\begin{code}
\KatlaSpace{}\IdrisFunction{TransitionRelation}\KatlaSpace{}\IdrisKeyword{:}\KatlaSpace{}\IdrisKeyword{(}\IdrisBound{shape}\KatlaSpace{}\IdrisKeyword{:}\KatlaSpace{}\IdrisType{Type}\IdrisKeyword{)}\KatlaSpace{}\IdrisKeyword{\KatlaDash{}>}\KatlaNewline{}
\KatlaSpace{}\KatlaSpace{}\KatlaSpace{}\IdrisKeyword{(}\IdrisBound{state}\KatlaSpace{}\IdrisKeyword{:}\KatlaSpace{}\IdrisType{Type}\IdrisKeyword{)}\KatlaSpace{}\IdrisKeyword{\KatlaDash{}>}\KatlaSpace{}\IdrisKeyword{(}\IdrisBound{stateShape}\KatlaSpace{}\IdrisKeyword{:}\KatlaSpace{}\IdrisBound{state}\KatlaSpace{}\IdrisKeyword{\KatlaDash{}>}\KatlaSpace{}\IdrisFunction{Shape}\IdrisKeyword{)}\KatlaSpace{}\IdrisKeyword{\KatlaDash{}>}\KatlaSpace{}\IdrisType{Type}\KatlaNewline{}
\KatlaSpace{}\IdrisFunction{TransitionRelation}\KatlaSpace{}\IdrisBound{shape}\KatlaSpace{}\IdrisBound{state}\KatlaSpace{}\IdrisBound{stateShape}\KatlaNewline{}
\KatlaSpace{}\KatlaSpace{}\KatlaSpace{}\IdrisKeyword{=}\KatlaSpace{}\IdrisKeyword{(}\IdrisBound{st}\KatlaSpace{}\IdrisKeyword{:}\KatlaSpace{}\IdrisBound{state}\IdrisKeyword{)}\KatlaNewline{}
\KatlaSpace{}\KatlaSpace{}\KatlaSpace{}\IdrisKeyword{\KatlaDash{}>}\KatlaSpace{}\IdrisKeyword{(}\IdrisBound{c}\KatlaSpace{}\IdrisKeyword{:}\KatlaSpace{}\IdrisType{Char}\IdrisKeyword{)}\KatlaNewline{}
\KatlaSpace{}\KatlaSpace{}\KatlaSpace{}\IdrisKeyword{\KatlaDash{}>}\KatlaSpace{}\IdrisType{List}\KatlaSpace{}\IdrisType{(}\IdrisBound{st'}\KatlaSpace{}\IdrisKeyword{:}\KatlaSpace{}\IdrisType{Maybe}\KatlaSpace{}\IdrisBound{state}\KatlaSpace{}\IdrisType{**}\KatlaSpace{}\IdrisType{Routine}\KatlaSpace{}\IdrisKeyword{(}\IdrisBound{stateShape}\KatlaSpace{}\IdrisBound{st}\IdrisKeyword{)}\KatlaNewline{}
\KatlaSpace{}\KatlaSpace{}\KatlaSpace{}\KatlaSpace{}\KatlaSpace{}\KatlaSpace{}\KatlaSpace{}\KatlaSpace{}\KatlaSpace{}\KatlaSpace{}\KatlaSpace{}\KatlaSpace{}\KatlaSpace{}\KatlaSpace{}\KatlaSpace{}\KatlaSpace{}\KatlaSpace{}\KatlaSpace{}\KatlaSpace{}\KatlaSpace{}\KatlaSpace{}\KatlaSpace{}\KatlaSpace{}\KatlaSpace{}\KatlaSpace{}\KatlaSpace{}\KatlaSpace{}\KatlaSpace{}\KatlaSpace{}\KatlaSpace{}\KatlaSpace{}\KatlaSpace{}\KatlaSpace{}\KatlaSpace{}\KatlaSpace{}\KatlaSpace{}\KatlaSpace{}\KatlaSpace{}\KatlaSpace{}\KatlaSpace{}\KatlaSpace{}\IdrisKeyword{(}\IdrisFunction{shapeOf}\KatlaSpace{}\IdrisBound{stateShape}\KatlaSpace{}\IdrisBound{shape}\KatlaSpace{}\IdrisBound{st'}\IdrisKeyword{)}\IdrisType{)}\KatlaNewline{}
\end{code}
Thus, we parameterise the type of transition relations by:
\begin{itemize}
\item \IdrisBound{shape}: the result parse-tree type;
\item \IdrisBound{state}: the type of the non-accepting NFA states; and
\item \IdrisBound{stateShape}: the assumed stack-shape for each state.
\end{itemize}
Given these parameters, a transition relation, given input
\begin{itemize}
\item \IdrisBound{st}: starting state; and
\item \IdrisBound{c}: \IdrisType{Char}acter
\end{itemize}
maps them to a list, representing a finite set, of transitions
\IdrisBound{st}${}\xrightarrow[\text{\IdrisBound{c}}]{\text{\IdrisBound{r}}}{}$ \IdrisBound{st'}:
\begin{itemize}
\item \IdrisBound{st'}: the target state, which may be the accepting
  state \IdrisData{Nothing}; and
\item \IdrisBound{r}: the labelling routine for this transition, which
  assumes the stack shape associated to \IdrisBound{st} and guarantees
  the stack shape associated to \IdrisBound{st'}, given by
  \IdrisFunction{shapeOf} (cf.~page~\pageref{shapeOf location}):
  \begin{itemize}
  \item When \IdrisBound{st'} is a non-accepting state \IdrisData{Just }\IdrisBound{s'}, it is \IdrisBound{stateShape s'}.
  \item When \IdrisBound{st'} is a the state \IdrisData{Nothing}, it is the given \IdrisBound{shape}.
  \end{itemize}
\end{itemize}

Similarly, if we view a starting state as being the target state for a
transition without a source, then we represent a subset, represented
as a list, of initial states as follows:
\ignore{
\begin{code}
\KatlaSpace{}\IdrisKeyword{0}\KatlaNewline{}
\end{code}
}
\begin{code}
\KatlaSpace{}\IdrisFunction{InitStatesType}\KatlaSpace{}\IdrisKeyword{:}\KatlaSpace{}\IdrisKeyword{(}\IdrisBound{shape}\KatlaSpace{}\IdrisKeyword{:}\KatlaSpace{}\IdrisType{Type}\IdrisKeyword{)}\KatlaSpace{}\IdrisKeyword{\KatlaDash{}>}\KatlaNewline{}
\KatlaSpace{}\KatlaSpace{}\KatlaSpace{}\IdrisKeyword{(}\IdrisBound{state}\KatlaSpace{}\IdrisKeyword{:}\KatlaSpace{}\IdrisType{Type}\IdrisKeyword{)}\KatlaSpace{}\IdrisKeyword{\KatlaDash{}>}\KatlaSpace{}\IdrisKeyword{(}\IdrisBound{stateShape}\KatlaSpace{}\IdrisKeyword{:}\KatlaSpace{}\IdrisBound{state}\KatlaSpace{}\IdrisKeyword{\KatlaDash{}>}\KatlaSpace{}\IdrisFunction{Shape}\IdrisKeyword{)}\KatlaSpace{}\IdrisKeyword{\KatlaDash{}>}\KatlaSpace{}\IdrisType{Type}\KatlaNewline{}
\KatlaSpace{}\IdrisFunction{InitStatesType}\KatlaSpace{}\IdrisBound{shape}\KatlaSpace{}\IdrisBound{state}\KatlaSpace{}\IdrisBound{stateShape}\KatlaNewline{}
\KatlaSpace{}\KatlaSpace{}\KatlaSpace{}\IdrisKeyword{=}\KatlaSpace{}\IdrisType{List}\KatlaSpace{}\IdrisType{(}\IdrisBound{st}\KatlaSpace{}\IdrisKeyword{:}\KatlaSpace{}\IdrisType{Maybe}\KatlaSpace{}\IdrisBound{state}\KatlaNewline{}
\KatlaSpace{}\KatlaSpace{}\KatlaSpace{}\KatlaSpace{}\KatlaSpace{}\KatlaSpace{}\KatlaSpace{}\KatlaSpace{}\KatlaSpace{}\KatlaSpace{}\KatlaSpace{}\IdrisType{**}\KatlaSpace{}\IdrisType{Subset}\KatlaSpace{}\IdrisKeyword{(}\IdrisType{Routine}\KatlaSpace{}\IdrisData{[<]}\KatlaSpace{}\IdrisKeyword{(}\IdrisFunction{shapeOf}\KatlaSpace{}\IdrisBound{stateShape}\KatlaSpace{}\IdrisBound{shape}\KatlaSpace{}\IdrisBound{st}\IdrisKeyword{))}\KatlaNewline{}
\KatlaSpace{}\KatlaSpace{}\KatlaSpace{}\KatlaSpace{}\KatlaSpace{}\KatlaSpace{}\KatlaSpace{}\KatlaSpace{}\KatlaSpace{}\KatlaSpace{}\KatlaSpace{}\KatlaSpace{}\KatlaSpace{}\KatlaSpace{}\KatlaSpace{}\KatlaSpace{}\KatlaSpace{}\KatlaSpace{}\KatlaSpace{}\KatlaSpace{}\KatlaSpace{}\IdrisFunction{IsInitRoutine}\IdrisType{)}\KatlaNewline{}
\end{code}
Thus, we represent each initial state
$\mathsf{Start}\xrightarrow{\text{\IdrisBound{r}}}{}$\IdrisBound{st} by:
\begin{itemize}
\item \IdrisBound{st}: a state, which may be the accepting state
  \IdrisData{Nothing}; and
\item \IdrisBound{r}: an initialisation routine, and so we require a
  runtime-erased proof that the routine contains only
  initialisation-safe instructions.
\end{itemize}

\begin{figure}
\begin{code}
\KatlaSpace{}\IdrisKeyword{record}\KatlaSpace{}\IdrisType{MooreMachine}\KatlaSpace{}\IdrisKeyword{(}\IdrisBound{shape}\KatlaSpace{}\IdrisKeyword{:}\KatlaSpace{}\IdrisType{Type}\IdrisKeyword{)}\KatlaSpace{}\IdrisKeyword{where}\KatlaNewline{}
\KatlaSpace{}\KatlaSpace{}\KatlaSpace{}\IdrisKeyword{constructor}\KatlaSpace{}\IdrisData{MkMooreMachine}\KatlaNewline{}
\KatlaSpace{}\KatlaSpace{}\KatlaSpace{}\IdrisKeyword{0}\KatlaSpace{}\IdrisFunction{s}\KatlaSpace{}\IdrisKeyword{:}\KatlaSpace{}\IdrisType{Type}\KatlaNewline{}
\KatlaSpace{}\KatlaSpace{}\KatlaSpace{}\IdrisKeyword{0}\KatlaSpace{}\IdrisFunction{lookup}\KatlaSpace{}\IdrisKeyword{:}\KatlaSpace{}\IdrisBound{s}\KatlaSpace{}\IdrisKeyword{\KatlaDash{}>}\KatlaSpace{}\IdrisFunction{Shape}\KatlaNewline{}
\KatlaSpace{}\KatlaSpace{}\KatlaSpace{}\IdrisKeyword{\{auto}\KatlaSpace{}\IdrisFunction{isEq}\KatlaSpace{}\IdrisKeyword{:}\KatlaSpace{}\IdrisType{Eq}\KatlaSpace{}\IdrisBound{s}\IdrisKeyword{\}}\KatlaNewline{}
\KatlaSpace{}\KatlaSpace{}\KatlaSpace{}\IdrisFunction{init}\KatlaSpace{}\IdrisKeyword{:}\KatlaSpace{}\IdrisFunction{InitStatesType}\KatlaSpace{}\KatlaSpace{}\KatlaSpace{}\KatlaSpace{}\KatlaSpace{}\IdrisBound{shape}\KatlaSpace{}\IdrisBound{s}\KatlaSpace{}\IdrisBound{lookup}\KatlaNewline{}
\KatlaSpace{}\KatlaSpace{}\KatlaSpace{}\IdrisFunction{next}\KatlaSpace{}\IdrisKeyword{:}\KatlaSpace{}\IdrisFunction{TransitionRelation}\KatlaSpace{}\IdrisBound{shape}\KatlaSpace{}\IdrisBound{s}\KatlaSpace{}\IdrisBound{lookup}\KatlaNewline{}
\end{code}
\caption{TyRE's Moore machine type}
\label{fig:Moore machine implementation}
\end{figure}
\figref{fig:Moore machine implementation} presents the TyRE's record
declaration for Moore machines. Its fields:
\begin{itemize}
\item \IdrisFunction{s}: a runtime-erased type of non-accepting states
\item \IdrisFunction{lookup}: a runtime-erased association of a shape
  to each non-accepting state;
\item \IdrisFunction{isEq}: an implicit equality-predicate on
  non-accepting states, elaborated using Idris's proof-search
  mechanism;
\item \IdrisFunction{init}: the initial states with their
  initialisation routines;
\item \IdrisFunction{next}: the transition relation labelled with the
  parsing routines;
\end{itemize}
We use the equality-predicate \IdrisFunction{isEq} in two places:
\begin{itemize}
\item during automata minimisation; and
\item during execution, to maintain a bounded number of threads. In
  detail, if two threads meet in the same state, that means both have
  parsed the same substring and have the same paths available to them
  going forward. They may only differ in the parts of the parse tree
  constructed so far. Being in the same state means that the parse
  trees constructed so far must have the same shape. Since we aren't
  striving to construct all the possible parse trees but return one
  possibility, we keep only one of such two threads going
  forward. Doing so ensures the number of thread is bounded by the
  number of states in the automaton, guaranteeing linear time and
  space complexity.
\end{itemize}

\subsection{Thompson's construction}
\cite{mcnaughton-yamada:nfa-construction} and
\cite{thompson:nfa-construction} describe algorithms for turning a
regex into an equivalent automaton.  Our compilation scheme is a
standard variation on these well-known constructions, augmented with
code routine labels to perform the parsing. When compiling a
\IdrisData{Group} sub-regex, we use a separate mechanism to build the
Moore machine for regexes wrapped in a group. Figure
\ref{fig:sm-constructions} shows how we build a state machine for all
\IdrisType{TyRE} constructors except \IdrisData{Group}. We describe
this remaining case in \S\ref{subsec:group minimisation}.

\begin{figure}
\input{thompsons}
\caption{\textbf{Moore machine construction.}}
\label{fig:sm-constructions}
\end{figure}

\newcommand\machine[1]{#1_{\mathrm m}}

\ignore{
\begin{code}
\IdrisKeyword{namespace}\KatlaSpace{}\IdrisNamespace{Thompsons}\KatlaNewline{}
\end{code}
}
\paragraph*{\figref{fig:sm-constructions:pred}: Predicate.}
The state machine accepting a language of single character
words has:
\begin{itemize}
\item
  a
single start state with an empty list of initial instructions;
\item
  a single transition to the accepting state, guarded by the
given predicate; and
\item
  a single instruction labelling this transition:
\IdrisData{PushChar}.
\end{itemize}

We use this contruction uniformly by defining when each
\IdrisType{CharCond} is satisfied:
\begin{itemize}
\begin{code}
\KatlaSpace{}\IdrisFunction{satisfies}\KatlaSpace{}\IdrisKeyword{:}\KatlaSpace{}\IdrisType{CharCond}\KatlaSpace{}\IdrisKeyword{\KatlaDash{}>}\KatlaSpace{}\IdrisType{Char}\KatlaSpace{}\IdrisKeyword{\KatlaDash{}>}\KatlaSpace{}\IdrisType{Bool}\KatlaNewline{}
\KatlaSpace{}\IdrisFunction{satisfies}\KatlaSpace{}\IdrisKeyword{(}\IdrisData{OneOf}\KatlaSpace{}\IdrisBound{xs}\KatlaSpace{}\KatlaSpace{}\KatlaSpace{}\KatlaSpace{}\IdrisKeyword{)}\KatlaSpace{}\IdrisKeyword{=}\KatlaSpace{}\IdrisKeyword{\textbackslash{}}\IdrisBound{c}\KatlaSpace{}\IdrisKeyword{=>}\KatlaSpace{}\IdrisFunction{contains}\KatlaSpace{}\IdrisBound{c}\KatlaSpace{}\IdrisBound{xs}\KatlaNewline{}
\KatlaSpace{}\IdrisFunction{satisfies}\KatlaSpace{}\IdrisKeyword{(}\IdrisData{Range}\KatlaSpace{}\IdrisKeyword{(}\IdrisBound{x}\IdrisData{,}\KatlaSpace{}\IdrisBound{y}\IdrisKeyword{))}\KatlaSpace{}\IdrisKeyword{=}\KatlaSpace{}\IdrisKeyword{\textbackslash{}}\IdrisBound{c}\KatlaSpace{}\IdrisKeyword{=>}\KatlaSpace{}\IdrisBound{x}\KatlaSpace{}\IdrisFunction{<=}\KatlaSpace{}\IdrisBound{c}\KatlaSpace{}\IdrisFunction{\&\&}\KatlaSpace{}\IdrisBound{c}\KatlaSpace{}\IdrisFunction{<=}\KatlaSpace{}\IdrisBound{y}\KatlaNewline{}
\KatlaSpace{}\IdrisFunction{satisfies}\KatlaSpace{}\IdrisKeyword{(}\IdrisData{Pred}\KatlaSpace{}\IdrisBound{f}\KatlaSpace{}\KatlaSpace{}\KatlaSpace{}\KatlaSpace{}\KatlaSpace{}\KatlaSpace{}\IdrisKeyword{)}\KatlaSpace{}\IdrisKeyword{=}\KatlaSpace{}\IdrisKeyword{\textbackslash{}}\IdrisBound{c}\KatlaSpace{}\IdrisKeyword{=>}\KatlaSpace{}\IdrisBound{f}\KatlaSpace{}\IdrisBound{c}\KatlaNewline{}
\end{code}
\end{itemize}

\paragraph*{\figref{fig:sm-constructions:eps}: Empty string.}
In the state machine for the \IdrisData{Empty} string $\epsilon$, the accepting
state is a starting state and the only existing state. It has a single
initialisation instruction: \IdrisData{Push ()}.

\paragraph*{\figref{fig:sm-constructions:alt}: Alternation.}
To build a state machine for $R\vert S$ (\IdrisData{Alt}
\IdrisBound{R}\KatlaSpace\IdrisBound{S}), first we build the state
machines $\machine R$ and $\machine S$ for $R$ and $S$, respectively.
The starting states in the alternation state machine will be the union
of the starting states from both $\machine R$ and $\machine S$. We
will keep all the transitions and merge the accepting states into a
single one, so to get to the accepting state we traverse either $\machine R$ or
$\machine S$. We track which one we have traversed and lift collected parse
tree to the \IdrisType{Either} type for all the transitions from
$\machine R$ to the accepting state. To do so, we add a
\IdrisData{Transform Left} instruction at the end of the routine, and,
similarly, a \IdrisData{Transform Rigth} instruction for transitions
from $\machine S$ to the accepting state.

\paragraph*{\figref{fig:sm-constructions:star}: Kleene star.}
We start building a state machine for $R^*$ (\IdrisData{Star}
\IdrisBound{R}) by building the state machine $\machine R$ for
$R$. The starting states in the new state machine are the starting
states from $\machine R$ together with the accepting state. We prepend
a \IdrisData{Push [<]} instruction to each initialisation routine,
pushing an accumulating empty \IdrisType{SnocList} on the stack. Thus, if we go
straight to the accepting state the parse tree will be the empty
\IdrisType{SnocList}. We keep all the transitions from the prefix parser $\machine R$. For each
transition into the accepting state, we:
\begin{itemize}
\item add a (potentially new) transition back to the starting states; and
\item for both the existing and new transitions, append a
  \IdrisData{ReducePair }\IdrisKeyword{(}\IdrisData{:<}\IdrisKeyword{)}
  instruction to the routine, pushing the newest $R$-parse tree on the
  accumulating \IdrisType{SnocList}.
\end{itemize}

\paragraph*{\figref{fig:sm-constructions:concat}: Concatenation.}
To build a state machine for $RS$ (\IdrisData{Concat}
\IdrisBound{R}\KatlaSpace\IdrisBound{S}) we start by building the
state machines $\machine R$ and $\machine S$ for $R$ and $S$,
respectively. The starting states in the state machine for concatenation will be
the starting states from $\machine R$. We keep all the transitions
from $\machine R$, except for the ones that lead to the accepting
state. For each of these latter transitions, we also add new
transitions to each of the starting states from $\machine S$, leaving
all the $\machine S$ unchanged. So now we will first traverse
$\machine R$ and then $\machine S$, and at the end the top of the
stack will contain a parse tree for $R$ and a parse tree for $S$. We
combine them by prepending a \IdrisData{ReducePair MkPair} instruction
to each routine label.

\paragraph*{\figref{fig:sm-constructions:conv}: Conversion.}
The state machine for the conversion operation (\IdrisData{Conv}
$R$\KatlaSpace\IdrisBound{f}) is the machine for $R$, with a
\IdrisData{Transform} \IdrisBound{f} instruction appended to each
accepting transition.

\paragraph*{An optimization.}
Often, when recursively constructing the Moore machine we add an
instruction at the very end of an existing routine. This happens for
\IdrisData{<|>}, \IdrisData{<*>}, \IdrisData{Rep}, and
\IdrisData{Conv}. On the other hand we add an element to the begining
of a routine only for \IdrisData{Rep} (initial \IdrisData{[Push
    [<]]}). Adding to the end of a list requires traversing the whole
list, which is suboptimal. For this reason we actually use snoc-list
accumulators for routines while constructing the Moore machine,
ensuring the pre and post stack-shapes of their constituent
instructions are aligned, and reverse the lists into
\IdrisType{Routine}s for execution.

\subsection{Group Minimization}\label{subsec:group minimisation}
For a typed regex \IdrisData{Group} \IdrisBound{a} the shape of the
parse tree is a \IdrisType{String}. The corresponding state machine is
then a regex \emph{matcher} rather than a parser.  We take advantage
of this property and create a completely separate mechanism for
building a state machine for groups. First we build an NFA; then we
perform some optimizations that shrink the NFA; and at last we build a
\IdrisType{MooreMachine String} from the NFA.

In the \emph{deterministic} case, automata minimisation can be solved
efficiently~\citep{hopcroft:nlogn-dra-minimisation}.  However, in the
non-deterministic case,
\cite{jiang-ravikumar:minima-NFA-problems-are-hard} show that NFA
minimisation is hard. Specifically, they show it is
\textsc{pspace}-complete.  We therefore use some simple heuristic to
minimise the NFA, and we leave to future work the task of implementing
optimal minimisation, such as
\citeauthor{kameda-weiner:NFA-minimisation}'s algorithm
\citeyearpar{kameda-weiner:NFA-minimisation}.

We build the NFA using a variant of Thompson's construction for our
type of NFA. This is the same construction as used for building Moore
machine for other typed regexes, but without the routines. To
represent an NFA here we use a dedicated data structure, which has a
list of the starting states and a mapping, for each state, of a list of pairs consisting of:
\begin{itemize}
\item a neighbouring state in the automaton; and
\item a \IdrisType{CharCond} character-predicate guarding the
  transition to this neighbouring state.
\end{itemize}
This is a simplification to a normal NFA, grouping the transition
relation by the target state.

\paragraph*{Optimizations.}
We shrink the size of the NFA by merging states that have the same
outgoing transitions, meaning that under the same condition we get the
same set of next states. We repeat this operation until there are no
more states that can be merged. We compare conditions by comparing the
constructor and arguments. Mind that we compare only the arguments in
\IdrisData{OneOf} and \IdrisData{Range}, but we cannot compare
functions passed to \IdrisData{Pred}, thus we always assume them to be
different. This choice can be optimised in the future using more
efficient data structures such as binary decision
diagrams~\citep{lee:bdds}.
\paragraph*{Building a Moore machine from the NFA.}
We build a state machine from the NFA. Each state, except the
accepting state, has the shape \IdrisData{[<]}. We define the
following code routines:
\begin{itemize}
\item We initialise each
starting state with the routine \IdrisData{[Record]}
\item We label all the
  transitions leading to the accepting state with \IdrisData{[EmitString]}.
\item If the accepting state is a starting state, we
  initialise it with \IdrisData{[Record, EmitString]}.
\item We label all other transitions with the empty routine
  \IdrisData{[]}.
\end{itemize}

\section{Evaluation}\label{evaluation}
\ignore{
\begin{code}
\IdrisKeyword{import}\KatlaSpace{}\IdrisModule{Data.Regex}\KatlaNewline{}
\IdrisKeyword{import}\KatlaSpace{}\IdrisModule{Data.Stream}\KatlaNewline{}
\IdrisKeyword{import}\KatlaSpace{}\IdrisModule{Data.SortedSet}\KatlaNewline{}
\IdrisKeyword{import}\KatlaSpace{}\IdrisModule{Data.List.Elem}\KatlaNewline{}
\IdrisKeyword{import}\KatlaSpace{}\IdrisModule{Data.DPair}\KatlaNewline{}
\IdrisKeyword{import}\KatlaSpace{}\IdrisModule{Sedris}\KatlaNewline{}
\KatlaNewline{}
\end{code}
}
We evaluate the TyRE library in two ways. Our first evaluation is a
quantitative comparison with the existing parser combinator library
and whether and how the TyRE design improves the state-of-the-art in
the Idris~2 ecosystem. We use pathological tests that emphasise
performance trade-offs between TyRE and Idris~2's parser combinator
library. Our second evaluation is qualitative, reporting on our
experience using TyRE as part of a bigger application: a stream
editing library.

\subsection{Comparing TyRE with Idris~2 parser combinators}

Since Idris~2 has no existing regex parser library, we evaluate TyRE
by comparing it with Idris~2's standard parser combinator
library. Idris~2's standard parser combinator library parses context
free languages, making it more powerful than a regex parser, although
without the same run-time complexity guarantees.

We have compared run-time for both TyRE and Idris 2's standard parser
combinator library for equivalent regex parsers and the same inputs,
for four pathological example regexes. These pathological examples
highlight the strengths and weaknesses of the two libraries.  Note
that in the examples in which we increase the size of the regex along
the horizontal axis, the resulting NFA is larger, and TyRE's run-time
scales non-linearly. For a \emph{fixed} regex, execution is guaranteed
to be linear in the size of the parsed string.

We ran\footnotemark{} the test files from a Python program, timing the
execution for each input size from the moment we supply the
input until the parser produces output. Each measurement for each
input size consists of 20 samples.  \footnotetext{We used
  an Intel Core i7 8700 3.2 2666MHz 6C, 65W CPU with 32GB (2x16GB)
  DDR4 2666 DIMM RAM and 7200RPM SATA-6G ROM.}

\begin{figure}
  \centering
  \begin{subfigure}[b]{0.49\textwidth}
      \centering
      \includegraphics[width=\textwidth]{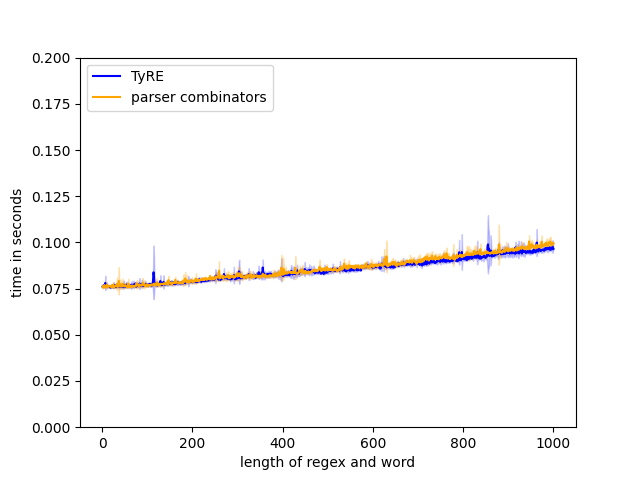}
      \caption{Regex: $a^n$; string: $a^n$}
      \label{fig:performence:concat}
  \end{subfigure}
  \hfill
  \begin{subfigure}[b]{0.49\textwidth}
    \centering
    \includegraphics[width=\textwidth]{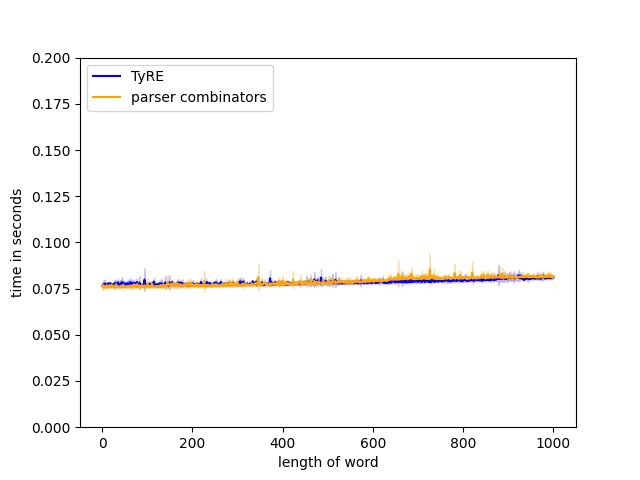}
    \caption{Regex: $a*$; string: $a^n$}
    \label{fig:performence:star}
  \end{subfigure}
  \hfill
  \begin{subfigure}[b]{0.49\textwidth}
      \centering
      \includegraphics[width=\textwidth]{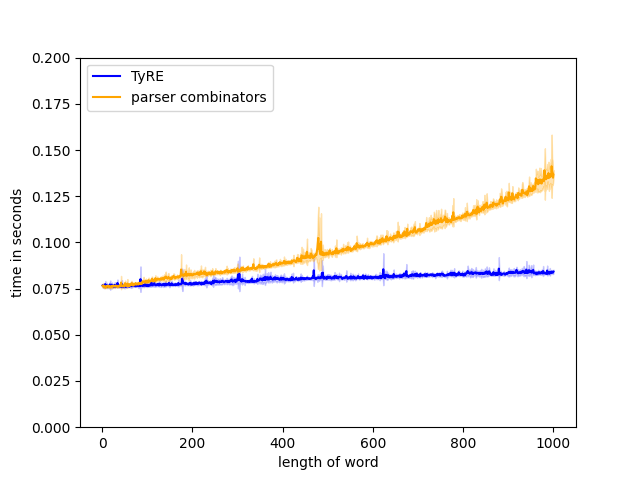}
      \caption{Regex: $((a*c)\vert a)*b$; string: $a^nb$}
      \label{fig:performence:star2}
  \end{subfigure}
  \hfill
  \begin{subfigure}[b]{0.49\textwidth}
      \centering
      \includegraphics[width=\textwidth]{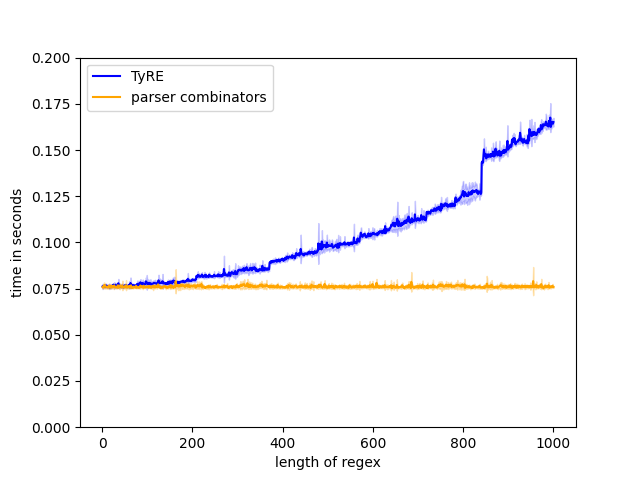}
      \caption{Regex: $a(\vert a)^{n-1}$; string: $a$}
      \label{fig:performence:alt}
  \end{subfigure}
  \caption{Parsing times for TyRE and Idris 2's stdlib parser combinator library.}
  \label{fig:performence}
\end{figure}

Figure \ref{fig:performence} compares parsing times for the parser
combinator library and TyRE. On each chart, we plot the mean execution
time for Idris~2 stdlib parser combinators ({\color{orange}orange})
and for TyRE ({\color{blue}blue}). We add a fainter band of the same
colour, whose magnitude is $1$ standard deviation of the sampled
run-time for this input size to each side.

\paragraph*{Figure~\ref{fig:performence:concat}: regex: $a^n$ parsee: $a^n$.}
We compare the times for parsing a string of $n$ consecutive $a$'s
with a the regex $a^n$ for $n$ from $1$ to $1000$. The execution times
are similar for both libraries and approximately linear with respect
to the length $n$ of the regex and string. Since at each step there is
only one possible next state, TyRE maintains a single thread as it
traverses the Moore machine. Similarly, the parser-combinator parser
consume characters one by one without any need for backtracking.
\paragraph*{Figure~\ref{fig:performence:star}: regex: $a*$ parsee: $a^n$.}
Here we parse the same family of strings $a^n$, but uniformly with the
regex $a*$. Again both libraries run in approximately linear time
with respect to $n$ and the execution times are similar across the two
libraries.
\paragraph*{Figure~\ref{fig:performence:star2}: regex: $((a*c)\vert a)*b$
  parsee: $a^nb$.}
In this test TyRE scales better than Idris~2's parser combinator library. The
parser that we built with parser combinators will always first choose
the left branch when presented with alternation, which for this
example is always the incorrect choice. This example thus forces the
parser to backtrack repeatedly, resulting in a quadratic
complexity. TyRE searches both branches simultaneously. Thanks to
the trimming of the search space by merging threads, TyRE maintains
only a constant number of threads resulting in the linear complexity.

To highlight TyRE's limitations, here is a pathological case for TyRE:
\paragraph*{Figure~\ref{fig:performence:alt}: regex: $a(\vert a)^{n-1}$
  parsee: $a$.}
TyRE searching all possibilities at once is not always
the best option.  Here, we parse a single character word $a$ with
$a(\vert a)^{n-1}$. The parser that we built with
parser combinators parses this string in an almost constant time. It
chooses one of the possible branches and quickly finds a match for the
single character. On the other hand, TyRE creates $n$ threads for each
possible path and creates a parse tree for each one. Because each
choice is encoded by the \IdrisType{Either} type, the built parse tree
has up to $O(n)$-many \IdrisType{Either} constructors. This simultaneous
traversal results in the quadratic execution time of TyRE in this
test.

\begin{figure}
  \centering
  \includegraphics[width=0.49\textwidth]{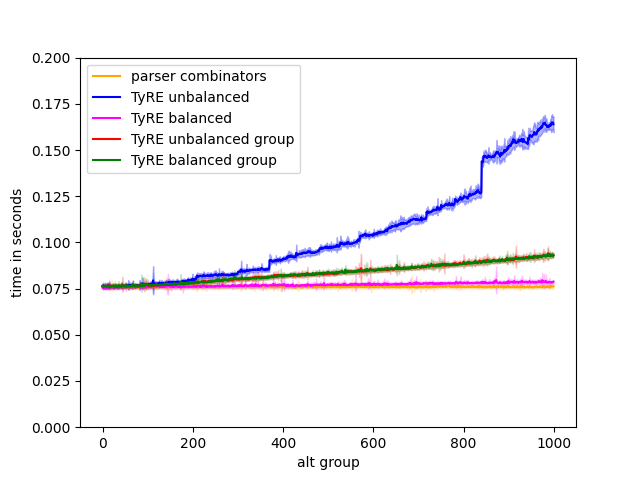}
  \caption{Parsing $a$ with different alternation patterns of $a(\vert a)^{n-1}$.}
  \label{fig:group}
\end{figure}

This example is however pathological, and we created it by meta
programming the regex and its associated type. Figure~\ref{fig:group}
evaluates three workarounds for this issue. We
repeat ({\color{blue}blue}) the evaluation from Figure \ref{fig:performence:alt}: an
unbalanced spine of regex alternations.

We can instead balance the alternation in the
regex to get shorter encoding of the chosen branch.  For example, for
$n = 8$:
\begin{center}
$\left(\left(a\mathbin\vert a\right)\mathbin\vert\left(a\mathbin\vert a\right)\right)\mathbin\vert\left(\left(a\mathbin\vert
a\right)\mathbin\vert\left(a\mathbin\vert a\right)\right)$ instead of $\left(\left(\left(\left(\left(\left(a\mathbin\vert a\right)\mathbin\vert a\right)\mathbin\vert
a\right)\mathbin\vert a\right)\mathbin\vert a\right)\mathbin\vert a\right)\mathbin\vert a$.
\end{center}
This balancing ({\color{magenta}magenta}) improves the runtime to
O($n\log n$).  This improved performance is slower than the
non-backtracking parser combinator counterpart
({\color{orange}orange}).

An easier `fix' is to disregard tracking the matching alternative. We
apply the \IdrisFunction{ignore} smart constructor to the regex, which
applies the \IdrisData{Group} constructor to the resulting
TyRE. Grouping the unbalanced regex ({\color{red}red}) improves the
parsing time significantly. This grouping means that the NFA will be
minimized and all the alternations will collapse into a single
path. However, grouping the balanced regex ({\color{Green4}green})
degrades its performance in this case. This visible overhead is the
cost of minimization, where at one point or another during NFA
traversal we compare each state with all the other ones, again getting
a flatter quadratic slope. The real benefit of the \IdrisData{Group}
constructor is in minimizing the NFA automatically, staging the
minimized NFA and reusing it.

More qualitatively, a big difference for users in practice is the
interface to both libraries. To support regex literals, TyRE only
parses character strings, whereas the parser combinator library is much
more generic. On the other hand, writing even a simple regex matcher
with the parser combinator library requires the user to define more
entities, e.g. a set of tokens and a lexer.  In fact, Idris~2's stdlib
supports a copy of the parser combinator library for dealing solely
with strings for this reason.  In addition, the interface to the
parser combinator library, even the streamlined character string one,
might be less familiar to newcomers compared to the more familiar
regex literals.

\subsection{Sedris}
We used TyRE in a stream-editing embedded DSL --- \emph{sedris}.
We base sedris on the GNU sed program\footnotemark~\citep{sed-manual}.%
\footnotetext{Lee E.~McMahon first developed sed around
  1973~\citep[Ch.~9]{unix-history}.}
A sedris script is a list of commands executed from top to bottom.  A
script can invoke `file scripts'. A file script operates on a file and
is executed for each consecutive line.

We demonstrate
how to use sedris to rename variables in code\footnotemark{}. For example, in a
hypothetical previous version of TyRE, we call the typed `Moore
machine' a `state machine'. We can use sedris to substitute in all the
variables the substrings `\IdrisBound{sm}'/`\IdrisBound{SM}' --- standing for state machine --- to the substrings `\IdrisBound{mm}'/`\IdrisBound{MM}' ---
standing for Moore machine. For example, `\IdrisBound{asSM}' $\mapsto$ `\IdrisBound{asMM}' and `\IdrisBound{sm}'
$\mapsto$ `\IdrisBound{mm}'. We'll match all the words that
either end with capital `\IdrisBound{SM}' (\IdrisFunction{endsWithSM}) or begin
with `\IdrisBound{sm}' or `\IdrisBound{SM}' followed by an optional string that starts with a
capital letter (\IdrisFunction{startsWithSM}). To match only whole
words our pattern must be delimited by a space. To deal with
the first and last word of a line we'll artificially add spaces on
both ends of each line before matching and later strip them before
writing the line to the file.
\footnotetext{We use this example for demonstration purposes only. We
  do not condone the use of simple regex-based text substitution for
  renaming binding and bound occurrences of program variables. We firmly
  believe substitution ought to be done by scope-aware tools that are
  well-equipped for this task. We only recommend regex-based
  substitution in the absence of convenient scope-aware tools.}

\begin{figure}
  \input{generated/SedrisCode}
  \caption{Sedris script replacing occurrences of `sm' to `mm'.}
  \label{fig:sedris-script}
\end{figure}

\ignore{
\begin{code}
\IdrisKeyword{namespace}\KatlaSpace{}\IdrisNamespace{SedrisExplain1}\KatlaNewline{}
\KatlaSpace{}\KatlaSpace{}\IdrisKeyword{data}\KatlaSpace{}\IdrisType{ReplaceCommand}\KatlaSpace{}\IdrisKeyword{:}\KatlaSpace{}\IdrisType{Type}\KatlaSpace{}\IdrisKeyword{where}\KatlaNewline{}
\end{code}
}

\paragraph*{Sedris with TyRE.}
Figure~\ref{fig:sedris-script} presents the example sedris script.
The full sedris DSL is beyond the scope of this manuscript,
and we explain only the relevant sedris fragments.

At the heart of sedris is the type \IdrisType{ReplaceCommand} whose
values tell sedris how to replace substrings. Here (line 16) we use
the following variant, which takes a \IdrisType{TyRE} to match and a
substitution sending each match to its replacement string:
\begin{code}
\KatlaSpace{}\KatlaSpace{}\KatlaSpace{}\KatlaSpace{}\IdrisData{AllRe}\KatlaSpace{}\IdrisKeyword{:}\KatlaSpace{}\IdrisKeyword{(}\IdrisBound{re}\KatlaSpace{}\IdrisKeyword{:}\KatlaSpace{}\IdrisType{TyRE}\KatlaSpace{}\IdrisBound{a}\IdrisKeyword{)}\KatlaSpace{}\IdrisKeyword{\KatlaDash{}>}\KatlaSpace{}\IdrisKeyword{\{auto}\KatlaSpace{}\IdrisKeyword{0}\KatlaSpace{}\IdrisBound{consuming}\KatlaSpace{}\IdrisKeyword{:}\KatlaSpace{}\IdrisType{IsConsuming}\KatlaSpace{}\IdrisBound{re}\IdrisKeyword{\}}\KatlaSpace{}\IdrisKeyword{\KatlaDash{}>}\KatlaNewline{}
\KatlaSpace{}\KatlaSpace{}\KatlaSpace{}\KatlaSpace{}\KatlaSpace{}\KatlaSpace{}\KatlaSpace{}\KatlaSpace{}\KatlaSpace{}\KatlaSpace{}\IdrisKeyword{(}\IdrisBound{a}\KatlaSpace{}\IdrisKeyword{\KatlaDash{}>}\KatlaSpace{}\IdrisType{String}\IdrisKeyword{)}\KatlaSpace{}\IdrisKeyword{\KatlaDash{}>}\KatlaSpace{}\IdrisType{ReplaceCommand}\KatlaNewline{}
\end{code}
\ignore{
\begin{code}
\KatlaSpace{}\KatlaSpace{}\IdrisKeyword{\%hide}\KatlaSpace{}Main.SedrisExplain1.ReplaceCommand\KatlaNewline{}
\end{code}
}
To guarantee productivity, \IdrisData{AllRe} also requires a proof that
the \IdrisType{TyRE} is \emph{consuming}, i.e., successful matches
consume at least one input character.

Sedris applies these transformations by storing each consecutive line
from its input in a designated \emph{pattern space}.  Some sedris
commands only make sense in the context of such line-by-line
processing, and so we index commands by the following type:
\begin{code}
\KatlaSpace{}\KatlaSpace{}\IdrisKeyword{data}\KatlaSpace{}\IdrisType{ScriptType}\KatlaSpace{}\IdrisKeyword{=}\KatlaSpace{}\IdrisData{Total}\KatlaSpace{}\IdrisKeyword{|}\KatlaSpace{}\IdrisData{LineByLine}\KatlaNewline{}
\end{code}
\ignore{
\begin{code}
\KatlaSpace{}\KatlaSpace{}\IdrisKeyword{\%hide}\KatlaSpace{}Main.SedrisExplain1.ScriptType\KatlaNewline{}
\end{code}
} More sophisticated sedris commands may bind and refer to sedris
variables --- which we will not discuss here.  We index commands by their relationship to the variables in context:
their scope, whose type is \IdrisFunction{Variables} --- a \IdrisType{SnocList} of
\IdrisFunction{Variable}s; and the \IdrisType{List} of
\IdrisFunction{Variable}s that they bind.  Some commands assume they
interact with a file in memory or on disk, and we express this type of
assumption using the type \IdrisType{FileScriptType}.
Thus, sedris indexes its \IdrisType{Command} type by four arguments:
\begin{code}
\KatlaSpace{}\KatlaSpace{}\IdrisKeyword{data}\KatlaSpace{}\IdrisType{Command}\KatlaSpace{}\IdrisKeyword{:}\KatlaSpace{}\IdrisKeyword{(}\IdrisBound{scope}\KatlaSpace{}\IdrisKeyword{:}\KatlaSpace{}\IdrisFunction{Variables}\IdrisKeyword{)}\KatlaSpace{}\IdrisKeyword{\KatlaDash{}>}\KatlaSpace{}\IdrisKeyword{(}\IdrisBound{binding}\KatlaSpace{}\IdrisKeyword{:}\KatlaSpace{}\IdrisType{List}\KatlaSpace{}\IdrisFunction{Variable}\IdrisKeyword{)}\KatlaSpace{}\IdrisKeyword{\KatlaDash{}>}\KatlaNewline{}
\KatlaSpace{}\KatlaSpace{}\KatlaSpace{}\KatlaSpace{}\KatlaSpace{}\KatlaSpace{}\KatlaSpace{}\KatlaSpace{}\KatlaSpace{}\KatlaSpace{}\KatlaSpace{}\KatlaSpace{}\KatlaSpace{}\KatlaSpace{}\IdrisType{ScriptType}\KatlaSpace{}\IdrisKeyword{\KatlaDash{}>}\KatlaSpace{}\IdrisType{FileScriptType}\KatlaSpace{}\IdrisKeyword{\KatlaDash{}>}\KatlaSpace{}\IdrisType{Type}\KatlaNewline{}
\end{code}
\ignore{
\begin{code}
\KatlaSpace{}\KatlaSpace{}\KatlaSpace{}\KatlaSpace{}\IdrisKeyword{where}\KatlaNewline{}
\end{code}
}
We call replacement commands such as \IdrisData{AllRe}
(line 15) with the \IdrisData{Replace} constructor:
\begin{code}
\KatlaSpace{}\KatlaSpace{}\KatlaSpace{}\KatlaSpace{}\IdrisData{Replace}\KatlaSpace{}\IdrisKeyword{:}\KatlaSpace{}\IdrisType{ReplaceCommand}\KatlaSpace{}\IdrisKeyword{\KatlaDash{}>}\KatlaSpace{}\IdrisType{Command}\KatlaSpace{}\IdrisBound{sx}\KatlaSpace{}\IdrisData{[]}\KatlaSpace{}\IdrisBound{st}\KatlaSpace{}\IdrisBound{io}\KatlaNewline{}
\end{code}
whose indices express the following guarantees:
\begin{itemize}
  \item \IdrisBound{sx}: the command assumes nothing about its
    enclosing variable scope, and so this command is valid in any
    scope;
  \item \IdrisData{[]}: the command does not bind any new variables;
  \item \IdrisBound{st}: this command may appear in any script type; and
  \item \IdrisBound{io}: this command may interact with any kind of file.
\end{itemize}
Sedris file scripts may also invoke arbitrary Idris string transformations
(lines 14 and 18):
\begin{code}
\KatlaSpace{}\KatlaSpace{}\KatlaSpace{}\KatlaSpace{}\IdrisData{Exec}\KatlaSpace{}\IdrisKeyword{:}\KatlaSpace{}\IdrisKeyword{(}\IdrisType{String}\KatlaSpace{}\IdrisKeyword{\KatlaDash{}>}\KatlaSpace{}\IdrisType{String}\IdrisKeyword{)}\KatlaSpace{}\IdrisKeyword{\KatlaDash{}>}\KatlaSpace{}\IdrisType{Command}\KatlaSpace{}\IdrisBound{sx}\KatlaSpace{}\IdrisData{[]}\KatlaSpace{}\IdrisBound{st}\KatlaSpace{}\IdrisBound{io}\KatlaNewline{}
\end{code}

We write the pattern space out to a file with the following command (line 22),
which requires the file script's stream to support output:
\begin{code}
\KatlaSpace{}\KatlaSpace{}\KatlaSpace{}\KatlaSpace{}\IdrisData{WriteTo}\KatlaSpace{}\IdrisKeyword{:}\KatlaSpace{}\IdrisKeyword{\{0}\KatlaSpace{}\IdrisBound{t}\KatlaSpace{}\IdrisKeyword{:}\KatlaSpace{}\IdrisType{FileScriptType}\IdrisKeyword{\}}\KatlaSpace{}\IdrisKeyword{\KatlaDash{}>}\KatlaSpace{}\IdrisKeyword{\{auto}\KatlaSpace{}\IdrisBound{isIO}\KatlaSpace{}\IdrisKeyword{:}\KatlaSpace{}\IdrisType{NeedsIO}\KatlaSpace{}\IdrisBound{t}\IdrisKeyword{\}}\KatlaNewline{}
\KatlaSpace{}\KatlaSpace{}\KatlaSpace{}\KatlaSpace{}\KatlaSpace{}\KatlaSpace{}\KatlaSpace{}\KatlaSpace{}\KatlaSpace{}\KatlaSpace{}\KatlaSpace{}\KatlaSpace{}\IdrisKeyword{\KatlaDash{}>}\KatlaSpace{}\IdrisFunction{IOFile}\KatlaSpace{}\IdrisKeyword{\KatlaDash{}>}\KatlaSpace{}\IdrisType{Command}\KatlaSpace{}\IdrisBound{sx}\KatlaSpace{}\IdrisData{[]}\KatlaSpace{}\IdrisData{LineByLine}\KatlaSpace{}\IdrisBound{t}\KatlaNewline{}
\end{code}

We can create a new output file/delete an existing output file's
contents in preparation for writing to it line-by-line (line 10):
\begin{code}
\KatlaSpace{}\KatlaSpace{}\KatlaSpace{}\KatlaSpace{}\IdrisData{ClearFile}\KatlaSpace{}\IdrisKeyword{:}\KatlaSpace{}\IdrisKeyword{\{0}\KatlaSpace{}\IdrisBound{t}\KatlaSpace{}\IdrisKeyword{:}\KatlaSpace{}\IdrisType{FileScriptType}\IdrisKeyword{\}}\KatlaSpace{}\IdrisKeyword{\KatlaDash{}>}\KatlaSpace{}\IdrisKeyword{\{auto}\KatlaSpace{}\IdrisBound{isIO}\KatlaSpace{}\IdrisKeyword{:}\KatlaSpace{}\IdrisType{NeedsIO}\KatlaSpace{}\IdrisBound{t}\IdrisKeyword{\}}\KatlaNewline{}
\KatlaSpace{}\KatlaSpace{}\KatlaSpace{}\KatlaSpace{}\KatlaSpace{}\KatlaSpace{}\KatlaSpace{}\KatlaSpace{}\KatlaSpace{}\KatlaSpace{}\KatlaSpace{}\KatlaSpace{}\KatlaSpace{}\KatlaSpace{}\IdrisKeyword{\KatlaDash{}>}\KatlaSpace{}\IdrisFunction{IOFile}\KatlaSpace{}\IdrisKeyword{\KatlaDash{}>}\KatlaSpace{}\IdrisType{Command}\KatlaSpace{}\IdrisBound{sx}\KatlaSpace{}\IdrisData{[]}\KatlaSpace{}\IdrisData{LineByLine}\KatlaSpace{}\IdrisBound{t}\KatlaNewline{}
\end{code}
\ignore{
\begin{code}
\KatlaSpace{}\KatlaSpace{}\IdrisKeyword{\%hide}\KatlaSpace{}Main.SedrisExplain1.Command\KatlaNewline{}
\end{code}
}
As we compose commands into file scripts, we can contextualise some of them
to only take effect in specific \emph{addresses}:
\begin{itemize}
\item explicitly given line numbers;
\item explicit line ranges;
\item lines that match a regex guard --- given by a \IdrisType{TyRE}; or
\item the last line in the file.
\end{itemize}
We specify an address by tagging the command with it:
\begin{code}
\KatlaSpace{}\KatlaSpace{}\IdrisKeyword{data}\KatlaSpace{}\IdrisType{CommandWithAddress}\KatlaSpace{}\IdrisKeyword{:}\KatlaSpace{}\IdrisKeyword{(}\IdrisBound{scope}\KatlaSpace{}\IdrisKeyword{:}\KatlaSpace{}\IdrisFunction{Variables}\IdrisKeyword{)}\KatlaSpace{}\IdrisKeyword{\KatlaDash{}>}\KatlaNewline{}
\KatlaSpace{}\KatlaSpace{}\KatlaSpace{}\KatlaSpace{}\KatlaSpace{}\KatlaSpace{}\IdrisKeyword{(}\IdrisBound{binding}\KatlaSpace{}\IdrisKeyword{:}\KatlaSpace{}\IdrisType{List}\KatlaSpace{}\IdrisFunction{Variable}\IdrisKeyword{)}\KatlaSpace{}\IdrisKeyword{\KatlaDash{}>}\KatlaNewline{}
\KatlaSpace{}\KatlaSpace{}\KatlaSpace{}\KatlaSpace{}\KatlaSpace{}\KatlaSpace{}\IdrisType{FileScriptType}\KatlaSpace{}\IdrisKeyword{\KatlaDash{}>}\KatlaSpace{}\IdrisType{Type}\KatlaSpace{}\IdrisKeyword{where}\KatlaNewline{}
\KatlaSpace{}\KatlaSpace{}\KatlaSpace{}\KatlaSpace{}\IdrisData{(>)}\KatlaSpace{}\KatlaSpace{}\IdrisKeyword{:}\KatlaSpace{}\IdrisType{Command}\KatlaSpace{}\IdrisBound{sx}\KatlaSpace{}\IdrisBound{ys}\KatlaSpace{}\IdrisData{LineByLine}\KatlaSpace{}\IdrisBound{t}\KatlaSpace{}\IdrisKeyword{\KatlaDash{}>}\KatlaNewline{}
\KatlaSpace{}\KatlaSpace{}\KatlaSpace{}\KatlaSpace{}\KatlaSpace{}\KatlaSpace{}\KatlaSpace{}\KatlaSpace{}\KatlaSpace{}\KatlaSpace{}\KatlaSpace{}\IdrisType{CommandWithAddress}\KatlaSpace{}\IdrisBound{sx}\KatlaSpace{}\IdrisBound{ys}\KatlaSpace{}\IdrisBound{t}\KatlaNewline{}
\KatlaSpace{}\KatlaSpace{}\KatlaSpace{}\KatlaSpace{}\IdrisData{(?>)}\KatlaSpace{}\IdrisKeyword{:}\KatlaSpace{}\IdrisType{Address}\KatlaSpace{}\IdrisKeyword{\KatlaDash{}>}\KatlaSpace{}\IdrisType{Command}\KatlaSpace{}\IdrisBound{sx}\KatlaSpace{}\IdrisData{[]}\KatlaSpace{}\IdrisData{LineByLine}\KatlaSpace{}\IdrisBound{t}\KatlaSpace{}\IdrisKeyword{\KatlaDash{}>}\KatlaNewline{}
\KatlaSpace{}\KatlaSpace{}\KatlaSpace{}\KatlaSpace{}\KatlaSpace{}\KatlaSpace{}\KatlaSpace{}\KatlaSpace{}\KatlaSpace{}\KatlaSpace{}\KatlaSpace{}\IdrisType{CommandWithAddress}\KatlaSpace{}\IdrisBound{sx}\KatlaSpace{}\IdrisData{[]}\KatlaSpace{}\IdrisBound{t}\KatlaNewline{}
\end{code}
\ignore{
\begin{code}
\KatlaSpace{}\KatlaSpace{}\IdrisKeyword{\%hide}\KatlaSpace{}Main.SedrisExplain1.CommandWithAddress\KatlaNewline{}
\KatlaSpace{}\KatlaSpace{}\IdrisKeyword{\%hide}\KatlaSpace{}Sedris.Lang.FileScript\KatlaNewline{}
\end{code}
}
In our example, we only clear the output file on the first line of the
input file (line 12) and we use the \IdrisData{Line} address
constructor. All other file script commands apply every line.

We compose these potentially contextualised commands together to form
scripts that apply to a file using ornate lists:
\begin{code}
\KatlaSpace{}\KatlaSpace{}\IdrisKeyword{data}\KatlaSpace{}\IdrisType{FileScript}\KatlaSpace{}\IdrisKeyword{:}\KatlaSpace{}\IdrisFunction{Variables}\KatlaSpace{}\IdrisKeyword{\KatlaDash{}>}\KatlaSpace{}\IdrisType{FileScriptType}\KatlaSpace{}\IdrisKeyword{\KatlaDash{}>}\KatlaSpace{}\IdrisType{Type}\KatlaSpace{}\IdrisKeyword{where}\KatlaNewline{}
\KatlaSpace{}\KatlaSpace{}\KatlaSpace{}\KatlaSpace{}\IdrisData{Nil}\KatlaSpace{}\IdrisKeyword{:}\KatlaSpace{}\IdrisType{FileScript}\KatlaSpace{}\IdrisBound{sx}\KatlaSpace{}\IdrisBound{t}\KatlaNewline{}
\KatlaSpace{}\KatlaSpace{}\KatlaSpace{}\KatlaSpace{}\IdrisData{(::)}\KatlaSpace{}\IdrisKeyword{:}\KatlaSpace{}\IdrisType{CommandWithAddress}\KatlaSpace{}\IdrisBound{sx}\KatlaSpace{}\IdrisBound{ys}\KatlaSpace{}\IdrisBound{t}\KatlaSpace{}\IdrisKeyword{\KatlaDash{}>}\KatlaSpace{}\IdrisType{FileScript}\KatlaSpace{}\IdrisKeyword{(}\IdrisBound{sx}\KatlaSpace{}\IdrisFunction{<><}\KatlaSpace{}\IdrisBound{ys}\IdrisKeyword{)}\KatlaSpace{}\IdrisBound{t}\KatlaNewline{}
\KatlaSpace{}\KatlaSpace{}\KatlaSpace{}\KatlaSpace{}\KatlaSpace{}\KatlaSpace{}\KatlaSpace{}\KatlaSpace{}\IdrisKeyword{\KatlaDash{}>}\KatlaSpace{}\IdrisType{FileScript}\KatlaSpace{}\IdrisBound{sx}\KatlaSpace{}\IdrisBound{t}\KatlaNewline{}
\end{code}
\ignore{
\begin{code}
\KatlaSpace{}\KatlaSpace{}\IdrisKeyword{\%unhide}\KatlaSpace{}Sedris.Lang.FileScript\KatlaNewline{}
\KatlaSpace{}\KatlaSpace{}\IdrisKeyword{\%hide}\KatlaSpace{}Main.SedrisExplain1.FileScript\KatlaNewline{}
\end{code}
}
The cons constructor (\IdrisData{::}) uses the `fish' operator --- the action of \IdrisType{List}s on \IdrisType{SnocList}s:
\begin{code}
\KatlaSpace{}\KatlaSpace{}\IdrisFunction{(<><)}\KatlaSpace{}\IdrisKeyword{:}\KatlaSpace{}\IdrisType{SnocList}\KatlaSpace{}\IdrisBound{a}\KatlaSpace{}\IdrisKeyword{\KatlaDash{}>}\KatlaSpace{}\IdrisType{List}\KatlaSpace{}\IdrisBound{a}\KatlaSpace{}\IdrisKeyword{\KatlaDash{}>}\KatlaSpace{}\IdrisType{SnocList}\KatlaSpace{}\IdrisBound{a}\KatlaNewline{}
\end{code}
A script comprises of \IdrisType{ScriptCommand}s:
\begin{code}
\KatlaSpace{}\KatlaSpace{}\IdrisKeyword{data}\KatlaSpace{}\IdrisType{ScriptCommand}\KatlaSpace{}\IdrisKeyword{:}\KatlaSpace{}\IdrisKeyword{(}\IdrisBound{scope}\KatlaSpace{}\IdrisKeyword{:}\KatlaSpace{}\IdrisFunction{Variables}\IdrisKeyword{)}\KatlaSpace{}\IdrisKeyword{\KatlaDash{}>}\KatlaSpace{}\IdrisKeyword{(}\IdrisBound{binding}\KatlaSpace{}\IdrisKeyword{:}\KatlaSpace{}\IdrisType{List}\KatlaSpace{}\IdrisFunction{Variable}\IdrisKeyword{)}\KatlaNewline{}
\KatlaSpace{}\KatlaSpace{}\IdrisKeyword{\KatlaDash{}>}\KatlaSpace{}\IdrisType{FileScriptType}\KatlaSpace{}\IdrisKeyword{\KatlaDash{}>}\KatlaSpace{}\IdrisType{Type}\KatlaNewline{}
\end{code}
\ignore{
\begin{code}
\KatlaSpace{}\KatlaSpace{}\KatlaSpace{}\KatlaSpace{}\KatlaSpace{}\KatlaSpace{}\IdrisKeyword{where}\KatlaNewline{}
\end{code}
}
For example, we apply a file script to a file using this constructor:
\begin{code}
\KatlaSpace{}\KatlaSpace{}\KatlaSpace{}\KatlaSpace{}\IdrisData{(*)}\KatlaSpace{}\IdrisKeyword{:}\KatlaSpace{}\IdrisType{List}\KatlaSpace{}\IdrisFunction{IOFile}\KatlaSpace{}\IdrisKeyword{\KatlaDash{}>}\KatlaSpace{}\IdrisType{FileScript}\KatlaSpace{}\IdrisBound{sx}\KatlaSpace{}\IdrisData{IO}\KatlaSpace{}\IdrisKeyword{\KatlaDash{}>}\KatlaSpace{}\IdrisType{ScriptCommand}\KatlaSpace{}\IdrisBound{sx}\KatlaSpace{}\IdrisData{[]}\KatlaSpace{}\IdrisData{IO}\KatlaNewline{}
\end{code}
\ignore{
\begin{code}
\KatlaSpace{}\KatlaSpace{}\IdrisKeyword{\%hide}\KatlaSpace{}Main.SedrisExplain1.ScriptCommand\KatlaNewline{}
\IdrisKeyword{namespace}\KatlaSpace{}\IdrisNamespace{SedrisExplain2}\KatlaNewline{}
\KatlaSpace{}\KatlaSpace{}\IdrisKeyword{\%hide}\KatlaSpace{}Sedris.Lang.Script.Script\KatlaNewline{}
\end{code}
}
We compose top-level script commands into scripts using another type of ornate lists:
\begin{code}
\KatlaSpace{}\KatlaSpace{}\IdrisKeyword{data}\KatlaSpace{}\IdrisType{Script}\KatlaSpace{}\IdrisKeyword{:}\KatlaSpace{}\IdrisFunction{Variables}\KatlaSpace{}\IdrisKeyword{\KatlaDash{}>}\KatlaSpace{}\IdrisType{FileScriptType}\KatlaSpace{}\IdrisKeyword{\KatlaDash{}>}\KatlaSpace{}\IdrisType{Type}\KatlaSpace{}\IdrisKeyword{where}\KatlaNewline{}
\KatlaSpace{}\KatlaSpace{}\KatlaSpace{}\KatlaSpace{}\IdrisData{Nil}\KatlaSpace{}\IdrisKeyword{:}\KatlaSpace{}\IdrisType{Script}\KatlaSpace{}\IdrisBound{sx}\KatlaSpace{}\IdrisBound{t}\KatlaNewline{}
\KatlaSpace{}\KatlaSpace{}\KatlaSpace{}\KatlaSpace{}\IdrisData{(::)}\KatlaSpace{}\IdrisKeyword{:}\KatlaSpace{}\IdrisType{ScriptCommand}\KatlaSpace{}\IdrisBound{sx}\KatlaSpace{}\IdrisBound{ys}\KatlaSpace{}\IdrisBound{t}\KatlaSpace{}\IdrisKeyword{\KatlaDash{}>}\KatlaSpace{}\IdrisType{Script}\KatlaSpace{}\IdrisKeyword{(}\IdrisBound{sx}\KatlaSpace{}\IdrisFunction{<><}\KatlaSpace{}\IdrisBound{ys}\IdrisKeyword{)}\KatlaSpace{}\IdrisBound{t}\KatlaSpace{}\IdrisKeyword{\KatlaDash{}>}\KatlaSpace{}\IdrisType{Script}\KatlaSpace{}\IdrisBound{sx}\KatlaSpace{}\IdrisBound{t}\KatlaNewline{}
\end{code}
\ignore{
\begin{code}
\KatlaSpace{}\KatlaSpace{}\IdrisKeyword{\%hide}\KatlaSpace{}Main.SedrisExplain2.Script\KatlaNewline{}
\KatlaSpace{}\KatlaSpace{}\IdrisKeyword{\%unhide}\KatlaSpace{}Sedris.Lang.Script.Script\KatlaNewline{}
\end{code}
}

\paragraph*{Experience report.}
When writing this small sedris script, we had no issues with
expressing the desired substitution using TyRE. Seeing that the
computed regex type matches the expected one gave us confidence in the
pattern. Later, when writing the \IdrisFunction{toStr} function, we
used type-driven style, interactively covering all the cases.

There were two inconvenient aspects to TyRE. First, the lack of
special marks matching end or beginning of a string in TyRE
tokens. Such tokens are a common extension in regexes
engines. In this case they would have allow us to skip artificially
adding spaces on both sides of each line. Second is the number of
brackets needed in the pattern. For example, in
\IdrisData{"((sm)|(SM))!"} all the brackets are needed, because in
TyRE alternation and concatenation have the same precedence, though
commonly concatenation binds stronger.

\section{Related work}\label{related_work}
Our work is based directly on \citeauthor{Radanne2019TypedEngines}'s
tyre design \citeyearpar{Radanne2019TypedEngines}. Radanne presents a
typed regex layer that can be added on top of an existing regex parser
to ensure safety. He implements his design in the OCaml \texttt{tyre}
library. We extend his work by ensuring type safety throughout the
layers, including our custom regex engine. An feature in Radanne's
design that we have excluded is the parse-tree unparsing. To
support unparsing, we can use Radanne's original \IdrisData{Conv}ersion constructor
that has bi-directional conversion functions.

Other work also ensure safety for regex matching by adding a safety
layer for an existing regex engine. \cite{java-regex-safe} implement a
type system for regular expressions in Java. The type system checks
the validity of a regex statically and ensures the correct number of
capture group results, eliminating an `IndexOutOfBoundsException'
exception. The implementation is available as a part of the Checker
Framework -- a plugin for the JVM.  \cite{scala-type-safe} provides
similar safety guarantees while also lifting possible \texttt{null}'s
coming from optinal parts into an \texttt{Option} type. Blanvillain
takes advantage of Scala 3's \emph{match types}~\citep{match-types} to
perform the neccassary type level computation.
\cite{typescript-typed-regex} takes a similar approach in the
Typescript `typed-regex' library which provides a safety layer for an
existing regex engine. This design uses conditional types (Typescript
Handbook, \citeyear{conditional-types}) for type level computation.
This libary provides named capture groups, and it guarantees to return a
directory with the relevant names.

\cite{dth} implements a type-safe regex parser in Haskell. While her
work is a proof of concept that aims to show possible uses of
dependent types in Haskell, her aims are similar to ours. Her
implementation uses Brzozowski derivatives for matching.
\cite{regex-applicative} implements `regex-applicative', a Haskell
type-safe regex parser. It uses parser combinators, allowing for an
easy transformation of the result. It does not yet support regex
string literals.

\cite{oregano} implement \emph{Oregano}, a typed regex parser in
Haskell. Like our implementation, Oregano parses using a typed Moore
machine with a heterogenous stack. Oregano implements the Moore
machine by generating a function that guarantees to return the correct
shape, rather than an explicit virtual-machine implementation. One
concrete difference is that TyRE removes duplicate threads that meet
in the same state and thus ensures linear performance. In comparison,
Oregano implements a backtracking parser, and counts all possible
ambiguous results.  Oregano also uses more advanced staging techniques
to improve performance.

While TyRE is type \emph{safe}, its types do not ensure neither:
\begin{itemize}
\item \emph{soundness}: only matches return a parse-tree;
\item \emph{completeness}: all matches will parse correctly.
\end{itemize}
We believe our design is not far from achieving soundess of parsing
for core untyped regexes. If we deleted the `Group' constructor and
the condition for character predicates changed to matching exactly
one character, we believe we could implement unparsing for core
untyped regexes and prove soundness.  For implementations of certified
regex parsers that do not focus on performance, see
\cite{Firsov2013CertifiedLanguages},
\cite{Ribeiro2017CertifiedParsing}, and \cite{Idris_certified}.

\section{Conclusion and further work}\label{conclusion}
We described a typed regex parser library. Using dependent types we
take advantage of the information in regex patterns at compile time to
create additional safety layers, both facing users and at the level of
the state machine. Static analysis of regex literals to validate the
literal is well-formed and compute its result type has been recently a
subject of interests
\citep{scala-type-safe,java-regex-safe,oregano,typescript-typed-regex}. Often,
the safety checks are bolted onto an untyped parser. We maintain the
parse-tree shape as an intrinsic characteristic of the built parser.

Interestingly, such powerful type system characteristics as dependent
types are not necessary to provide all of these guarantees. Haskell's
Oregano parser \citep{oregano} monitors the result type. It seems
type-level computation, extensive preprocessing or macro capabilities,
or bespoke language support seem necessary to support regex literals
with static guarantees. Finally, as we traverse the NFA in TyRE, when
two threads are in the same state we merge them. This kind of analysis
uses a deep-embedding of the typed NFA, which is particularly
straightforward with dependent types. This optimisation is an
important performance improvement, that guarantees linear time parsing
for all patterns.

There are many possible extensions to this work.
First, we could prove soundness or even completeness of the
parser. We conjecture soundness to be straightforward to achive with
our architecture under some restriction. These would exclude
user-defined transformations, groups, and non-equality predicates on
characters. We would keep proofs in the parse tree that the character
satisfies the predicate. As a consequence, the parse tree itself
would almost be the soundness proof. We would only have to prove
additionally that the parse tree really represents the input
word. This fact should naturally follow from the Moore machine
construction. Achieving completeness would require much more
effort. Not only we would have to prove that if there is no accepting
path in our Moore machine, then the word doesn't match the pattern,
but also that getting rid of some threads as we do doesn't affect
completeness.

There's a useful consequence to proving soudness --- it provides an
unparsing feature for free. Unparsing is possible in Radanne's
design. However, we can achieve unparsing more directly.  One adds a
bi-directional conversion constructor for TyRE shape
transformations. Then, parse trees for TyREs without uni-directional
conversion support unparsing.

Finally, it could be interesing to integrate TyRE with Idris 2's
stdandard library parser combinators, both for tokenizing, and parsing,
or extending the family of languages that TyRE supports to
$\mu$-regular expressions~\citep{algebraic-parsing} or nested
words~\citep{Alur2009AddingWords}. These extensions would allow us for
bootstraping TyRE and use the integrated parser library or TyRE itself
for parsing regex literals.

\paragraph*{Acknowledgements.}
Supported by a Royal Society University Research Fellowship and
Enhancement Award, and an Alan Turing Institute seed-funding grant. We are grateful to
Guillaume X.~Allais,
Edwin C.~Brady,
Donovan Crichton,
Paul Jackson,
James McKinna,
Michel Steuwer, and
Jeremy Yallop
for helpful suggestions and fruitful discussions.

\bibliographystyle{agsm}
\bibliography{TyRE}
\label{lastpage01}
\end{document}